\def\ersteseite{1} 
\def\letzteseite{60}
\documentclass[editor]{artjlt-29-arx}
\usepackage{amsmath}
\usepackage{latexsym}
\usepackage{amssymb}
\usepackage{hyperref} 
\usepackage{cancel}

\font\tengoth=eufm10 at 10pt
\font\sevengoth=eufm7 at 6pt
\newfam\gothfam
\textfont\gothfam=\tengoth
\scriptfont\gothfam=\sevengoth

\newcommand{\g}{{\mathfrak g}}

\newcommand{\h}{{\mathfrak h}}

\newcommand{\fa}{{\mathfrak a}}
\newcommand{\fb}{{\mathfrak b}}
\newcommand{\fc}{{\mathfrak c}}

\newcommand{\fe}{{\mathfrak e}}
\newcommand{\ff}{{\mathfrak f}}
\newcommand{\fg}{{\mathfrak g}}
\newcommand{\fh}{{\mathfrak h}}

\newcommand{\fk}{{\mathfrak k}}
\newcommand{\fl}{{\mathfrak l}}

\newcommand{\fq}{{\mathfrak q}}
\newcommand{\fp}{{\mathfrak p}}

\newcommand{\fs}{{\mathfrak s}}
\newcommand{\ft}{{\mathfrak t}}
\newcommand{\fu}{{\mathfrak u}}

\newcommand{\fz}{{\mathfrak z}}

\renewcommand\sp{\mathfrak {sp}}

\renewcommand{\:}{\colon}
\newcommand{\1}{\mathbf{1}}

\newcommand{\cA}{\mathcal{A}}
\newcommand{\cB}{\mathcal{B}}
\newcommand{\cC}{\mathcal{C}}
\newcommand{\cD}{\mathcal{D}}
\newcommand{\cE}{\mathcal{E}}

\newcommand{\cH}{\mathcal{H}}

\newcommand{\cM}{\mathcal{M}}
\newcommand{\cN}{\mathcal{N}}
\newcommand{\cO}{\mathcal{O}}
\newcommand{\cP}{\mathcal{P}}

\newcommand{\cT}{\mathcal{T}}

\newcommand{\cW}{\mathcal{W}}

\newcommand\bx{{\bf{x}}}

\newcommand{\bO}{\mathbf{O}}

\newcommand{\eset}{\emptyset}

\newcommand{\trile}{\trianglelefteq}
\newcommand{\subeq}{\subseteq}
\newcommand{\supeq}{\supseteq}

\newcommand{\into}{\hookrightarrow}
\newcommand{\eps}{\varepsilon}

\newcommand{\shalf}{{\textstyle{\frac{1}{2}}}}

\def\onto{\to\mskip-14mu\to}

\newcommand{\Z}{{\mathbb Z}}
\newcommand{\R}{{\mathbb R}}
\newcommand{\C}{{\mathbb C}}

\renewcommand{\H}{{\mathbb H}}
\newcommand{\T}{{\mathbb T}}

\newcommand{\bH}{{\mathbb H}}
\newcommand{\bS}{{\mathbb S}}

\renewcommand{\hat}{\widehat}

\renewcommand{\tilde}{\widetilde}

\newcommand{\GL}{\mathop{{\rm GL}}\nolimits}
\newcommand{\SL}{\mathop{{\rm SL}}\nolimits}

\newcommand{\PSL}{\mathop{{\rm PSL}}\nolimits}
\newcommand{\SO}{\mathop{{\rm SO}}\nolimits}
\newcommand{\SU}{\mathop{{\rm SU}}\nolimits}
\newcommand{\OO}{\mathop{\rm O{}}\nolimits}

\newcommand{\U}{\mathop{\rm U{}}\nolimits}

\newcommand{\Sp}{\mathop{{\rm Sp}}\nolimits}
\newcommand{\Sym}{\mathop{{\rm Sym}}\nolimits}

\newcommand{\Skew}{\mathop{{\rm Skew}}\nolimits}

\newcommand{\gl}  {\mathop{{\mathfrak{gl} }}\nolimits}

\newcommand{\fsl} {\mathop{{\mathfrak{sl} }}\nolimits}
\newcommand{\fsp} {\mathop{{\mathfrak{sp} }}\nolimits}
\newcommand{\su}  {\mathop{{\mathfrak{su} }}\nolimits}
\newcommand{\so}  {\mathop{{\mathfrak{so} }}\nolimits}

\newcommand{\Exp}{\mathop{{\rm Exp}}\nolimits}

\newcommand{\ad}{\mathop{{\rm ad}}\nolimits}
\newcommand{\Ad}{\mathop{{\rm Ad}}\nolimits}

\newcommand{\tr}{\mathop{{\rm tr}}\nolimits}

\newcommand{\coker}{\mathop{{\rm coker}}\nolimits}

\newcommand{\pr}{\mathop{{\rm pr}}\nolimits}

\newcommand{\Herm}{\mathop{{\rm Herm}}\nolimits}
\newcommand{\Aherm}{\mathop{{\rm Aherm}}\nolimits}

\newcommand{\Aut}{\mathop{{\rm Aut}}\nolimits}

\newcommand{\diag}{\mathop{{\rm diag}}\nolimits}

\newcommand{\id}{\mathop{{\rm id}}\nolimits}
\newcommand{\rk}{\mathop{{\rm rank}}\nolimits}

\renewcommand{\dim}{\mathop{{\rm dim}}\nolimits}

\newcommand{\im}{\mathop{{\rm im}}\nolimits}

\newcommand{\Inn}{\mathop{{\rm Inn}}\nolimits}
\newcommand{\Int}{\mathop{{\rm int}}\nolimits}

\newcommand{\cone}{\mathop{{\rm cone}}\nolimits}
\newcommand{\conv}{\mathop{{\rm conv}}\nolimits}

\newcommand{\Spann}{\mathop{{\rm span}}\nolimits}

\newcommand{\dS}{\mathop{{\rm dS}}\nolimits}
\newcommand{\PSO}{\mathop{{\rm PSO}}\nolimits}

\newcommand{\Rarrow}{\Rightarrow}
\newcommand{\nin}{\noindent} 
\newcommand{\oline}{\overline}

\newcommand{\la}{\langle}
\newcommand{\ra}{\rangle}

\newcommand{\up}{\mathop{\uparrow}}

\newcommand{\res}{\vert}

\newcommand{\Spec}{{\rm Spec}}

\newcommand{\ssssarr}{\hbox to 15pt{\rightarrowfill}}
\newcommand{\sssarr}{\hbox to 20pt{\rightarrowfill}}
\newcommand{\ssarr}{\hbox to 30pt{\rightarrowfill}}
\newcommand{\sarr}{\hbox to 40pt{\rightarrowfill}}
\newcommand{\arr}{\hbox to 60pt{\rightarrowfill}}
\newcommand{\larr}{\hbox to 60pt{\leftarrowfill}}
\newcommand{\Arr}{\hbox to 80pt{\rightarrowfill}}

\newcommand{\ssmapright}[1]{\smash{\mathop{\ssarr}\limits^{#1}}}
\newcommand{\smapright}[1]{\smash{\mathop{\sarr}\limits^{#1}}}

\newcommand{\pmat}[1]{\begin{pmatrix} #1 \end{pmatrix}}

\renewcommand{\phi}{\varphi}

\usepackage{color}

\newcommand\be{{\bf{e}}}

\renewcommand\up{{\uparrow}}

\renewcommand{\rk}{\mathop{{\rm rk}}\nolimits}

\newcommand{\Cart}{\mathop{{\rm Cart}}\nolimits}

\newcommand{\PSU}{\mathop{{\rm PSU}}\nolimits}
\newcommand{\PdS}{\mathop{{\rm PdS}}\nolimits}

\renewcommand{\bO}{\mathbb O}

\renewcommand{\phi}{\varphi} 

\newcommand{\AdS}{\mathop{{\rm AdS}}\nolimits}

\title{From Euler elements and $3$-gradings to \\ 
non-compactly causal symmetric spaces} 
\author{Vincenzo Morinelli, Karl-Hermann Neeb, Gestur \'Olafsson} 
\lastname{Morinelli, Neeb, \'Olafsson}



  \address{Vincenzo Morinelli \\ 
Department of Mathematics \\
University of Rome Tor Vergata \\
Via della Ricerca Scientifica 1 \\
00133 Rome, Italy \\
   morinell@mat.uniroma2.it \\[3mm]
Karl-Hermann Neeb \\ 
Department Mathematik \\ 
Friedrich-Alexander-Universit\"at \\ 
Erlangen-N\"urnberg \\ 
Cauerstrasse 11 \\ 
91058 Erlangen, Germany \\ 
neeb@math.fau.de \\[3mm]
Gestur \'Olafsson \\ 
 Department of mathematics \\ 
Louisiana State University \\ 
Baton Rouge, LA 70803 \\ 
olafsson@math.lsu.edu
}
  
\begin{document}

\maketitle
\begin{center}
{\bf Dedicated to Karl Heinrich Hofmann\\ on the occasion 
of his 90th birthday} \\[5mm]
\end{center}

\begin{abstract}
  In this article we discuss the interplay between causal structures
  of symmetric spaces and geometric aspects
of Algebraic Quantum Field Theory (AQFT). The central focus is the set of Euler elements in a Lie algebra, i.e., elements whose adjoint action defines a
$3$-grading. In the first half of this article we survey the classification of reductive 
causal symmetric spaces from the perspective of Euler elements.
This point of view is motivated by recent
applications in AQFT.
In the second half we obtain several results that prepare
the exploration of the deeper connection between the structure of causal
symmetric spaces and AQFT. In particular, we explore the technique
of strongly orthogonal roots and corresponding systems
of $\fsl_2$-subalgebras. Furthermore, we exhibit real Matsuki crowns in the adjoint
orbits of Euler elements and we describe the group of connected components of the 
stabilizer group of Euler elements.
\end{abstract}

\tableofcontents

\section{Introduction} 
\label{sec:1}

From the group theoretic perspective,
symmetric spaces are quotients $M = G/H$, where
$G$ is a Lie group, $\tau$ is an involutive automorphism of $G$
and $H \subeq G^\tau$ is an open subgroup.
Symmetric spaces subsume quadrics on which pseudo-orthogonal
groups act and Lie groups $G$ on which the product group $G \times G$
acts by left and right translations. For an axiomatic
approach to symmetric spaces we refer to O.~Loos' monograph
\cite{Lo69}.

Causal symmetric spaces $G/H$ carry a $G$-invariant field
of pointed generating closed convex cones $C_m \subeq T_m(M)$
in their tangent spaces. They subsume time-orientable Lorentzian
symmetric spaces, but it is not required that the cones come from
an invariant Lorentzian metric. They permit to study causality
aspects of spacetimes in a highly symmetric environment.
On some of these spaces the causal curves define a global order structure
with compact intervals (they are called globally hyperbolic)
and in this context one can also prove the existence of a global
``time function'' with group theoretic methods (see \cite{Ne91}). 
We refer to the monograph \cite{HO97}
for more details and a complete exposition
of the classification of irreducible symmetric spaces.

Recent interest in causal symmetric spaces in relation with 
representation theory arose from their role as analogs of
spacetime manifolds in the context of
Algebraic Quantum Field Theory (AQFT) 
in the sense of Haag--Kastler, where one considers 
{\it nets} of von Neumann algebras $\cM(\cO)$
of operators on a fixed Hilbert space $\cH$,
associated to  regions $\cO$ in some spacetime manifold~$M$ 
(\cite{Ha96}). 
The hermitian elements of the algebra $\cM(\cO)$ represent
observables  that can be measured in the ``laboratory'' $\cO$.
One typically requires the following properties:
\begin{itemize}
\item[\rm(I)] Isotony: 
  $\cO_1 \subeq \cO_2$ implies $\cM(\cO_1) \subeq \cM(\cO_2)$ 
\item[\rm(L)] Locality: 
  $\cO_1 \subeq \cO_2'$ implies $\cM(\cO_1) \subeq \cM(\cO_2)'$,
  where $\cO'$ is the ``causal complement'' of $\cO$, i.e., the maximal
  open subset that cannot be connected to $\cO$ by causal curves.
\item[\rm(RS)] Reeh--Schlieder property: There exists a unit vector
  $\Omega\in \cH$ that is cyclic for $\cM(\cO)$ if $\cO \not=\eset$.
\item[\rm(Cov)] Covariance: 
  There is a Lie group $G$ acting on $M$ and a unitary representation
  $U \: G \to \U(\cH)$ such that 
  $U_g \cM(\cO) U_g^{-1} = \cM(g\cO)$ for $g \in G$. 
\item[\rm(BW)] Bisognano--Wichmann property: 
  $\Omega$ is separating for some ``wedge region'' $W \subeq M$
and there exists an element
$h \in \g$ with $\Delta^{-it/2\pi} = U(\exp th)$ for $t\in \R$,
where $\Delta$ is the modular operator
corresponding to $(\cM(W),\Omega)$
in the sense of the Tomita--Takesaki Theorem \cite[Thm.~2.5.14]{BR87}.
\item[\rm(Vac)] Invariance of vacuum: $U(g)\Omega = \Omega$   for every $g \in G$. 
\end{itemize}
The (BW) property gives a geometrical meaning to the dynamics
  provided by the  modular group $\Delta^{it}$ of von Neumann algebras
  associated to wedge regions with respect to the vacuum state.
  This also holds under the weaker hypothesis of the
  modular covariance property (MC)\footnote{$\Delta_{\cM(W),\Omega}^{-it}\cM(\cO)
    \Delta_{\cM(W),\Omega}^{it}=\cM(\exp(2\pi th)\cO)$, $\cO\subset M$.}. Under one of these assumptions, the relations among the modular groups of the wedge algebras can be used to reconstruct a positive energy representation
  of the Poincar\'e group,
  acting covariantly on the net of von Neumann algebra on Minkowski
  spacetime (\cite{GL95,Bo98}). In particular, one can
  start with a finite configuration of von Neumann algebras  with a cyclic and separating vector in some specific relative position to determine a large group of symmetries generated by their modular groups whose action on the family of von Neumann algebras is generating an AQFT on the  spacetime manifold.
  For instance, in the chiral theories, such configurations 
  are represented by half-sided modular inclusions
  (\cite{Wiesbrock93-1,AZ05}): An inclusion
  $\cA\subset \cB\subset\cB(\cH)$ of von Neumann algebras
  with a common cyclic and separating vector $\Omega\in\cH$,
  such that $\Delta^{-it}_{\cB,\Omega}\cA\subset\cA$ for all $t\geq0$.
  On Minkowski spacetime these structures   have
  been discussed by \cite{Wiesbrock98}. 
  The homogeneous spacetimes occurring naturally in AQFT
  are causal symmetric spaces associated to their symmetry groups 
    (Minkowski spacetime for the Poincar\'e group,
  de Sitter space for the Lorentz group and
  anti-de Sitter space for $\SO_{2,d}(\R)$) and the localization
  in wedge regions is ruled by the acting group.

In our abstract context a natural question is, given a symmetry group $G$,
to which extent such nets of von Neumann algebras exist on causal symmetric spaces.
For representations $(U,\cH)$ of $G$ for which the positive cone
\begin{equation}\label{eq:poscone}
 C_U := \{ x \in \g \: -i \cdot \partial U(x) \geq 0\} \end{equation}
spans $\g$ such nets can be constructed via Second Quantization
from nets of so-called standard subspaces. We refer
to \cite{NO21} for left invariant nets on reductive Lie groups,
to \cite{Oeh21} for left invariant nets on non-reductive Lie groups,
and to \cite{NO22a} for invariant nets on compactly causal symmetric spaces.
In all these constructions the non-triviality of the cone $C_U$ is a
crucial assumption, but this restricts the class of representations considerably.
These papers construct so-called
  one-particle nets on symmetric spaces from which nets of von Neumann
  algebras can be obtained by second quantization functors.
  An abstract description of wedge spaces is introduced 
  in \cite{MN21}.  Here the spectral condition 
  $C_U \not=\{0\}$ is only needed to encode non-trivial
  inclusions among the wedge regions.
For instance, $C_U = \{0\}$ for non-trivial representations 
of the Lorentz groups $\SO_{1,d}(\R)_e$ and in de Sitter space
there are no proper inclusions for wedge regions. On the other hand,
\cite{BM96} shows that covariant nets for the Lorentz group exist for de
Sitter space. Moreover, according to \cite{MN22} 
the potential generators $h \in \g$ of the modular groups
in (BW) are {\it Euler elements}, i.e.,
$\ad h$ defines a $3$-grading
\[ \g = \g_1(h) \oplus \g_0(h) \oplus \g_{-1}(h), 
\quad \mbox{ where } \quad \g_\lambda(h)
= \ker(\ad h - \lambda \1).\]
This leads to  the question how the existence and
  the choice of the Euler element affects the geometry of
  symmetric spaces.

The goal of this article is twofold.
First, we present an approach to reductive causal symmetric
spaces and their classification
from the perspective of Euler elements that should be accessible
to a large readership beyond the Lie group community
\break (Sections~\ref{sec:2}-\ref{sec:4}).
Second, we intend to lay the foundation for the exploration
of the deeper connection between the structure of causal
symmetric spaces and AQFT (Sections~\ref{sec:5}-\ref{sec:7}).
In particular, a better understanding of the
locality condition (L) and ``causal complements'' is under development;
see in particular \cite{MNO22a} and \cite{MNO22b} on
wedge regions in causal symmetric spaces. \\

We recall some basic terminology concerning symmetric spaces and
symmetric Lie algebras: 
\begin{itemize}
\item A  {\it symmetric Lie algebra}
is a pair $(\g,\tau)$, where $\g$ is a finite-dimensional real Lie algebra 
and $\tau$ is an involutive automorphism of~$\g$. 
We write 
\[ \g = \fh \oplus \fq \quad \mbox{ with } \quad 
\fh = \g^\tau= \ker(\tau -\1) \quad \mbox{ and } \quad 
\fq = \g^{-\tau}= \ker(\tau +\1).\] 
\item A {\it causal symmetric Lie algebra}
is a triple $(\g,\tau,C)$, where $(\g,\tau)$ is a symmetric Lie algebra 
and $C \subeq \fq$ is a pointed generating 
closed convex cone invariant under the group $\Inn_\g(\fh) := \la e^{\ad \fh}\ra
\subeq \Aut(\g)$. 
We call $(\g,\tau,C)$
\begin{itemize}
\item {\it compactly causal~(cc)} if 
$C$ is {\it elliptic} in the sense that, for $x \in C^\circ$
(the interior of $C$), the 
operator $\ad x$ is semisimple with purely imaginary spectrum. 
\item {\it non-compactly causal (ncc)} if 
$C$ is {\it hyperbolic} in the sense that, for $x \in C^\circ$, the 
operator $\ad x$ is diagonalizable. 
\end{itemize}
\end{itemize}

Let $(\g,\tau,C)$ be an ncc symmetric Lie algebra, 
$G$ a connected Lie group with Lie algebra $\g$, 
$\tau_G$~an involution on $G$ integrating the involution $\tau$ on 
$G$ and $H \subeq G^{\tau_G}$ be an open subgroup with $\Ad(H)C = C$. 
Then we obtain the structure of a  {\it causal symmetric space} 
on $M = G/H$, specified by the $G$-invariant field of open convex cones 
$g.C^\circ \subeq T_{gH}(M), g \in G$. 

If $G$ is semisimple with finite center,
then, for compactly causal spaces, there are closed causal curves,
so that no global causal order exists on $M$,
but non-compactly causal spaces carry a 
global order which is {\it globally hyperbolic} in the sense that
all order intervals are compact (\cite[Thm.~5.3.5]{HO97}).
In general the semi-Riemannian
  metric on semisimple causal symmetric spaces is not 
  Lorentzian and the cone $C$ may not have a smooth boundary away from~$0$.

  The two main examples of non-flat Lorentzian
  symmetric spaces are de Sitter space and anti de Sitter space.
    {\it De Sitter space}
\begin{equation}
  \label{eq:desitter1}
  \dS^d := \{ (x_0,x_1, \ldots, x_d) \in \R^{1,d} \: x_0^2
  -  x_1^2 - \cdots - x_d^2 = - 1\} 
\end{equation}
is a non-compactly causal irreducible symmetric space with 
$G = \SO_{1,d}(\R)_e$, $H = G_{\be_1} = \SO_{1,d-1}(\R)_e$, and 
$C \subeq T_{\be_1}(\dS^d) \cong  \be_1^\bot$ given by 
\[ C = \{  (x_0,0, x_2, \ldots, x_d) \: x_1 = 0, x_0 \geq 0, x_0^2 \geq
x_2^2 + \cdots + x_d^2\}.\]
Likewise {\it anti-de Sitter space} is the compactly causal irreducible
symmetric space 
\begin{equation}
  \label{eq:adesitter1}
  \AdS^d := \{ (x_0,x_1, \ldots, x_d)
  \in \R^{2,d-1} \: x_0^2 + x_1^2 - x_2^2 - \cdots
  - x_d^2= 1\} 
\end{equation}
with 
$G = \SO_{2,d-1}(\R)_e$, $H = G_{\be_1} \cong \SO_{1,d-1}(\R)_e$, and 
$C \subeq T_{\be_1}(\AdS^d) \cong  \be_1^\bot$ given by 
\[ C = \{  (x_0,0, x_2, \ldots, x_d)\: x_0 \geq 0, x_0^2 \geq x_2^2 + \cdots
+ x_d^2\}.\]
  In Appendix~\ref{subsec:lorentz} we recall the well-known result 
  that any {\bf irreducible} Lorentzian $d$-dimensional causal
  symmetric space  
  is either locally isomorphic to anti-de Sitter space 
  (if compactly causal) 
  or to de Sitter space (if non-compactly causal).
  However, there are many {\bf reducible}
  Lorentzian symmetric spaces (see Appendix~\ref{subsec:lorentz}).

The {\bf contents of this paper} is as follows: 
We start in Section~\ref{sec:2} by introducing
Euler elements and their classification, as presented in \cite{MN21}.
Since Euler elements correspond to $3$-gradings of
Lie algebras, their classification is also contained
implicitly in the work of S.~Kaneyuki
(cf.~\cite[p.~600]{Kan98} or \cite{Kan00}). 

In Section~\ref{sec:3} we explore the close relation between
Euler elements, H-elements (\cite{Sa80}) and invariant cones.
Recall that an element $z$ of a reductive Lie algebra $\g$ 
is called an {\it H-element} if 
$\ker(\ad z)$ is maximal compactly embedded in $\g$ and $iz \in \g_\C$ is an 
Euler element. 
In Section~\ref{subsec:3.1} we recall the classification
of simple hermitian Lie algebras in terms of Euler elements
of their complexification. 
Using maximal sets of strongly orthogonal
roots, we explore in Section~\ref{subsec:3.2} how the symmetry of an Euler element
is related to the projection of the root system 
to the subspace spanned by the strongly orthogonal
roots: These projections are either of type $C_r$ or $BC_r$, and
the first case occurs if and only if $h$ is symmetric, i.e.,
to $-h \in \Inn(\g)h$. 
This implies in particular that hermitian
Lie algebras of tube type correspond to symmetric Euler elements
in complex simple Lie algebras. This
is closely related to their characterization in terms of the existence 
of a ``morphism of hermitian Lie algebras'' $\fsl_2(\R) \to \g$ whose
range contains $h$ (\cite[Cor.~III.1.6]{Sa80}). 
We  recall in Section~\ref{subsec:3.3} how H-elements
are related to Cartan involutions
and in Section~\ref{subsec:3.4} we discuss the duality between 
H-elements and Euler elements. 
We conclude Section~\ref{sec:3} 
with a review concerning
invariant cones in hermitian Lie algebras
and irreducible symmetric spaces (Section~\ref{subsec:3.5}). 

In Section~\ref{sec:4} we present a classification 
of irreducible non-compactly causal symmetric spaces based
on Euler elements. Here a key concept is that of a {\bf causal Euler element} 
that is explored in Section~\ref{subsec:4.1}: 
If $(\g,\tau)$ is a symmetric Lie algebra,
an Euler element $h \in \fq$ is said to be causal 
if it is contained in the interior of an $\Inn_\g(\fh)$-invariant
pointed convex cone in $\fq$. Theorem~\ref{thm:2.3} asserts that, for 
any pair $(\theta, h)$ of a Cartan involution $\theta$ and an Euler
element satisfying $\theta(h) = -h$,
the involution $\tau = \tau_h \theta$ with
$\tau_h = e^{\pi i \ad h}$ makes $h$ causal for $(\g,\tau)$.
In Section~\ref{subsec:4.3} we use this construction to classify
irreducible non-compactly causal symmetric Lie algebras in
terms of $\Inn(\g)$-orbits of Euler elements (Theorem~\ref{thm:classif}).
From the dual perspective, focusing on H-elements, this classification
goes back to \cite{Ol91}, which ties in naturally with
\cite[Prop.~II.3]{HNO94}, where it is shown that the 
maximal  generating invariant cones in $\g$ 
(and by duality the minimal ones) are parametrized by 
adjoint orbits of H-elements.

In general there are many locally isomorphic causal symmetric spaces
$G/H$ corresponding to the same triple $(\g,\tau,C)$. The maximal one
is the universal covering space of $\Inn (\g ) /\Inn (\g)^\tau$, but
the ``minimal model'' is not as obvious.  
In Section~\ref{subsec:4.2} we show that the minimal
symmetric space $\Inn(\g)/\Inn(\g)^\tau$
for the irreducible  symmetric Lie algebra $(\g,\tau)$ is this
minimal model for $(\g,\tau,C)$
if and only if the corresponding causal Euler element is not symmetric;
otherwise we have to pass to a two-fold covering. This can
  be understood in terms of wedge regions in non-compactly causal symmetric spaces
  as described in \cite[Sect.~1]{NO22b}.  The wedge region associated to
  an Euler element can be defined 
  as the open subset of the causal space where the generator of the
  modular flow takes values in the open cones. 
  This generalizes the fact that the flux of a uniformly accelerated observer in the Rindler wedge is indeed timelike. 
  If $h$ is a symmetric Euler element
  ($h$ and $-h$ belong to the same orbit), then, on the
    \textit{minimal symmetric space }$M=\Inn(\g)/\Inn(\g)^\tau$
    we cannot distinguish between $C$ and $-C$. Hence the causality of the manifold $M$ is lost. In order to preserve causality one has to consider the \textit{minimal causal symmetric space} $\Inn(\g)/H_C$, where $H_C\subeq \Inn(\g)^\tau$ is
  the subgroup  of those elements $g$ satisfying $gC=C$.
  In this case causally complementary
  wedge regions correspond to opposite Euler elements:   if $h$ determines
  the wedge region $W_h\subset M$, then $-h$ determine $W_h'\subset M$, where
  the prime refers to the locality condition (L). A prominent example is
  de Sitter spacetime which is discussed in Remark \ref{rmk:dessym}.

%

In Section~\ref{subsec:4.4} we present a structured table with the 
classification of the irreducible non-compactly causal symmetric
Lie algebras from the perspective of causal Euler elements. Together with Theorem \ref{thm:classif} it provides a complete classification of the local structure of non-compactly causal symmetric spaces given by
  non-compactly causal simple symmetric Lie algebras.

In Section~\ref{sec:5} we introduce a key technical tool in the structure theory
of causal symmetric spaces: maximal $\tau$-invariant 
sets of strongly orthogonal long roots.
In Section~\ref{subsec:5.1} we recall the construction of such sets
from \cite{Ol91} and discuss its basic properties.
In particular we connected Euler elements with
strongly orthogonal roots and the corresponding $\fsl_2$-subalgebras
(Proposition~\ref{prop:testing}).
An important application of this technique is Theorem~\ref{thm:3.4}
that characterizes irreducible non-compactly causal symmetric Lie algebras
$(\g,\tau,C)$ for which $\fh=\g^\tau$ contains Euler elements as those
for which the causal Euler element is symmetric.
Quadruples $(\g,\tau,C,h')$ with
an Euler element $h'\in \fh$  are called {\it modular
  causal symmetric Lie algebras}.
For these symmetric Lie algebras 
wedge regions in $G/H$ corresponding to $h'$
are studied in \cite{NO22a, NO22b}.
In this context there are several equivalent ways to define wedge regions
  (cf.~\cite[Thm.~7.1]{NO22a}, \cite[\S\S 6,7]{NO22b}).
  In particular, a characterization by a KMS like condition
  implies that the Euler conjugation $\tau_{h'}$
  implements a ``wedge reflection''.
In a minimal irreducible causal symmetric space,
the  existence of a complementary wedge region,
with respect to the causal structure, also
corresponds to the property of the Euler element associated to the wedge to be symmetric (cf.\ \cite{MN21} and \cite{MNO22a}).


In Section~\ref{sec:6} we extend some of the results obtained 
in \cite{NO22b} for the special class of modular 
ncc spaces to general
semisimple non-compactly causal symmetric spaces. 
Our main result is that, if the cone $C$ is maximal 
 and $(\g,\tau)$ is semisimple without Riemannian 
 ideals, then the connected component  of $h$ in the intersection of the 
adjoint orbit $\cO_h = \Inn(\g)h$ with the 
real tube domain $\cT_{C} = \fh + C^\circ \subeq \g$ 
is the Matsuki crown of the Riemannian symmetric space 
$\cO_h^\fq := \Inn_\g(\fh)h \cong e^{\ad \fh_p}h$, i.e., 
\[ (\cT_C \cap \cO_h)_h
  = \Inn_\g(\fh)e^{\ad \Omega_{\fq_\fk}}h \subset \cO_h  \quad \mbox{ for } \quad 
\Omega_{\fq_\fk} = \Big\{ x \in \fq_\fk \: \rho(\ad x) < \frac{\pi}{2}\Big\}.\] 
Here $\rho(\ad x)$ is the spectral radius of $\ad x$ 
(Theorem~\ref{thm:crownchar-gen}) and $\fq_\fk = \fq \cap \fk$ for 
a Cartan decomposition $\g = \fk \oplus \fp$ with~$h \in \fp$. 
In \cite{MNO22a} we actually show that $\cE(\g)  \cap \cT_C$ 
is connected.

Section~\ref{sec:7} is devoted to an analysis of the group
$\pi_0(G^h)$ of connected components of the centralizer $G^h$
of an Euler element~$h$ in a simple real Lie algebra.
By the polar decomposition
$G^h = K^h \exp(\fh_\fp)$, this group equals $\pi_0(K^h)$.
As $K/K^h \cong \cO_h^K := \Ad(K)h$ is a compact symmetric space, we discuss
this problem in Section~\ref{subsec:7.1} in the context of
compact symmetric spaces, where
$\pi_0(K^h)$ appears as a quotient of $\pi_1(K/K^h)$
in the long exact homotopy sequence
\[ \pi_1(K) \to \pi_1(K/K^h) \onto \pi_0(K^h) \to \pi_0(K) = \1. \] 

In Section~\ref{subsec:7.2} we explore this situation further,
using that $\cO_h^K$ actually is a symmetric R-space (cf.\ \cite{Lo85}). Here
the strongly orthogonal roots come in handy and permit us
 to show that
$\Ad(G)^h$ is connected if $(\g,\tau)$ is either of complex type or
non-split type (cf.\ Section~\ref{subsec:4.4}),
and if it is of split type or Cayley type, then
it either is trivial or $\Z_2$
(see Theorem~\ref{thm:7.8} for details). In particular $\Ad(G)^h$
has at most two connected components.
In Section~\ref{subsec:7.3} we finally collect some consequences
of this result such as the identity 
\[ \Inn_{\g_\C}(\g^c) \cap \Inn_{\g_\C}(\g)  = \Inn_{\g_\C}(\fk)^h e^{\ad \fh_\fp}\]
for the c-dual Lie algebra $\g^c := \fh + i \fq$. 

We conclude this paper with three short appendices containing
some calculations in $\fsl_2(\R)$ (Appendix~\ref{subsec:sl2b}),
some general facts on invariant cones and their extensions
(Appendix~\ref{app:a}) and on Lorentzian symmetric spaces
(Appendix~\ref{subsec:lorentz}). \\

\nin {\bf Notation:} 
\begin{itemize}
\item We write $e \in G$ for the identity element in the Lie group~$G$ 
and $G_e$ for its identity component. 
\item For $x \in \g$, we write $G^x := \{ g \in G \: \Ad(g)x = x \}$ 
for the stabilizer of $x$ in the adjoint representation 
and $G^x_e = (G^x)_e$ for its identity component. 
\item For $h \in \g$ and $\lambda \in \R$, we write 
$\g_\lambda(h) := \ker(\ad h - \lambda \1)$ for the corresponding eigenspace 
in the adjoint representation.
\item If $\g$ is a Lie algebra, we write $\cE(\g)$ for the set of 
{\it Euler elements} $h \in \g$, i.e., $\ad h$ is non-zero and diagonalizable 
with $\Spec(\ad h) \subeq \{-1,0,1\}$. We write $\cO_h = \Inn(\g)h$ for the
adjoint orbit of $h$ and call it {\it symmetric} if
$-h \in\cO_h$.
\item For a Lie subalgebra $\fs \subeq \g$, we write 
$\Inn_\g(\fs)= \la e^{\ad \fs} \ra \subeq \Aut(\g)$ for the subgroup 
  generated by $e^{\ad \fs}$. We call $\fs$ {\it compactly embedded} if
  the group $\Inn_\g(\fs)$ has compact closure. 
\item For a convex cone $C$ in a vector space $V$, we write
  $C^\circ := \Int_{C-C}(C)$ for the relative interior of $C$ in its span.
\item For a symmetric space $M = G/H$ we write $\Exp \: T(M) \to M$
  for the exponential function which, in the base point $eH$ takes
  the form $\Exp_{eH}(x) = \exp x H$ if we identity $\fq$ with $T_{eH}(M)$. 
\end{itemize}

\section{The classification of Euler elements} 
\label{sec:2}

In this section we introduce Euler elements and their
classification, as presented in \cite{MN21}.
Since Euler elements correspond to $3$-gradings of
Lie algebras, their classification is also contained
implicitly in the work of S.~Kaneyuki
(cf.~\cite[p.~600]{Kan98} or \cite{Kan00}).

\begin{Definition} \label{def:euler}
We call an element $h$ of the finite dimensional 
real Lie algebra $\g$ an
{\it Euler element} if $\ad h$ is non-zero and diagonalizable with 
$\Spec(\ad h) \subeq \{-1,0,1\}$. Then the eigenspace 
decomposition of $\g$ with respect to $\ad h$ defines a $3$-grading 
of~$\g$: 
\[ \g = \g_1(h) \oplus \g_0(h) \oplus \g_{-1}(h), \quad \mbox{ where } \quad 
\g_\nu(h) = \ker(\ad h - \nu \id_\g)\] 
Then $\tau_h(y_j) = (-1)^j y_j$ for $y_j \in \g_j(h)$ 
defines an involutive automorphism of $\g$ that can also be written
as $e^{\pi i \ad h}$ such that
\[\g^{\tau_h} = \g_0(h)\quad\text{and}\quad \g^{-\tau_h} = \g_1 (h) \oplus \g_{-1}(h).\]

We write $\cE(\g)$ for the set of Euler elements in~$\g$. 
The orbit of an Euler element  $h$ under the group 
$\Inn(\g) = \la e^{\ad \g} \ra$ of 
{\it inner automorphisms} 
is denoted with $\cO_h = \Inn(\g)h \subeq \g$.
We say that $h$ is {\it symmetric} if $-h \in \cO_h$. 
\end{Definition}

\begin{Definition} Let $\theta$ be a Cartan involution of the semisimple
  Lie algebra $\g$ and $\fa \subeq \fp$ maximal abelian,
  so that we obtain the restricted root system
  $\Sigma := \Sigma(\g,\fa)$. Then the Cartan--Killing form
  $\kappa(x,y) = \tr(\ad x \ad y)$ restricts to a scalar product
  on~$\fa$. 
  For $\alpha \in \Sigma$, we define the {\it coroot}
  $\alpha^\vee \in \fa$ as the unique element which is orthogonal to
  $\ker \alpha$ and satisfies
  \[ \alpha(\alpha^\vee) = 2.\]
  Then the corresponding {\it reflection} is given on $\fa$ by
  \[  s_\alpha \: \fa\to \fa, \quad s_\alpha(x) = x - \alpha(x) \alpha^\vee,\]
  and on $\fa^*$ by 
  \[  s_\alpha \: \fa^*\to \fa^*, \quad s_\alpha(\beta) =
  \beta  - \beta(\alpha^\vee) \alpha.\]
  The Weyl group $\cW := \cW(\g,\fa)$ is the finite subgroup
  of $\GL(\fa)$, generated by these reflections.

  If $\g$ is simple, we call a root {\it long} if its length
  is maximal (\cite[\S 2.9]{Hu90}). 
  \end{Definition}

\begin{Remark}   An element $h \in \fa$ is an Euler element
if and only if it represents a so-called {\it minuscule weight}
of the root system $\Sigma^\vee = \{ \alpha^\vee \: \alpha \in \Sigma\}$
dual to $\Sigma = \Sigma(\g,\fa)$.
This follows from the
remark at the end of \S 7, no.~3 in \cite[Ch.~8]{Bo90b},
that also provides an interesting relation with
the corresponding affine root system 
(see also Remark~\ref{rem:2.17}). 
\end{Remark}

If $h\not= 0$ is hyperbolic, then there exists a maximal abelian hyperbolic subspace $\fa$ containing $h$.
Then $h$ is an Euler element if and only if 
\[ \Sigma(\g,\fa)(h) \subeq \{-1,0,1\}\]
where $\Sigma (\g,\fa)$ is the system of restricted roots of $\fa$ in $\g$. 
Therefore Euler elements are easy to detect in terms of
the structure of $\Sigma(\g,\fa)$.
For a given set $\{\alpha_1, \ldots, \alpha_r\}$ of simple roots
contained in the half-space $\{\lambda \in \fa^*\: \lambda (h)\ge 0\}$
and with the highest root $\alpha = \sum_{j = 1}^r n_j \alpha_j$, 
the different $3$-gradings correspond to those indices $j$
with $n_j = 1$.

The following theorem describes the conjugacy classes
of Euler elements in simple real Lie algebras.

\begin{theorem} \label{thm:classif-symeuler} {\rm(\cite[Thm.~3.10]{MN21})}
Suppose that $\g$ is a non-compact simple 
real Lie algebra and that  $\fa \subeq \g$ is maximal $\ad$-diagonalizable 
with restricted root system 
$\Sigma = \Sigma(\g,\fa) \subeq \fa^*$ of type $X_n$. 
We follow the conventions of the tables in {\rm\cite{Bo90a}}
for the classification of irreducible root systems and the enumeration 
of the simple roots $\alpha_1, \ldots, \alpha_n$. 
For each $j \in \{1,\ldots, n\}$, we consider the uniquely determined element 
$h_j \in \fa$ satisfying $\alpha_k(h_j) =\delta_{jk}$.
Then every Euler element in $\g$ is conjugate under inner automorphism 
to exactly one~$h_j$. For every irreducible root system, 
the Euler elements among the $h_j$ are  the following: 
\begin{align} 
&A_n: h_1, \ldots, h_n, \quad 
\ \ B_n: h_1, \quad 
\ \ C_n: h_n, \quad \ \ \ D_n: h_1, h_{n-1}, h_n, \quad 
E_6: h_1, h_6, \quad 
E_7: h_7.\label{eq:eulelts2}
\end{align}
For the root systems $BC_n$, $E_8$, $F_4$ and $G_2$ no Euler element exists 
(they have no $3$-grading). 
The symmetric Euler elements are 
\begin{equation}
  \label{eq:symmeuler}
A_{2n-1}: h_n, \qquad 
B_n: h_1, \qquad C_n: h_n, \qquad 
D_n: h_1, \qquad 
D_{2n}: h_{2n-1},h_{2n}, \qquad 
E_7: h_7.  
\end{equation}
\end{theorem}

\section{Euler elements, H-elements and invariant cones}
\label{sec:3}

In this section we turn to the close relation between
Euler elements, H-elements (\cite{HNO94, Sa80}) and invariant cones.
In Section~\ref{subsec:3.1} we recall the classification
of simple hermitian Lie algebras in terms of Euler elements. 
Using arguments involving maximal sets of strongly orthogonal
roots, we explore in Section~\ref{subsec:3.2} how the symmetry of an Euler element
is related to the projection of the root system 
to the subspace spanned by the strongly orthogonal
roots: These projections are either of type $C_r$ or $BC_r$, and
the first case is equivalent to $h$ being symmetric. 
This implies in particular that hermitian
Lie algebras of tube type correspond to symmetric Euler elements
in complex simple Lie algebras.
We  recall in Section~\ref{subsec:3.3} how H-elements
are related to Cartan involutions
and in Section~\ref{subsec:3.4} the duality between 
H-elements and Euler elements. 
We conclude this section with a review on 
invariant cones in hermitian Lie algebras
and irreducible symmetric Lie algebras  in Section~\ref{subsec:3.5}.

\subsection{Hermitian real forms of complex simple Lie algebras}
\label{subsec:3.1}

\begin{Definition} 
Let $\g$ be a reductive Lie algebra and let
$z \in \g$ be an Element for which $iz$ is an Euler element of $\g_\C$.
Then $\theta_z := e^{\pi \ad z}$ is an involution of $\g$ whose fixed point 
set is $\ker(\ad z)$. We call $z$  an {\it H-element} if 
$\ker(\ad z)$ is maximal compactly embedded, i.e., if 
$\theta_z$ restricts to a Cartan involution on the commutator algebra $[\g,\g]$
of~$\g$.

We call a pair
$(\g,z)$, where $\g$ is a reductive Lie algebra and 
$z \in \g$ an H-element a {\it reductive Lie algebra of hermitian type}
(\cite{Sa80}). A reductive Lie algebra containing an H-element
is called {\it quasihermitian}.
\end{Definition}

\begin{Remark}\label{rem:herm}
  If $\g$ is a simple non-compact real Lie algebra,
  then an H-element exists if and only if $\g$ is {\it hermitian}, 
  i.e., for any Cartan decomposition $\g = \fk \oplus \fp$, the center
  $\fz(\fk)$ of $\fk$ is non-zero. Then $\fz(\fk) = \R z$ for an H-element~$z$.
  In particular $iz$ is an Euler element in $\g_\C$ and, for the conjugation 
  $\sigma$ with respect to the real form $\g$, the involution
    $\theta = \sigma \tau_{iz} $ is a  Cartan involution of $\g_\C$
  with $\theta (iz) = -iz$.

If, conversely, $\g$ is a complex simple Lie algebra and
  $h \in \g$ an Euler element, then there exists a Cartan involution
  $\theta$ of $\g$ (automatically antilinear)
  with $\theta(h) = -h$, and then $\sigma := \theta \tau_h$
  defines the real form $\fh := \g^\sigma$ with $\fh_\C \cong \g$. Then
  $\sigma(h) = -h$ implies that $z := ih \in  \fh$ is an H-element, and thus
  $\fh$ is hermitian. By this construction,
  the Euler elements in the root systems
  of simple complex Lie algebras, listed in \eqref{eq:eulelts2}, specify
  the simple hermitian Lie algebras as the associated real forms:\\


  \begin{tabular}{||l|l|l|l|l||}\hline
{} $\fh$ \mbox{(hermitian)} & $\Sigma(\fh, \fa_\fh)$  & $\g = \fh_\C$ & $\Sigma(\g,\fa)$ & 
{\rm Euler elt.} \\ 
\hline\hline 
$\su_{p,q}(\C), 1 \leq p\not=q$ & $BC_p\ (p < q)$ & $\fsl_{p+q}(\C)$ & $A_{p+q-1}$ & 
$h_p$ \\ 
 &  $C_p\ (p=q)$ &  &  &  \\ 
$\sp_{2n}(\R)$ & $C_n$ & $\fsp_{2n}(\C)$ & $C_n$ & $h_n$ \\ 
 $\so_{2,n}(\R), n > 2$ & $C_2$ & $\so_{2+n}(\C)$ & $B_m\ (n = 2m-1)$ & $h_1$ \\ 
  & &  & $D_m\ (n = 2m-2)$ &  \\ 
 $\so^*(2n)$ & $BC_m\ (n = 2m+1)$ & $\so_{2n}(\C)$  & $D_{n}$  & $h_{n-1}, h_n$ \\
    & $C_m\ (n = 2m)$ &&&\\
$\fe_{6(-14)}$ & $BC_2$ & $\fe_6$ & $E_6$ & $h_1, h_6$ \\  
$\fe_{7(-25)}$ & $C_3$ & $\fe_7$ & $E_7$ & $h_7$ \\ 
\hline
  \end{tabular} \\[2mm] {\rm Table 1: Simple hermitian Lie algebras $\fh$
  and corresponding Euler elements.}\\

\nin Note that $\fsl_2(\R) \cong \so_{2,1}(\R) \cong \su_{1,1}(\C)$. More exceptional isomorphisms 
are discussed in \cite[\S 17]{HN12}. 
We recall that the hermitian real form $\so^*(2n)$ of $\so_{2n}(\C)$ is given by 
\[ \so^*(2n) =\Big\{ \pmat{ a & b \\ -\oline b & \oline a}  \in
\gl_{2n}(\C) \: a^* = -a, b^* = b\Big\}.\]
\end{Remark}

For hermitian Lie algebras $\fh$, 
the restricted root system $\Sigma = \Sigma(\fh,\fa_\fh)$ 
is either of type $C_r$ or $BC_r$ (cf.\ 
\cite[Lemmas~13-16]{HC56}, \cite[Thm.~2]{Mo64}, 
\cite[Thm.~XII.1.14]{Ne00} or Proposition~\ref{prop:hc} below).
We say that $\g$ is 
{\it of tube type} if the restricted root system is of type $C_r$
(cf.\ \cite{KW65}). 
A comparison of Table~$1$ with the list \eqref{eq:symmeuler} reveals
that simple hermitian Lie algebras of
tube type $\fh$ correspond to symmetric Euler elements in~$\fh_\C$.
Since the Euler elements $h_{n-1}$ and $h_n$ for the root system of type 
$D_n$ are conjugate under a diagram automorphism, they correspond to 
isomorphic hermitian real forms. These Euler elements are symmetric if and only if
$n$ is even (Theorem~\ref{thm:classif-symeuler}). 
\\[2mm]

\begin{tabular}{||l|l|l|l|l||}\hline
{} $\fh$ \mbox{(hermitian)}  & $\Sigma(\fh, \fa_\fh)$ & $\g = \fh_\C$ & $\Sigma(\g,\fa)$ & {\rm symm.\ Euler element} \\ 
\hline\hline 
$\su_{n,n}(\C)$ & $C_n$ & $\fsl_{2n}(\C)$ & $A_{2n-1}$ & $h_n$ \\ 
$\sp_{2n}(\R)$ & $C_n$ & $\fsp_{2n}(\C)$ & $C_n$ & $h_n$ \\ 
 $\so_{2,n}(\R), n > 2$ & $C_2$ & $\so_{2+n}(\C)$ & $B_m\ (n = 2m-1)$ & $h_1$ \\ 
  & &  & $D_m\ (n = 2m-2)$ &  \\ 
 $\so^*(4n)$ & $C_n$ & $\so_{4n}(\C)$  & $D_{2n}$ & $h_{2n-1},  h_{2n}$ \\ 
$\fe_{7(-25)}$ & $C_3$ & $\fe_7$ & $E_7$ & $h_7$ \\ 
\hline
\end{tabular} \\[2mm] {\rm Table 2: Simple hermitian Lie algebras $\fh$ 
of tube type}\\

\subsection{Strongly orthogonal roots and symmetric
  Euler elements}
\label{subsec:3.2}

In this subsection we consider maximal systems of long strongly
orthogonal roots in $\Sigma_1$, where $\Sigma$ is a $3$-graded
irreducible root system. As we shall see in
Section~\ref{subsec:5.1}, strongly orthogonal
roots provide a powerful tool to reduce problems
on groups and symmetric spaces to the Lie algebra~$\fsl_2(\R)$. 

\begin{Proposition} {\rm(Harish--Chandra)} \label{prop:hc}
  Suppose that $\g$ is simple with restricted root system
  ${\Sigma = \Sigma(\g,\fa)}$.
  Let $h \in \fa$ be an Euler element and consider the corresponding
  $3$-grading of the restricted root system
  $\Sigma = \Sigma(\g,\fa)$, defined by
\[ \Sigma_j := \{ \alpha \in \Sigma \: \alpha(h) = j\}.\] 
  Let
  \[ \Gamma = \{ \gamma_1, \ldots, \gamma_r \} \subeq \Sigma_1 \]
  be a maximal set of long
  strongly orthogonal roots (the sums and differences are no roots)
  and $\Sigma^+ \subeq \Sigma$ a positive system containing $\Sigma_1$.
  The orthogonal projection $\pr_{\Gamma} $ from $\fa$ to the span of $\{\gamma_1,\ldots , \gamma_r\}$
  is given by  
\[  \pr_\Gamma(\alpha) :=\sum_{j=1}^r \frac{\alpha(\gamma_j^\vee)}{2} \gamma_j,\quad \alpha \in \fa^*.\]
We put
$\cC_0  := \Sigma_0 \cap\pr_\Gamma^{-1}(0)$ and consider the subsets  
\[ \cC_j := \Sigma_0 \cap \pr_\Gamma^{-1}\Big(\frac{\gamma_j}{2}\Big), \quad
\cP_j :=\Sigma_1 \cap \pr_\Gamma^{-1}\Big(\frac{\gamma_j}{2}\Big),
\quad \mbox{ for }\quad j = 1,\ldots, r, \]
and 
\[  \cC_{jk} := \Sigma_0 \cap
\pr_\Gamma^{-1}\Big(\frac{\gamma_j -\gamma_k}{2}\Big), \quad
\cP_{jk} := \Sigma_1 \cap
\pr_\Gamma^{-1}\Big(\frac{\gamma_j +\gamma_k}{2}\Big)
\quad \mbox{ for } \quad j < k. \] 
Then the following assertions hold:
\begin{equation}
  \label{eq:hc1}
 \Sigma_1 = \Gamma \cup \bigcup_j \cP_j \cup\bigcup_{j < k} \cP_{jk} 
\quad \mbox{ and } \quad 
\Sigma_0^+  \subeq \cC_0 \cup \bigcup_j \cC_j \cup\bigcup_{j < k}
\cC_{jk}
\end{equation}
where all the unions are disjoint.
Further, $\pr_\Gamma(\Sigma)\setminus \{0\}$ either is a root system
of type $BC_r$ or of type $C_r$.
\end{Proposition}

\begin{Proof}   Since every $3$-graded root system 
arises in the complexification of a hermitian simple 
Lie algebra $\g$ (cf.~Theorem~\ref{thm:classif-symeuler}) \eqref{eq:hc1} 
can be derived from \cite[Lemmas~13-16]{HC56}
or \cite[\S 2]{Mo64}.

However, we have to show that our choice of $\Gamma$
is equivalent to the construction of the set
$\Gamma'$ of strongly orthogonal roots in \cite{HC56}, which
proceeds as follows:
Let $\gamma_1'$ be the maximal root (with respect to $\Sigma^+$),
and choose $\gamma_j'$ maximal in $\Sigma_1$ orthogonal to 
$\gamma_1', \ldots, \gamma_{j-1}'$.

The set $\Sigma_1$ is the weight set of the irreducible
representation of the Lie subalgebra $\g_{\C,0}(h)$ on
$\g_{\C,1}(h)$. The long roots in $\Sigma_1$ are the
extremal weights of this representation. Therefore 
the Weyl group  $\cW_0$ of the subsystem $\Sigma_0$
acts transitively on this set.
Harish--Chandra's results (\cite[Lemmas~13-16]{HC56}) show in particular that, for
$r > 1$, the root system
$(\gamma_1')^\bot$ is irreducible. Here the main point is that,
for any $\alpha \in \cC_0$, there exists a $\beta \in \Sigma_1$
with $\alpha + \beta \in \Sigma_1$. This implies the existence
of $\beta' \in \Sigma_1 \setminus \alpha^\bot$.

Now we show that Harish--Chandra's set $\Gamma'$ of strongly
orthogonal roots is conjugate under $\cW_0$ to $\Gamma$. 
As all long roots in $\Sigma_1$ are conjugate under the Weyl group
$\cW_0$, 
we may assume that $\gamma_1 = \gamma_1' \in \Gamma$.
Now one proceeds inductively to see that in the
irreducible root system $\gamma_1^\bot \cap \Sigma$, the roots $\gamma_2$
and $\gamma_2'$ are conjugate under the Weyl group~$\cW_0$.
We conclude that
$\Gamma$ is conjugate under the Weyl group $\cW_0$ to $\Gamma'$. 

The set $\pr_\Gamma(\Sigma) \setminus \{0\}$
is a root system of type $BC_r$ if all sets
$\cC_j, \cP_j, \cC_{jk}, \cP_{jk}$ are non-empty,
and of type $C_r$ if the sets $\cC_j$ and $\cP_j$ are empty
(\cite[\S 2]{Mo64}).
\end{Proof}

\begin{Example} (a) For $\g = \fsl_3(\R)$, the restricted root system
  is of type $A_2$: 
  \[ \Sigma = \{ \pm (\eps_1 - \eps_2), \pm (\eps_2 - \eps_3), \pm (\eps_1-\eps_3)\}.\]
  For the $3$-grading defined by the Euler element
  $h = \frac{1}{3}(2,-1,-1)$, we have
  \[ \Sigma_1 =\{ \eps_1 - \eps_2, \eps_1 - \eps_3\}, \quad
    \Sigma_0 = \{ \pm (\eps_2 - \eps_3) \}, \quad
  \cW_0 \cong S_2, \]
and $\Gamma$ is $\{ \eps_1 - \eps_2\}$ or $\{ \eps_1 - \eps_3\}$. The projection
  $\pr_\Gamma(\Sigma) \setminus \{0\}$ is a root system of
  type $BC_1$. 

  \nin  (b) For $\g = \sp_4(\R)$, the restricted root system is o f type $C_2$:
  \[ \Sigma = \{ \pm 2\eps_1, \pm 2\eps_2, \pm \eps_1 \pm \eps_2\}. \] 
  For the $3$-grading defined by $h := \frac{1}{2}(1,1)$, we then have
  \[ \Sigma_1 =\{ 2\eps_1, 2 \eps_2, \eps_1 + \eps_2\}
    \quad \mbox{ and } \quad \Sigma_0 = \{ \pm (\eps_1 - \eps_2)\}. \]
  Here $\Gamma = \{ 2 \eps_1, 2 \eps_2\}$ is a maximal subset of
  long strongly orthogonal roots. The subset
  $\{\eps_1 + \eps_2\}$ is also maximal
  strongly orthogonal in $\Sigma_1$, but it consists of a short root.   
\end{Example}

\begin{corollary} \label{cor:hc}
  In the notation of {\rm Proposition~\ref{prop:hc}}, consider
$h_s := \frac{1}{2} \sum_{j = 1}^r \gamma_j^\vee.$ 
  Then the following are equivalent: 
  \begin{itemize}
  \item[\rm(a)] The Euler element $h$ is symmetric. 
  \item[\rm(b)] $h = h_s$. 
  \item[\rm(c)] The sets $\cP_j$ and $\cC_j$ are empty. 
  \item[\rm(d)] $\pr_\Gamma(\Sigma) \setminus \{0\}$ is a root system
    of type $C_r$. 
  \end{itemize}
\end{corollary}

\begin{Proof} Let $s_\alpha(x) = x - \alpha(x) \alpha^\vee$ be the reflection
  corresponding to the root $\alpha$. 
  We consider the Euler element
  \[ h'
  := s_{\gamma_1} \cdots s_{\gamma_r}(h)
  = h  - \sum_{j = 1}^r \gamma_j(h) \gamma_j^\vee
  = h  - \sum_{j = 1}^r \gamma_j^\vee.\]
  Since the Weyl group acts by automorphisms of the root system
  (\cite[\S 1.2]{Hu90}),
  $h'$ also is an Euler element. Next we observe that
  $\gamma_j(h') = -1$ for each $j$,
  $\alpha(h') = 1 - 1 =  0$ for $\alpha \in \cP_j$ and 
  $\alpha(h') = 1 - 2 =  -1$ for $\alpha \in \cP_{jk}$.
  Moreover, $\alpha(h') = 0$ for $\alpha \in \cC_0$, 
  $\alpha(h') = - 1$ for $\alpha \in \cC_j$ and 
  $\alpha(h') = 0$ for $\alpha \in \cC_{jk}$. This shows that
  $\alpha(h') \leq 0$ for all $\alpha \in \Sigma^+$, so that
  $h' \in \cW.h$ is the unique element with this property.
  It follows that $h$ is symmetric if and only if $h' = -h$.

  \nin (a) $\Leftrightarrow$ (b):
  The Euler element $h$ is symmetric if and only if $h' = -h$,
  which is equivalent to $2h = \sum_{j = 1}^r \gamma_j^\vee = 2h_s$, i.e.,
  to $h = h_s$.

  \nin (b) $\Rarrow$ (c): For $\alpha \in \cP_j \cup \cC_j$ we have
  $\alpha(h_s) = \frac{1}{2}$, which is impossible if $h_s$ is an
  Euler element. 

  \nin (c) $\Rarrow$ (b): If all sets $\cP_j$ and $\cC_j$ are empty,
  then the discussion above shows that $\alpha(h)= -\alpha(h')$ for
  all roots $\alpha$, hence that $h' = -h$ and thus $h = h_s$.

  \nin (c) $\Leftrightarrow$ (d) follows from Proposition~\ref{prop:hc}. 
  \end{Proof}

The following proposition translates the information
on symmetry of the Euler element to the Lie algebra context. 

\begin{Proposition} \label{prop:3.7} 
Let $\g$ be a simple real Lie algebra and 
$h \in \g$ be an Euler element. 
Let $\fa \ni h$ be a maximal abelian hyperbolic subspace,  
$\Sigma := \Sigma(\g,\fa)$ the corresponding set of restricted roots, 
$\Gamma = \{ \gamma_1, \ldots, \gamma_r \} 
\subeq \Sigma_1$ a maximal set of strongly orthogonal roots
and $\g(\gamma_j)\cong\fsl_2(\R)$ corresponding $\fsl_2(\R)$-subalgebras. 
Then the following are equivalent: 
\begin{itemize}
\item[\rm(a)] $h$ is symmetric. 
\item[\rm(b)] $h \in \sum_{j = 1}^r \g(\gamma_j)$. 
\end{itemize}
\end{Proposition}

\begin{Proof} (a) $\Rarrow$ (b): From Corollary~\ref{cor:hc}
  we know that $h$ is symmetric if and only if $h = h_s$,
 and $h_s$ is contained in $\sum_{j = 1}^r \g(\gamma_j)$ because
  $\gamma_j^\vee \in\g(\gamma_j)$.

\nin (b) $\Rarrow$ (a) follows from the fact that all Euler elements 
in $\fsl_2(\R)$ are symmetric, and this is inherited by $\fsl_2(\R)^r$.   
\end{Proof}

\subsection{Cartan involutions and H-elements}
\label{subsec:3.3}

In this subsection we briefly discuss the relation between
Cartan involutions and H-elements in quasihermitian semisimple
Lie algebras.

\begin{Lemma} \label{lem:cartancrit} 
  If $\g$ is semisimple without compact ideals, then the following assertions
  hold:
  \begin{itemize}
  \item[\rm(a)]  If $\sigma$ is an involutive automorphism for which $\g^\sigma$ 
is compactly embedded. Then $\sigma$ is a Cartan involution.
\begin{footnote}
  {Every non-trivial involution on a compact simple Lie algebra shows that this
  conclusion is false if $\g$ contains compact ideals.}  
\end{footnote}

\item[\rm(b)] If $C_\g \subeq \g$ is a pointed generating invariant cone 
and $z \in C_\fg^\circ$ such that 
$h := iz$ is an Euler element in $\g_\C$, then $z$ is an H-element. 
  \end{itemize}
\end{Lemma}

\begin{Proof} (a) Let $\fk \supeq \g^\sigma$ be a maximal compactly embedded 
subalgebra and $\theta$ the corresponding Cartan involution with 
$\fk = \g^\theta$. Then 
\[ \fp := \g^{-\theta} = \fk^\bot \subeq (\g^\sigma)^\bot = \g^{-\sigma}, \] 
where $\bot$ refers to the Cartan--Killing form. 
As $\g$ contains no compact ideals, we have 
\[ \g = \fp + [\fp,\fp] \] 
because the right hand side is an ideal containing all non-compact simple ideals. Since $\theta$ and $\sigma$ coincide on $\fp$, they coincide, i.e., 
$\sigma$ is a Cartan involution.

\nin (b)  First we observe that 
$\Spec(\ad z) = \{0,\pm i\}$ and that  
$\ker(\ad z)$ is compactly embedded because $z \in C_\g^\circ$ 
(\cite[Prop.~V.5.11]{Ne00}). 
Hence $\sigma := e^{\pi \ad z} \in \Aut(\g)$ is an involution for which 
$\g^\sigma = \ker(\ad z)$ is compactly embedded. 
Now (a) implies that $\theta_z$ 
is a Cartan involution, i.e., $z$ is an H-element.
\end{Proof}

\begin{Lemma}\label{lem:comtheta} Let $\fg$ be a simple Lie algebra and $\theta_1$ and $\theta_2$ two 
Cartan involutions. Then $\theta_1$ and $\theta_2$ commute if and only if 
$\theta_1=\theta_2$.
\end{Lemma}
\begin{Proof} If $\theta_1=\theta_2$ then they clearly commute. 
For the other direction write $\g_j = \fk_j\oplus \fp_j$, $j=1,2$, for the Cartan decomposition corresponding
the $\theta_j$. Then $\fk_2$ is $\theta_1$-stable and hence
\[ \fk_2= (\fk_2\cap \fk_1)\oplus (\fk_2 \cap \fp_1).\]
As $\fk_2$ is compactly embedded, we have $\fk_2 \cap \fp_1 = \{0\}$,
  hence $\fk_2 \subeq \fk_1$. We likewise obtain $\fk_1 \subeq \fk_2$,
  hence equality. Then $\fp_1 = \fp_2$ is the orthogonal space with
  respect to the Cartan--Killing form and thus $\theta_1= \theta_2$.
\end{Proof}

\begin{Lemma} \label{lem:unique-H}
If $\g$ is a simple hermitian Lie algebra with 
Cartan decomposition $\g = \fk \oplus \fp$ and 
$z \in \fz(\fk)$ is an H-element, then $\pm z$ are the only 
H-elements in $\fk$. 
\end{Lemma}

\begin{Proof} Clearly, $\pm z$ are both H-elements in $\fk$,
defining the same Cartan involution $\theta = \theta_z = e^{\pi \ad z}$. 
If $z' \in \fk$ is another H-element, then $[z,z'] = 0$, so that 
$\theta_z$ and $\theta_{z'}$ are two commuting Cartan involutions
and hence $\theta_z=\theta_{z^\prime}$ by Lemma \ref{lem:comtheta},
so that $z,z^\prime \in\fz(\fk)$. But $\dim \fz(\fk) =1$ and hence 
$z^\prime = \pm z$. 
\end{Proof}

\subsection{The duality between Euler elements and H-elements} 
\label{subsec:3.4}

The following lemma relates Euler elements in $\fp = \g^{-\theta}$ to 
symmetric Lie algebras $(\g^c = \fh + i \fq,\tau^c,z)$ of hermitian type. 

\begin{Lemma} \label{lem:HEuler} 
Let 
\[ \cA := \{ (\g, \theta,h) \: \g \ \text{\rm semisimple}, h 
\in \cE(\g), \theta \ \mbox{\rm Cartan involution with } \theta(h) = -h\}\]
and 
\[ \cB := \{ (\g, \tau,z) \: \g \ \text{\rm semisimple}, z \ 
\text{\rm H-element}, \tau\ \mbox{\rm involution with } \tau(z) = -z\}.\]
Then we have a bijection 
\[ \Phi \: \cA \to \cB, \quad 
\Phi(\g,\theta,h) := (\g^c,\tau^c,ih), \]
where
\[ 
\tau := \tau_h \theta, \quad  \g^c = \g^\tau + i \g^{-\tau} ,\quad
\tau^c(x + iy) = x-iy.\]
\end{Lemma}

\begin{Proof} Let $(\g,\theta,h) \in \cA$. 
As $\theta$ commutes with $\tau_h$, 
the product $\tau := \tau_h \theta$ 
is an involution on $\g$. It satisfies 
\[\fh:= \g^\tau = \fk^{\tau_h} \oplus \fp^{-\tau_h} \quad \mbox{ and } \quad 
\fq := \g^{-\tau} = \fk^{-\tau_h} \oplus \fp^{\tau_h},\] 
so that 
\begin{equation}
  \label{eq:hkqp}
 \g^{\tau_h} = \ker(\ad h) 
= \fh_\fk \oplus \fq_\fp \quad \mbox{ and } \quad h \in \fq_\fp.
\end{equation}
We associate to $(\g,\tau)$ the  dual symmetric Lie algebra 
$(\g^c, \tau^c)$. It 
contains $z:= i h$ with centralizer
$\ker(\ad z) = \fh_\fk + i \fq_\fp = \fk^c$ 
maximal compactly embedded in $\g^c$ 
(where $\fh = \g^\tau, \fq= \g^{-\tau}$). Therefore 
$z \in i \fq_\fp$ is an H-element and the involution 
$\tau^c$ on $\g^c$ satisfies $\tau^c(z) = -z$. 
This shows that $(\g^c,\tau^c,ih) \in \cB$. 

If, conversely, $(\g, \tau, z) \in \cB$, 
then the dual symmetric Lie algebra $(\g^c,\tau^c)$ contains 
the Euler element $h := -iz$ and 
$\theta := \tau^c \tau_h$ is a Cartan involution because its fixed point set 
\[ \fk := \fh^{\tau_h} + i\fq^{-\tau_h} 
= \fh_{\fk} + i \fq_{\fp} \] 
is maximal compactly embedded in $\g$. This shows that 
$(\g^c,\theta,h) \in \cA$. As $\tau^c = \theta \tau_h$, we have 
$\Phi(\g^c,\theta,h) = (\g,\tau,z)$. Therefore $\Phi$ is bijective. 
\end{Proof}

\begin{Remark} Since the Lie algebras $\g$ and $\g^c$ are real forms of the 
same complex Lie algebra $\g_\C$, one can also represent  the 
triples $(\g,\theta,h)$ by triples 
$(\sigma,\theta,h)$, where 
$\sigma$ is an antilinear involution of~$\g_\C$, 
$\theta$ is a linear involution of $\g_\C$ commuting with $\sigma$ 
such that $\sigma\theta$ is a Cartan involution of $\g_\C$, and 
$h\in \cE(\g_\C)$ satisfies $\sigma(h) = h$ and $\theta(h) = -h$.

Likewise elements of $\cB$ can be represented by  triples 
$(\sigma,\tau,z)$, where  
$\sigma$ is an antilinear involution of $\g_\C$, 
$\tau$ is a linear involution of $\g_\C$ commuting with $\sigma$ 
and $iz\in \cE(\g_\C)$ satisfies $\sigma(z) = z$ and $\tau(z) = -z$.
In these terms $\Phi$ maps $(\sigma,\theta,h)$ to 
$(\sigma\tau, \tau, ih) = (\sigma\tau_h \theta, \tau, ih)$.
\end{Remark}

\subsection{Basic facts on invariant cones} 
\label{subsec:3.5} 

\subsubsection{Invariant cones in hermitian Lie algebras}
\label{subsubsec:2.3.1}

Let $\g$ be a simple hermitian Lie algebra
and $\g = \fk \oplus \fp$ be a Cartan decomposition.
Then $\fz(\fk) = \R z$ is one-dimensional and generated by an
H-element $z$ satisfying~$\theta = e^{\pi \ad z}$
(Remark~\ref{rem:herm}).
Now every closed convex $\Inn(\g)$-invariant cone $C_\fg \subeq C$ satisfies
\begin{equation}
  \label{eq:czk}
 C_{\fz(\fk)} := p_{\fz(\fk)}(C_\fg) = C_\g \cap \fz(\fk)
  \quad \mbox{ and } \quad
  C_{\fz(\fk)}^\circ = C_\g^\circ \cap \fz(\fk)
\end{equation}
where the projection $p_{\fz(\fk)} \: \g \to \fz(\fk)$
is the composition of the fixed point projection
$\g \to \fk$ for the compact group $\Inn_\g(\fz(\fk)) \cong \T$
and the fixed point projection for the compact group $\Inn_\g(\fk)$
(Proposition~\ref{prop:project}).
Therefore every non-trivial invariant cone $C_\g$ either contains
$z$ or $-z$.

Conversely, it is easy to see with the Iwasawa decomposition that 
\[ C_\g^{\rm min}(z) := \oline{\R_+ \conv(\Inn_\g(z))} \]
is a pointed invariant cone (\cite[Thm.~7.25]{HN93}).
For any invariant cone $C_\g \subeq \g$ containing $z$ we then have
\begin{equation}
  \label{eq:minincl}
  C_\g^{\rm min}(z) \subeq C_\g.
\end{equation}
The Cartan--Killing form $\kappa(x,y) = \tr(\ad x\ad y)$
is negative definite on $\fk$ and positive
definite on $\fp$, so that
\[ C_\g^{\rm max}(z) := \{ x \in \g \: (\forall y \in C_\g^{\rm min}(z))\
  \kappa(x,y) \leq 0 \} \]
is a pointed generating invariant cone containing $z$,
hence also $C_\g^{\rm min}(z)$. Dualizing \eqref{eq:minincl}  
implies that every invariant cone $C_\g$ containing $z$ is contained
in $C_\g^{\rm max}(z)$:
\begin{equation}
  \label{eq:dual-cg}
  C_\g^{\rm min}(z) \subeq C_\g \subeq C_\g^{\rm max}(z).
\end{equation}

If $\g$ is a semisimple Lie algebra and $z \in \g$ an H-element,
then $\g = \g_k \oplus \g_n$, where $\g_k$ is the sum of all simple
ideals commuting with $z$ (these are compact)
and $\g_n$ is a sum of hermitian simple ideals.
Applying the preceding discussion to all simple ideals in $\g_n$, we obtain
pointed generating invariant cones
$C_{\g_n}^{\rm min}(z) \subeq C_{\g_n}^{\rm max}(z)$ such that any pointed generating
invariant cone $C_{\g_n}\subeq \g_n$ containing $z$ satisfies
\[ C_{\g_n}^{\rm min}(z) \subeq C_{\g_n} \subeq C_{\g_n}^{\rm max}(z).\]
We then put
\begin{equation}
  \label{eq:min-max-geng}
 C_\g^{\rm min}(z) := C_{\g_n}^{\rm min}(z) \quad \mbox{ and } \quad 
 C_\g^{\rm max}(z) := \g_k \oplus C_{\g_n}^{\rm max}(z). 
\end{equation}
In this context it is still true that, for every pointed generating
invariant cone $C_\g \subeq \g$, there exists an H-element $z$ with
\begin{equation}
  \label{eq:dual-cg-gen}
  C_\g^{\rm min}(z) \subeq C_\g \subeq C_\g^{\rm max}(z)
\end{equation}
(cf.~\cite[Prop.~II.3]{HNO94}).

\subsubsection{Invariant cones in irreducible symmetric Lie algebras}
\label{subsubsec:2.3.2}

Let $\g$ be a simple Lie algebra
with Cartan involution $\theta$  and $h \in \g^{-\theta}$ an
Euler element. For the involution $\tau  = \tau_h \theta$ we then have
for $\fh= \g^\tau$ and $\fq = \g^{-\tau}$ 
\[ \fh = \fh_\fk \oplus \fh_\fp, \quad
  \fq = \fq_\fk \oplus \fq_\fp \quad \mbox{ with } \quad
  \g^{\tau_h} = \ker(\ad h)  = \fh_\fk \oplus \fq_\fp.\]
It follows that $z := i h \in \g^c$ is an H-element and that its
centralizer $\fk^c = \fh_\fk + i \fq_\fp$ is a maximal compact subalgebra
of~$\g^c$ (Lemma~\ref{lem:HEuler}).
The closed convex $\Inn_\g(\fh)$-invariant cone
$C_\fq^{\rm min}(h) \subeq \fq$ generated by $h$ 
is pointed because it is contained in the pointed
cone $-i C_{\g^c}^{\rm min}(ih)$. By \eqref{eq:czk}  $h$ is contained
in its interior (cf.~also \cite[Prop.~3.1.3]{HO97}). 

\begin{Lemma} \label{lem:euler-exist} {\rm(cf.~\cite[Prop.~3.1.11]{HO97})}
  If $(\g,\tau,C)$ is irreducible ncc,
  then the following assertions hold:
  \begin{itemize}
  \item[\rm(a)] Every element in $C_\fq^\circ$ fixed by
    $\Inn_\g(\fh_\fk)$ is contained in $\fq_\fp$. 
  \item[\rm(b)] $\fz_{\fq_\fp}(\fh_\fk) = \R h$ for an Euler element~$h$.
 \item[\rm(c)] $C^\circ \cap \fq_\fp$ contains an Euler element $h$. 
  \end{itemize}
\end{Lemma}

\begin{Proof} The group $H := \Inn_\g(\fh)$ is the identity component
  of the group $\Aut(\g)^\tau$, hence closed. It has the polar decomposition
  $H = H_K e^{\ad \fh_\fp}$, where $H_K = \Inn_\g(\fh_\fk)$
  (\cite[Prop.~13.1.5]{HN12}).
  As the simplicity of $\g$ implies that $\fh = [\fq,\fq]$, the representation
  $\Ad_\fq$   of $H$ on $\fq$ is faithful and its image in $\GL(\fq)$ is closed with
  maximal compact subgroup $\Ad_\fq(H_K)$.

  \nin (a)  Let $x \in C^\circ \cap  \fq_\fp$ be fixed by the compact group~$H_K$. 
  As the stabilizer of $x$ in $\Ad_\fq(H)$ is compact (\cite[Prop.~V.5.11]{Ne00}), $H^x = H_K$.
Since $x$ is hyperbolic, there exists an element $g \in H$ with
$y := \Ad(g) x \in \fq_\fp$ (\cite[Cor.~II.9]{KN96}).
Then $\theta(y) = -y$ implies that the stabilizer
$H^y$ is $\theta$-invariant. So $g_1 = h_1 e^{\ad z} \in H^y$ with $h_1 \in H_K$
and $z \in \fh_\fp$ implies $\theta(g_1) g_1 = e^{2 \ad z} \in H^y$. The compactness
of $H^y = g H^x g^{-1} = g H_K g^{-1}$ now entails $z = 0$. Hence
  $H^y \subeq H_K$ and thus $H^y = H_K$ as $H^y$  is maximally compact.
  We conclude that $g H_K g^{-1} = H_K$ which further implies $g \in H_K$,
  the stabilizer group of the base point in the Riemannian symmetric space~$H/H_K$.
  This shows that $y = x \in \fq_\fp$.

  \nin (b) Let $x \in \fq_\fp$ be fixed by $H_K$, i.e., $x$ commutes with
  $\fh_\fk$.   Then $[x,\fq_\fp] \subeq \fh_\fk$, and
\[ \kappa([x,\fq_\fp], \fh_\fk) = \kappa(\fq_\fp, [x,\fh_\fk]) = \{0\} \] 
implies $[x,\fq_\fp] = \{0\}$, so that
$x$ is central in the maximal compactly embedded subalgebra
$\fk^c := \fh_\fk + i \fq_\fp$ of $\g^c = \fh +  i \fq$.
As $\g$ is simple, either $\g^c$ is simple as well
(this happens if $(\g,\tau)$ is not of complex type), or if
$(\g,\tau)$ is of complex type and then $(\g^c,\tau) \cong (\fh \oplus \fh,
\tau_{\rm flip})$. In the first case $\g^c$ is simple hermitian, 
so that $\fz(\fk^c) = \R i x$ contains an H-element and thus $x$
is a multiple of an Euler element. In the second case
$\fh$ is also simple hermitian and
$\fk^c = \fh_\fk \oplus \fh_\fk$ has $2$-dimensional center
with $\fz(\fk^c) \cap i \fq_\fp = \R (z,-z)$ for an H-element $z \in \fh_\fk$.
Again,  some positive multiple of $x$ is an Euler element
and $\fz_{\fq_\fp}(\fh_\fk) = \R x$.

\nin (c) As the fixed point projection for the compact group $H_K$ 
leaves the interior of $C_\fq$ invariant (Proposition~\ref{prop:project}),
$C_\fq^0$ contains an $H_K$-fixed point $x$. By (a) it is contained in $\fq_\fp$
and by (b) it is a multiple of some Euler element~$h$.
Hence $C_\fq^\circ$ contains $h$ or~$-h$. 
\end{Proof}

It follows from Lemma~\ref{lem:euler-exist}(c) 
that, for every $\Inn_\g(\fh)$-invariant pointed generating closed convex cone
$C_\fq \subeq \fq$, the interior 
$C_\fq^\circ$ contains $h$ or $-h$. 
Further, $h \in C_\fq^\circ$ implies
\[ C_\fq^{\rm min}(h) \subeq C_\fq.\]
Similar arguments as above, using the Cartan--Killing form on $\fq$,
now lead to a  maximal
$\Inn_\g(\fh)$-invariant cone $C_\fq^{\rm max}(h)$ with
\[ C_\fq^{\rm min}(h) \subeq C_\fq \subeq
  C_\fq^{\rm max}(h)
  = C_\fq^{\rm min}(h)^\star
  = \{ x \in \fq \: (\forall y \in C_\fq^{\rm min}(h))\ \kappa(x,y) \geq 0\}.\]

If $\g$ is only semisimple, we decompose it as
$\g = \g_r \oplus \g_s$, where $\g_s$ is the sum of all simple ideals
not commuting with $h$. We then obtain a pointed $\Inn_\g(\fh)$-invariant cone
$C_{\fq_s}^{\rm min}(h) \subeq \fq_s := \fq \cap \g_s$ whose dual cone 
$C_{\fq_s}^{\rm max}(h)$ with respect to the Cartan--Killing form satisfies
\begin{equation}
  \label{eq:min-max-q}
  C_{\fq_s}^{\rm min}(h) \subeq C_{\fq_s}^{\rm max}(h).
\end{equation}
Both cones are adapted to the decomposition of $(\g,\tau)$ 
into irreducible summands. 
Further, each pointed generating $\Inn_\g(\fh)$-invariant cone $C_\fq$ containing
$h$ satisfies
\begin{equation}
  \label{eq:min-max-q2}
  C_\fq^{\rm min}(h) \subeq C_\fq \subeq C_\fq^{\rm max}(h).
\end{equation}
Here the first inclusion is obvious, and the second one follows from the 
fact that $h$ is also contained in the dual cone
\[ C_\fq^{\star}
= \{ y \in \fq \: (\forall x \in C_\fq) \ \kappa(x,y) \geq 0\}.\]
This leads to
$C_\fq^{\rm min}(h) \subeq C_\fq^\star$, and thus to
$C_\fq \subeq  C_\fq^{\rm min}(h)^\star = C_\fq^{\rm max}(h).$

\section{Euler elements and   non-compactly causal symmetric spaces} 
\label{sec:4} 

In this section we present a classification 
of irreducible non-compactly causal symmetric spaces based
on Euler elements. Here a key concept is that of a {\bf causal} Euler element
that is explored in Section~\ref{subsec:4.1}: 
If $(\g,\tau)$ is a symmetric Lie algebra and
$h \in \fq$ an Euler element, then we call it causal
if it is contained in the interior of an $\Inn_\g(\fh)$-invariant
pointed convex cone in $\fq$. Theorem~\ref{thm:2.3} asserts that, 
for any pair $(\theta, h)$ of a Cartan involution $\theta$ and an Euler
element $h$ satisfying $\theta(h) = -h$,
the involution $\tau = \tau_h \theta$ makes $h$ causal for $(\g,\tau)$.
In Section~\ref{subsec:4.3} we use this construction to classify
irreducible non-compactly causal symmetric Lie algebras in
terms of $\Inn(\g)$-orbits of Euler elements (Theorem~\ref{thm:classif}),
which essentially goes back to \cite{Ol91}.

In general there are many locally isomorphic causal symmetric spaces
$G/H$ corresponding to a causal symmetric Lie algebra $(\g,\tau,C)$.
It is therefore natural to ask for a ``minimal model''.
In Section~\ref{subsec:4.2} we show that the minimal
symmetric space $\Inn(\g)/\Inn(\g)^\tau$
for the irreducible  symmetric Lie algebra $(\g,\tau)$ is this
minimal model for $(\g,\tau,C)$
if and only if the causal Euler element is not symmetric;
otherwise we have to pass to a two-fold covering.
In Section~\ref{subsec:4.4} we present a structured table with the 
classification of the irreducible non-compactly causal symmetric
Lie algebras, based on causal Euler elements. 

Throughout this section the Lie algebra $\g$ is assumed to be 
{\bf semisimple} if not stated otherwise.

\subsection{Causal Euler elements} 
\label{subsec:4.1}

\begin{Definition} Let $(\g,\tau)$ be a symmetric Lie algebra and
  $h \in \cE(\g) \cap \fq$ an Euler element.
  We say that $h$ is {\it causal} if there exists a pointed generating
  $\Inn_\g(\fh)$-invariant cone $C \subeq \fq$ with $h\in C^\circ$.
We write $\cE_c(\fq) \subeq \cE(\g) \cap \fq$ for the 
{\it set of causal Euler elements in $\fq$}.
\end{Definition}

We start with a {\bf reductive}  Lie algebra $\g$.
Let $h \in \cE(\g)$ be an Euler element. 
As $h$ is hyperbolic, there exists a Cartan involution 
$\theta$ with $\theta(h) = -h$, where we assume that
$\fz(\g) \subeq \g^{-\theta}$. 
To make $h$ causal, 
one has to find an appropriate involution $\tau$ on~$\g$. 
It turns out that $\tau := \tau_h \theta$ is the natural choice 
making $h$ a causal Euler element (cf.~Lemma~\ref{lem:HEuler}).

\begin{Theorem} \label{thm:2.3} 
{\rm (Euler elements vs.~causal symmetric spaces)}
Let $h \in \cE(\g)$ be an Euler element in the reductive 
Lie algebra $\g$ and $\theta$ a Cartan involution 
with $\theta(h) =-h$ and $\fz(\g) \subeq \g^{-\theta}$. Then the following assertions hold: 
\begin{itemize}
\item[\rm(a)] $\tau := \tau_h \theta$ defines an 
involution on $\g$.
\item[\rm(b)] The symmetric Lie algebra 
$(\g,\tau)$ is non-compactly causal and there exists a pointed 
generating $\Inn_\g(\fh)$-invariant cone $C \subeq \fq$ with 
$h\in C^\circ$. 
\item[\rm(c)] All ideals of $\g$ contained in $\g^\tau$ are compact.
\end{itemize}
\end{Theorem}

This theorem implies in particular that every Euler element 
is causal with respect to a suitably chosen involution $\tau$ on~$\g$.
In \cite{HO97} causal Euler elements in $\fq_\fp$ are called 
``cone generating elements'' because they generate pointed generating
 invariant cones (see also \cite{Ol91} and Section~\ref{subsubsec:2.3.2}). 

 \begin{Proof} Let $\g = \fz(\g) \oplus \g_1 \oplus \cdots \oplus \g_n$ denote
   the composition    of $\g$ into the center and  simple ideals.
   Recall that $\theta$ preservers all these ideals because all Cartan
   involutions  are conjugate under $\Inn(\g)$ (\cite[Cor.~13.2.13]{HN12})
   and all simple ideals possess Cartan involutions
   (\cite[Thm.~13.2.10]{HN12}). 
Now $h = h_z + h_1 + \cdots + h_n$, where 
either $h_j = 0$ or $h_j \in \g_j$ is an Euler element. 
As $\theta$ and $\tau_h$ preserve $\g_j$, the
involution $\tau := \tau_h\theta$ defines on $\g_j$ an involution 
$\tau_j$ which is Cartan if $h_j = 0$, 
and if this is not the case, then $h_j \in \fq_j := \g_j^{-\tau_j}$. 
Note that $\ad h \not=0$ implies that some $h_j$ is non-zero. 
As all ideals of $\g$ contained in $\g^\tau$ commute with $h$, 
they are contained in $\g^\theta$, hence are compact Lie algebras.

From the discussion in Section~\ref{subsubsec:2.3.2}
we know that $h_j$ is contained in the interior of 
a pointed generating $\Inn_{\g_j}(\fh_j)$-invariant 
cone~$C_j := C_{\fq_j}^{\rm min}(h_j) \subeq \fq_j$.
As $h_j$ is hyperbolic and contained in $C_j^\circ$, 
the causal symmetric Lie algebra $(\g_j, \tau_j, C_j)$ is ncc. 
The Lie algebra 
\[ \fh_r := \sum_{h_j = 0} \fh_j \] 
is compactly embedded and 
\[ C_s := \sum_{h_j \not=0} C_j = C_{\fq_s}^{\rm min}(h)
  \subeq  \fq_{s} := \sum_{h_j \not=0} \fq_j \] 
is pointed and generating. 
Further 
\[ \fh_{s} := \sum_{h_j \not= 0} \fh_j \] 
satisfies $\Inn_\g(\fh_{s})\res_\fq \subeq \SL(\fq)$. 
Hence Lemma~\ref{lem:coneext} 
implies the existence of a pointed generating 
$\Inn_\g(\fh)$-invariant cone $C \subeq \fq = \fz(\g)
\oplus \fq_r \oplus \fq_s$ with 
$C \cap \fq_s = C_s$ 
and $h \in C_s^\circ \subeq  C^\circ$. 
\end{Proof}

In the preceding proof, we have seen that it is
convenient to decompose the Lie algebra $\g$ as
\begin{equation}
  \label{eq:gdeco}
  \g = \g_k \oplus \g_r \oplus \g_s,
\end{equation}
where $\g_s$ is the sum of all simple ideals not commuting with $h$
(the strictly ncc part), $\g_r$ is the sum of all non-compact simple
ideals commuting with $h$ on which $\tau = \theta$
(the non-compact Riemannian part),
and $\g_k$ is the sum of all compact ideals (they commute with $h$).
All these ideals are invariant under $\theta$ and $\tau = \tau_h \theta$,
so that we obtain decompositions
\begin{equation}
  \label{eq:qs}
 \g_s = \fh_s\oplus \fq_s, \quad
  \g_r = \fh_r\oplus \fq_r \quad \mbox{ and } \quad
  \g_k = \fh_k,
\end{equation}
where $\fh_r \oplus \fh_k$ is a compact ideal of~$\fh$ and $\fh_r\oplus \fq_r$ is a Cartan decomposition of $\g_r$.

\begin{Remark} \label{rem:sncc-reduction}
Let $\fq = \fq_r\oplus\fq_{s}$ be the decomposition 
of $\fq$ into the Riemannian part $\fq_r$ and its orthogonal complement 
which is ncc. We write $p_{s} \: \fq \to \fq_{s}$ for the 
projection with kernel $\fq_r$, which is the fixed point projection
for the compact group $\Inn_\g(\fh_k \oplus \fh_r)$.
Then  every 
$\Inn_\g(\fh)$-invariant closed convex cone 
$C \subeq \fq$ satisfies 
\begin{equation}
  \label{eq:psc}
p_{s}(C) = C \cap \fq_s =: C_{s}
\quad \mbox{ and } \quad 
C_s^\circ = C^\circ\cap \fq_{s}
\end{equation}
(Proposition~\ref{prop:project}).
\end{Remark}

\begin{corollary} Let $(\g,\tau)$  be a reductive 
symmetric Lie algebra and write it as $\g = \g_k \oplus \g_r \oplus \g_s$, 
where $\g_k$ is the sum of all compact ideals of $\g$ contained
in $\fh = \g^\tau$, $\g_r$ is the sum of all Riemannian summands 
on which $\tau$ is a Cartan involution, and
$\g_s$ is the sum of all non-Riemannian irreducible summands.
Then $\g$ is non-compactly causal if and only if $\g_s\not=\{0\}$ 
and $\g_s$ is a sum of irreducible non-compactly causal 
symmetric Lie algebras.
\end{corollary}

\begin{Proof} If $\g_s$ is non-zero and non-compactly causal, 
  then the proof of Theorem~\ref{thm:2.3} implies that
  $(\g,\tau)$ is non-compactly causal.
  If, conversely, this is the case, then 
  Remark~\ref{rem:sncc-reduction} shows that $\fq_s \not=\{0\}$
  and that $(\g_s,\tau)$ is non-compactly causal. 
\end{Proof}

\begin{Theorem} {\rm(Uniqueness of the causal Euler elements)} \label{thm:uniqueinv} 
Let $(\g,\tau,C)$ be a semisimple ncc symmetric Lie algebra 
for which all ideals of $\g$ contained in $\fh$ are compact, 
and $\theta$ a Cartan  involution commuting with~$\tau$. 
Then the following assertions hold: 
\begin{itemize}
\item[\rm(a)] $C_s^\circ \cap \fq_\fp$ contains a unique Euler element~$h$ 
and $\tau = \tau_h \theta$. 
\item[\rm(b)] $\Inn_\g(\fh)$ acts transitively on $C_s^\circ \cap \cE(\g)$. 
\item[\rm(c)] For every Euler element $h \in C_s^\circ$, the involution 
$\tau\tau_h$ is Cartan. 
\end{itemize}
\end{Theorem}

\begin{Proof}  (a) We write 
$(\g,\tau)$ as a direct sum of irreducible symmetric Lie algebras 
$(\g_j, \tau_j)$, $j = 1,\ldots, n$ and a
compact ideal $\g_k$ of $\g$ contained in $\fh$
(cf.\ \cite[Prop.~2.14]{NO22b}).  
Let $x \in C$ be such that all components $x_j \in \fq_j = \fq \cap \g_j$
are non-zero. Then
\[ \cO_x := \Inn_\g(\fh)x = \cO_{x_1} + \cdots + \cO_{x_n} \subeq C \]
implies that all the orbits $\cO_{x_j}$ are contained in
$\Inn_\g(\fh_j)$-invariant closed convex subsets not containing affine lines.
If $\cO_{x_j}$ is bounded, then $(\g_j, \tau_j)$ is Riemannian, and if
it is unbounded, $(\g_j, \tau_j)$ is causal. Moreover, the fact that
the projection of $C^\circ$ to $\fq_j$ consists of hyperbolic elements
implies that $(\g_j,\tau_j)$ is non-compactly causal.
We enumerate the simple ideals in such a way that
$(\g_j,\tau_j)$ is Riemannian for
$j \leq m$ and causal for $j > m$.

The action of the compact group $\Inn_\g(\fh_\fk)$ on $C^\circ$ 
has a fixed point $x = \sum_{j = m+1}^n x_j \in~C^\circ$
(Proposition~\ref{prop:project}) with 
$x_j$ contained in $\fq_{\fp,j}$ (Lemma~\ref{lem:euler-exist}(a)).
By Lemma~\ref{lem:euler-exist}, there exist 
$\lambda_j > 0$ for $j = m+1,\ldots,n$, such that
$h_j := \lambda x_j$ is an Euler element in~$\g_j$.
Then 
\[ h  := \sum_{j = m+1}^n h_j \in \sum_{j = m+1}^n C_{\fq_j}^{\rm min}(h_j)^\circ    
\subeq C_s^\circ \]
(see \eqref{eq:min-max-q2} and \eqref{eq:psc} for the last inclusion) 
is a causal Euler element in $\g$.  
Further 
\begin{equation}
  \label{eq:zgh}
  \fz_\g(h) = \sum_{j = 1}^m \g_j + \sum_{j = m+1}^n \fz_{\g_j}(x_j) = \fh_\fk \oplus \fq_\fp
\end{equation}
implies $\tau = \tau_h \theta$. 

To verify the uniqueness of $h \in \fq_\fp \cap C_s^\circ$, 
let $h_1 \in \fq_\fp \cap C_s^\circ$ be an Euler element. 
We pick $\fa\subeq \fq_\fp$ maximal 
abelian containing $h_1$ and observe that 
$h \in \fa$ follows from $[h,\fq_\fp] = \{0\}$. 
We choose a positive system 
$\Sigma^+ \subeq \Sigma(\g,\fa)$ such that 
\[ C\cap \fa \subeq C_\fa^{\rm max} := C^{\rm max}_\fq  \cap \fa 
= \{ x \in \fa \: (\forall \alpha\in \Sigma_1)\ \alpha(x) \geq 0\}\] 
(\cite[Thm.~VI.6]{KN96}). 
As all positive non-compact roots are positive on the interior of 
$C_\fa^{\rm max}$, we must have $\alpha(h_1) = 1$ 
for every $\alpha \in \Sigma_1$. 
Next we observe that 
$\Sigma_1$ spans $(\fa\cap \fq_s)^*$ (\cite[\S V]{KN96}), 
so that we obtain $h = h_1$.
Therefore $h$ is the only Euler element in 
$\fq_\fp \cap C_s^\circ$. 

\nin (b) If $h_1 \in C_s^\circ$ is an Euler element, then it is in particular 
hyperbolic, hence conjugate under $\Inn_\g(\fh)$ to an element 
$h_2 \in \fq_\fp$ (\cite[Cor.~II.9]{KN96}). 
Then $h_2 \in \fq_\fp \cap C_s^\circ$ is an Euler element, hence equal to 
the Euler element $h$ from  (a). This implies that 
$h_1 \in \Inn_\g(\fh)h$, and thus $\Inn_\g(\fh)$ acts transitively 
on $C_s^\circ\cap\cE(\g)$.

\nin (c) The assertion holds for the 
unique Euler element $h \in C_s^\circ \cap \fq_\fp$ by (a). 
If $h_1\in C_s^\circ$ is another Euler element, then (b) implies the existence of 
$\phi \in \Inn_\g(\fh)$ with $h_1 = \phi(h)$. 
Then 
\[ \tau \tau_{h_1} = \tau \phi \tau_h \phi^{-1} 
= \phi \tau \tau_h \phi^{-1} = \phi \theta \phi^{-1} \]
is a Cartan involution. 
\end{Proof}

Assertion (b) in Theorem~\ref{thm:uniqueinv} has important consequences
for the possible choices of an open subgroup $H \subeq \Inn(\g)^\tau$ preserving
the cone~$C$.

  \begin{corollary} \label{cor:hhcompact} 
If $(\g,\tau,C)$ is a semisimple ncc symmetric 
Lie algebra for which $\fh$ contains no non-compact ideal of $\g$,
$h \in C_s^\circ$ is a causal Euler element,
and $H \subeq \Inn(\g)^\tau$ is an open subgroup 
preserving $C$, then the following assertions hold: 
\begin{itemize}
\item[\rm(a)] $H = H_e H^h$,
  i.e., every connected component of $H$ meets $H^h$,
  which is equivalent to $H.h$ being connected. 
\item[\rm(b)] $H$ is closed and $H^h$ is a maximal compact subgroup  of $H$. 
\item[\rm(c)] $H^h = H^{\tau_h}$ and  $\tau_h$ induces a Cartan involution on~$H$.
\item[\rm(d)] $\tau$ induces a Cartan involution on $H^h$ for which 
$(H^h_e)^\tau = e^{\ad \fh_\fk}$ is connected. 
\end{itemize}
  \end{corollary}

  \begin{Proof} (a) From Theorem~\ref{thm:uniqueinv}(b) and $H_e = \Inn_\g(\fh)$, 
we derive that 
$H.h \subeq C_s^\circ \cap \cE(\g) = H_e.h$, so that 
$H = H_e H^h$. 

\nin (b) Since $\Aut(\g)^\tau$ is an algebraic group, it is closed and has only 
finitely many connected components. It contains $H$ as an open subgroup, 
so that $H$ is also closed. Further, $H^h$ fixes the element 
$h$ in the interior of $C$, so that $H^h$ acts on $\fq$ 
as a relatively compact group. Next we use that 
$[\fq,\fq] + \fq \trile \g$ is an ideal of $\g$ and since $\fh$ contains 
no non-compact ideal, it follows that $\fh = [\fq,\fq] \oplus \fh_k$, 
where $\fh_k \subeq \fh \cap \fk$ is the sum of all compact ideals of
$\g$ contained in $\fh$. 
Therefore the closed subgroup $H^h$ of $\Aut(\g)$ is compact. 
Its maximality in $H$ now follows from the polar decomposition 
$H = H^h \exp(\fh_\fp)$ and (a). Note that Riemannian components
correspond to compact ideals of $\fh$, and the corresponding 
subgroups of $\Inn(\g)$ are compact, commute with $\tau$ and fix~$h$. 

\nin (c)  From $H^{\tau_h}.h \subeq C_s^\circ\cap \fq_\fp$ 
and Theorem~\ref{thm:uniqueinv}(a), it follows that 
$H^{\tau_h}$ fixes $h$,  so that 
$H^{\tau_h} \subeq H^h$. Conversely, 
every $h \in H^h \subeq \Aut(\g)$ commutes with $\ad h$, hence also 
with $\tau_h$, and thus $H^h = H^{\tau_h}$. 
In view of (b),  $\tau_h$ induces a Cartan 
involution on $H$. 

\nin (d) On $H^h = H^{\tau_h}$ the involution 
$\tau = \tau_h \theta$ acts like $\theta$. 
Restricting to the identity component $H^h_e$, we obtain (d)
because the group of fixed points of a Cartan involution is connected.
  \end{Proof}

  \begin{corollary} \label{cor:2.8} {\rm(Characterization of causal
      Euler elements)} 
 Let $(\g,\tau)$ be a semisimple symmetric Lie algebra 
for which all ideals of $\g$ contained in $\fh$ are compact  
and $h \in \cE(\g)$ (cf.~\eqref{eq:qs}).
Consider the following assertions: 
\begin{footnote}
  {If $h$ is not contained in $\fq_s$, then (b) does not follow from  (a).
      Remark~\ref{rem:4.5}(b) provides an example
      for $h \in (C^\lambda)^\circ$ with $\lambda > 1$.     
}\end{footnote}

\begin{itemize}
\item[\rm(a)] $h$ is causal, i.e., there exists a pointed generating 
  $\Inn_\g(\fh)$-invariant closed convex cone $C \subeq \fq$ with $h \in  C^\circ$.
\item[\rm(b)] $\tau \tau_h$ is a Cartan involution. 
\item[\rm(c)] $ih$ is an H-element in $\g^c = \fh + i \fq$. 
\end{itemize}
Then {\rm(a)} $\Leftarrow$ {\rm(b)} $\Leftrightarrow$ {\rm(c)}
and if $h \in \fq_s$, then {\rm(a),(b),(c)} are equivalent. 
\end{corollary}

\begin{Proof} (a) $\Rarrow$ (b):
  If $h \in \fq_s$, then this implication
  follows from Theorem~\ref{thm:uniqueinv}(c).
  
\nin (b) $\Rarrow$ (a) follows from Theorem~\ref{thm:2.3} 
and the fact that $\tau\tau_h(h) = \tau(h) = -h$.

\nin (b) $\Rarrow$ (c): Suppose that $\theta := \tau\tau_h$ is a 
Cartan involution of $\g$. As $\theta(h) = -h$, 
Lemma~\ref{lem:HEuler} implies that $ih$ is an 
H-element of $\g^c$. 

\nin (c) $\Rarrow$ (b) follows immediately from 
Lemma~\ref{lem:HEuler} because 
$(\g^c, \tau^c, ih) \in \cB$. 
\end{Proof}

\begin{Remark} \label{rem:4.5}
  In Theorem~\ref{thm:uniqueinv}(b) it
  is crucial to restrict to the cone~$C_s = C \cap \fq_s$ in~$\fq_s$.

\nin (a) First we observe that, for a 
Riemannian symmetric Lie algebra $(\g,\theta)$, 
where $\theta$ is a Cartan involution, we have $\fp = \fq_\fp$. If
$h\in \fp$ is an Euler element then $\tau_h\theta$ is never a Cartan involution.
The hyperbolic space $\bH^d \cong \SO_{1,d}(\R)/\SO_d(\R)$ is
an example showing that this situation does in fact occur.

Moreover, $\Inn_\g(\fk)$ need not act transitively on 
$\cE(\g) \cap \fp$. 
If the restricted root system $\Sigma(\g,\fa)$
($\fa \subeq \fp$ maximal abelian) is of type 
$A_n, n \geq 2$, $D_n, n \geq 4$ or $E_6$, 
then there exists more than one conjugacy class of Euler 
elements in $\fp$  (cf.\ Theorem~\ref{thm:classif-symeuler}). 

\nin (b) (cf.\ also Example~\ref{ex:3.3} below) 
We consider the example 
\[ \g = (\fsl_2(\R), \theta) \oplus (\fsl_2(\R), \tau_h),
\quad \mbox{ where } \quad 
h = \frac{1}{2}\pmat{1 & 0 \\ 0 & -1},\] 
which is ncc.
For $G := \SL_2(\R) \times \SL_2(\R)$ and
$H := \SO_2(\R) \times \SO_{1,1}(\R)$, the symmetric space
\[ M = G/H \cong \bH^2 \times \dS^2\] 
is Lorentzian. A maximal abelian hyperbolic subspace is 
\[ \fa := \R h \oplus \R h \subeq \fq  \cong \R^4, \] 
where 
\[ C^{\rm min}_\fa = \{0\} \oplus [0,\infty) h, \qquad 
 C^{\rm max}_\fa = \R h  \oplus [0,\infty) h,\] 
and the Riemannian Weyl group $\cW_c = \{ \1,\sigma\}$ 
(generated by the reflections corresponding to ``compact'' roots; 
see \cite{KN96}) acts by 
$\sigma(x,y) = (-x,y)$. For any $\lambda > 0$, we obtain a Weyl group invariant 
cone $C^\lambda_\fa$ between the minimal and the maximal cone by 
\[ C^\lambda_\fa := \cone( (\lambda,1),(-\lambda,1)).\] 
The cone  $C^\lambda_\fa$ contains 
a unique Euler element $(0,h)$ if and only if $\lambda < 1$. 
For $\lambda \geq 1$, it also contains the Euler elements $(\pm h,h)$. 
If $C^\lambda \subeq \fq$ is the corresponding $\Ad(H)$-invariant cone with 
$C^\lambda \cap \fa = C^\lambda_\fa$, then the conclusion of 
Theorem~\ref{thm:uniqueinv}(a)-(c) fail for $\lambda > 1$
and the cone $C^\lambda$ if it is not first intersected with~$\fq_s$. 
\end{Remark}

\begin{Remark}
Suppose that $h \in \cE(\g)$ and $\theta$ is a Cartan involution 
with $\theta(h) = -h$ and $\tau = \tau_h \theta$. 
Then 
\[ \g^{\tau_h} = \g^{\tau\theta} = \fh_\fk \oplus \fq_\fp.\]

Let $G$ be a connected Lie group with 
semisimple Lie algebra $\g$ 
on which $\tau_h^G$, hence also $\tau^G$, exist 
(cf.\ the discussion in \cite[Rem.~2.12]{NO22a}). 
From the polar decomposition $G = K \exp(\fp)$ we derive for the centralizer of $h$ 
the decomposition 
\begin{equation}
  \label{eq:gh1}
 G^h = K^h \exp (\fq_\fp).
\end{equation}
We also obtain 
\begin{equation}
  \label{eq:gh2}
 G^{\tau^G} = K^{\tau^G} \exp(\fh_\fp) = K^{\tau_h^G} \exp(\fh_\fp).
\end{equation}
In general $K^h \not= K^{\tau^G_h}$, as the simply connected
covering group $G = \tilde\SL_2(\R)$ of $\SL_2(\R)$ shows 
(\cite[Rem.~5.3]{NO22a}). 
If $Z(G) = \{e\}$, i.e., $G \cong \Ad(G) = \Inn(\g)$, then 
$G^h \subeq G^{\tau_h^G}$ implies  $K^h \subeq K^{\tau_h^G} = K^{\tau^G}$. 
\end{Remark}

\begin{Remark} For a semisimple symmetric Lie algebra $(\g,\tau)$ 
for which all ideals of $\g$ contained in $\fh$ are compact, 
Corollary~\ref{cor:2.8} provides a map 
\[ \Gamma \: \cE_c(\fq_s) \to \Cart(\g)^\tau,\quad 
h \mapsto \tau \tau_h \] 
to the set $\Cart(\g)^\tau$ of Cartan involutions on $\g$ commuting with~$\tau$. 
For a given Cartan involution $\theta$ commuting with~$\tau$, 
and $K := \Inn(\g)^\theta$, 
the transitivity of the action of $G := \Inn(\g)$ on $\Cart(\g)$
(\cite[Cor.~13.2.13]{HN12}) implies that 
\[ \Cart(\g) \cong G/K \quad \mbox{ and } \quad 
\Cart(\g)^\tau \cong \Exp_{eK}(\fh_\fp) = \Inn_\g(\fh).\theta.\] 
The equivariance of $\Gamma$ with respect to $\Inn_\g(\fh)$ thus shows 
that 
$\Gamma$ maps any $\Inn_\g(\fh)$-orbit in $\cE_c(\fq_s)$ surjectively onto 
$\Cart(\g)^\tau$. 

If $\Gamma(h) = \Gamma(h') = \theta$ for $h,h' \in \cE_c(\fq)$, 
then $h,h' \in \fq_\fp$. If $\g$ is simple, then the following lemma implies that 
$h' \in \{\pm h\}$. 
\end{Remark}

\begin{Lemma} \label{lem:hpm1} 
  Let $(\g,\tau)$ be a simple symmetric Lie algebra,
  $h\in \cE_c(\fq)$ and $\theta = \tau\tau_h$.
  Then the following assertions hold:
  \begin{itemize}
  \item[\rm(a)] $\cE_c(\fq) \cap \fq_\fp = \{\pm h\}.$
  \item[\rm(b)] $\Inn_\g(\fh)$ has 
two orbits in the set $\cE_c(\fq)$ of causal Euler elements in $\fq$. 
They  are represented by $\pm h$ for any $h \in \cE_c(\fq)$. 
  \end{itemize}
\end{Lemma}

\begin{Proof} (a) Let $\g^c := \fh + i \fq$ be the dual symmetric Lie algebra. 
Then $\fk^c = \fh_\fk + i \fq_\fp$ is maximally compact in $\g^c$ and 
$z := i h \in \fz(\fk^c)$ is an H-element with  
$\ker(\ad z) = \fk^c$.

We distinguish two cases according to $(\g,\tau)$ being of complex type or not. \\
\nin {\bf The complex case:} If $\g$ is complex and $\tau$ antilinear, 
then $\g \cong \fh_\C$, where $\fh$ is simple hermitian. 
Then $\fq = i \fh$, and $\cE_c(\fq) \cap \fq_\fp 
= \cE_c(\fq) \cap i \fh_\fk$ consists of all elements of the form 
$h = iz$, where $z \in \fk^c_\fq = i \fq_\fp$ is an H-element. 
Hence the assertion follows from the fact that $\pm z$ are the only H-elements of 
$\g^c$ contained in $\fk^c$ (Lemma~\ref{lem:unique-H}). 

\nin {\bf The real case:} If $\g$ is not complex, then $\g^c$ 
is a simple hermitian Lie algebra with 
$\fk^c = \fh_\fk + i \fq_\fp$ maximal compactly embedded. 
If $h,h' \in \fq_\fp$ are causal 
Euler elements, then $ih, ih' \in \fk^c$ are H-elements, 
and Lemma~\ref{lem:unique-H} implies that 
$h' \in \{\pm h\}$. 

\nin(b) Let $h \in \fq$ be a causal Euler element and consider 
the corresponding Cartan involution $\theta = \tau\tau_h$. 
Then $h \in \fq_\fp$ and any other causal Euler element $h_1 \in \fq$ is hyperbolic, 
hence conjugate under $\Inn_\g(\fh)$ to an element of 
$\fq_\fp$ (\cite[Cor.~II.9]{KN96}),  and therefore 
by (a) to $h$ or $-h$. 
\end{Proof}

\begin{Remark} \label{rem:9.7} 
For every irreducible ncc symmetric Lie algebra 
$(\g,\tau,C)$ and $h \in \cE(\g)$, 
the intersection $\cO_h \cap C^\circ$ is either empty or 
a single orbit of $\Inn_\g(\fh)$ (Theorem~\ref{thm:uniqueinv}(b)).
Therefore at most one $\Inn(\g)$-orbit in 
$\cE(\g)$ intersects~$C^\circ$. In general $\Inn(\g)$ does not act transitively 
on~$\cE(\g)$. A~typical example is $\g = \fsl_n(\R)$ with 
$n-1$ orbits of $\Inn_\g(\fh)$ in $\cO_h \cap \fq$ 
(cf.~Example~\ref{ex:3.3} below; see also
\cite[Rem.~4.15]{MNO22a}). 
\end{Remark}

\begin{Proposition} {\rm(Classification of $\Inn_\g(\fh)$-orbits in $\cO_h \cap \fq)$}
  \label{prop:orbitq}
  Let $(\g,\tau)$ be a reductive symmetric Lie algebra
  with $\fz(\g) \subeq \fq$. Further,
  let $h \in \fq_\fp$ be a causal Euler element with $\theta = \tau\tau_h$ 
and $\fa \subeq \fq_\fp$ maximal abelian. 
Let $\Sigma = \Sigma(\g,\fa)$ denote the corresponding 
set of restricted roots, 
$\Sigma_0 := \{ \alpha \in \Sigma \: \alpha(h) = 0\}$, and 
$\cW_0 \subeq \cW$ the corresponding Weyl group. Then the map 
\[  \cW_0\backslash \cW/\cW_0 \to (\cO_h \cap \fq)/\Inn_\g(\fh), \quad 
\cW_0 w \cW_0 \mapsto \Inn_\g(\fh).wh \] 
is a bijection from the set of $\cW_0$-double cosets in $\cW$ 
to the set of orbits of the group $\Inn_\g(\fh)$ in $\cO_h \cap \fq$.
\end{Proposition}

\begin{Proof} As $\fz_\g(h) \cap \fp = \fq_\fp$, 
the subspace $\fa$ is also maximal abelian in $\fp$. Clearly, 
$\fa$ contains $h$. Then 
$\cO_h \cap \fa = \cW.h$  
by \cite[Thm.~III.10]{KN96} 
and elementary Coxeter theory implies that the stabilizer of $h$ in $\cW$ is 
$\cW^h = \cW_0$ (\cite[Prop.~V.2.7]{Ne00}). 
Every $\Inn_\g(\fh)$-orbit in $\cO_h \cap \fq$ 
intersects $\fa$ (\cite[Cor.~II.9]{KN96},
\cite[Prop.~VII.2.10]{Ne00}) and 
for $x \in \fa$ we have 
\[ \Inn_\g(\fh)x \cap \fa = \Inn_\g(\fh_\fk)x \cap \fa = \cW_0x.\] 
Therefore 
\[ \Inn_\g(\fh) \backslash (\cO_h \cap \fq) \cong \cW_0\backslash \cW h 
\cong \cW_0\backslash \cW/\cW_0.\qedhere \] 
\end{Proof}

    \begin{Remark} Let $H := \Inn_\g(\fh) \supeq H_K = \Inn_\g(\fh_\fk)$ and
      $K := \Inn_\g(\fk)$. 
      We show that, if $H_K$ fixes the causal
      Euler element $h$, then
      \[ H_K.x \cap \fa = \cW_0 x \quad \mbox{ for }\quad x \in \fa.\]
      So let $x \in \fa \subeq \fq_\fp$ and $g \in H_K \subeq K^h$ with
      $x' := g.x \in \fa$. Then
      \[ \fa' := g.\fa \subeq \fq_\fp \cap \fz_\g(x') \]
      is maximal abelian in $\fq_\fp$, hence in particular in
      $\fq_\fp \cap \fz_\g(x')$, which also contains $\fa$.
      Therefore \cite[Thm.~III.3]{KN96} implies the existence of a
      $g_1 \in \Inn_\g(\fh_\fk \cap \fz_\g(x'))$ with
      $g_1.\fa' = \fa$. Then $g_2 := g_1 g \in H_K$ satisfies
      $x' =  g_1.x' = g_2.x$ and
      $g_2.\fa = \fa$. Therefore
      $g_2 \in N_{H_K}(\fa)\subeq N_K(\fa)$
      acts on $\fa$ as a Weyl group element $ w\in \cW$ 
      (\cite[Def.~III.9]{KN96}). As $H_K$ fixes $h$ by assumption,  
      we have $w \in \cW^h = \cW_0$ (\cite[Prop.~V.2.7]{Ne00}).
        \end{Remark}

\begin{Remark} \label{rem:2.17} (Supplement to Proposition~\ref{prop:orbitq}) 
  Let $\g$ be a simple real Lie algebra.
Euler elements in $\g$ are classified by 
their representatives in a closed Weyl chamber $\fa_+ \subeq \fa$, 
where $\g = \fk \oplus \fp$ is a Cartan involution, and  
$\fa \subeq \fp$ is maximal abelian.
Theorem~\ref{thm:classif-symeuler} provides a  concrete list for all 
types of root systems $\Sigma(\g,\fa)$ for which Euler elements exist. 
For $r := \dim \fa$, the representatives $h_j$ are labeled 
by a subset of $\{1,\ldots, r\}$. Any such Euler element 
defines an involution $\tau_j := \theta \tau_{h_j}$ for which 
$h_j \in \fa \subeq \fq_j = \g^{-\tau_j}$. The stabilizer group
$\cW^{h_j} \subeq \cW = \cW(\g,\fa)$
is generated by the reflections fixing $h_j$ (\cite[Prop.~V.2.7]{Ne00}). 
Now Proposition~\ref{prop:orbitq} applies to this 
situation and identifies the set of $\Inn_\g(\fh_j)$-orbits in 
$\cO_{h_j} \cap \fq_j$ with the $\cW^{h_j}$-double cosets in $\cW$. 
\end{Remark}

\begin{Example}
  \label{ex:orbitq} 
For $\g = \fsl_n(\R)$, $n = p + q$, $p \leq q$, the Euler element 
\[ h_p := \frac{1}{p+q} \pmat{ q \1_p & 0 \\ 0 & -p \1_q},\] 
and the Cartan involution $\theta(x) = -x^\top$, 
we obtain $\tau(x) = - \tau_h(x)^\top$, so that we may take 
\[ \fa = \Big\{ \diag(x_1,\ldots, x_n) \:  \sum_j x_j = 0\Big\}.\]
Now 
\[ \cW \cong S_n \supeq \cW_0 \cong S_p \times S_q,\] 
and the orbit space $\cW/\cW_0$ corresponds to the set of 
$p$-element subsets $F_p \subeq \{1,\ldots, n\}$. The orbits of $\cW_0$ 
on this set are represented by $F_p \cap \{1,\ldots,p\}$. 
Parametrized by the cardinality of this intersection, 
we have $p+1$ orbits, and these correspond to $p+1$ orbits 
of $\Inn_\g(\fh)$ in $\cO_h \cap \fq$. They are obtained by  
permuting the entries of the diagonal matrix~$h_p$. 
\end{Example}

\subsection{On the causality of $G/G^\tau$}
\label{subsec:4.2}

In this section we show that the minimal irreducible symmetric space
$\Inn(\g)/\Inn(\g)^\tau$ is a causal symmetric space for the triple
$(\g,\tau, C)$ 
if and only if the corresponding causal Euler element is not symmetric;
otherwise we have to pass to a two-fold covering to obtain a causal space.

\begin{Definition}   Suppose that $(\g,\tau)$ is a semisimple symmetric Lie algebra. 

  \nin(a) We call the corresponding symmetric space
  \[ \cM(\g,\tau) := \Inn(\g)/\Inn(\g)^\tau \]
  a {\it minimal symmetric space}
  associated to the symmetric Lie algebra $(\g,\tau)$.
  All other connected symmetric spaces $M = G/H$ corresponding to $(\g,\tau)$
  are equivariant coverings of $\cM(\g,\tau)$ by the map
  \[ g H \mapsto \Ad(g) \tau \Ad(g)^{-1} \in \Ad(G).\tau \cong
    \cM(\g,\tau).\]

    \nin (b)   If $(\g,\tau,C)$ is a causal semisimple symmetric Lie algebra,
    $G = \Inn(\g)$, and
    \[ H_C := \{ g \in G^\tau \: \Ad(g)C = C \}, \]
    then we call $G/H_C$ the
  {\it minimal causal symmetric space} associated to $(\g,\tau,C)$.
\end{Definition}

Minimal symmetric spaces corresponding to causal symmetric
Lie algebras $(\g,\tau,C)$ are not always causal because 
$G^\tau$ does not always leave a pointed generating cone 
$C \subeq \fq$ invariant, which is equivalent to 
$\cone(G^\tau.h)$ being pointed for a causal Euler element 
$h \in \cE_c(\fq)$. The following proposition and its corollary 
make this requirement easy to check.

\begin{Proposition} \label{prop:caus1}
Suppose that $(\g,\tau)$ is simple ncc and $h \in \cE_c(\fq)$ a causal 
Euler element, $G$ a connected Lie group with 
Lie algebra $\g$ to which $\tau$ integrates, and 
$H \subeq G^\tau$ an open subgroup. 
Then two mutually exclusive cases occur:
\begin{itemize}
\item[\rm(a)] $\Ad(H)h = \Ad(H_e)h$ and $G/H$ is causal.
\item[\rm(b)] $-h \in \Ad(H)h$ and $G/H$ is not causal.
\end{itemize}
\end{Proposition}

\begin{Proof} Clearly, $\Ad(H)h \subeq \cE_c(\g) \cap \fq$. In view of
  Lemma~\ref{lem:hpm1}(b), 
we either have $\Ad(H)h =\Ad(H_e)h$ or $-h \in \Ad(H)h$. 
In the first case $G/H$ is causal and in the second case it is not. 
\end{Proof}

The following corollary identifies the minimal causal symmetric space
in terms of the symmetry property of the corresponding Euler element. 
\begin{Theorem}  \label{thm:2.23} 
Let $(\g,\tau)$ be simple ncc with causal Euler element 
$h \in \cE_c(\fq)$ and  $G = \Inn(\g)$. 
Then $\cM(\g,\tau) \cong G/G^\tau$ is causal if and only if $h$
is not symmetric, i.e., $-h \not\in \cO_h$. 
\end{Theorem} 

\begin{Proof} Let $H := \Inn(\g)^\tau$, so that $G/H \cong \cM(\g,\tau)$ and consider 
the pointed generating closed convex cone $C \subeq \fq$ generated by~$H_e.h$. 
If $h$ is symmetric, then $-h = g.h$ for some $g\in G$.
Proposition~\ref{prop:orbitq} implies the existence of a
  Weyl group  element $w \in \cW$ with $w.h = -h$.
  As every $w \in \cW$ is a restriction of $\Ad(k)$ to $\fa$ for some
  $k \in K = G^\theta$ (cf.~\cite[Def.~III.9]{KN96}), we may assume that
  $g \in G^\theta$. Further $g.h = - h$ implies that
  $g$ commutes with $\tau_h$, hence also with $\tau$, i.e.,
  $g \in G^\tau = H$.
Therefore Proposition~\ref{prop:caus1} implies that $\cM(\g,\tau)$ is not causal. 

If, conversely, $\cM(\g,\tau)$
is not causal, then Proposition~\ref{prop:caus1} implies that 
$-h \in H.h \subeq \cO_h$, so that $h$ is symmetric. 
\end{Proof}

\begin{Remark}\label{rmk:dessym} (a) 
  Suppose that $(\g,\tau)$ is an irreducible ncc symmetric Lie algebra
  and $h$ an Euler element with 
  $\theta(h) = -h$ and $\tau = \tau_h \theta$. 
  Let $G$ be a connected Lie group with Lie algebra $\g$ on which
  $\tau$ integrates to an involution $\tau^G$ and let 
  $H \subeq G^{\tau_G}$ be an open $\theta$-invariant subgroup, so that 
\[ H = H_K \exp(\fh_\fp) \quad \mbox{ for } \quad H_K = H^\theta = H \cap K, \] 
and Lemma~\ref{lem:hpm1} implies that 
\[ H_K.h \subeq \cE_c(\fq) \cap \fq_\fp = \{ \pm h\}.\] 
Therefore the causality of $G/H$ is equivalent to $H_K \subeq G^h$ 
(cf.\ Proposition~\ref{prop:caus1}).

\nin (b) If $G = \Inn(\g)$, then $K^h \subeq K^{\tau_h} = K^\tau \subeq G^\tau$, so that 
\[H_{C^{\rm max}} := K^h e^{\ad \fh_\fp} \subeq G^\tau \] 
is the maximal open subgroup $H \subeq G^\tau$ for which $G/H_{C^{\rm max}}$ is causal,
i.e., $G/H_{C^{\rm max}}$ is a minimal causal symmetric space
associated to $(\g,\tau,C^{\rm max})$.

In view of Theorem~\ref{thm:2.23}, $H_{C^{\rm max}} = G^\tau$ if and only if 
$h$ is not symmetric.
We shall see in Theorem~\ref{thm:3.4} below that,
if $\fh$ contains Euler elements, then $h$ is symmetric.
Therefore $H_{C^{\rm max}}$ is an index $2$-subgroup of $G^\tau = \Inn(\g)^\tau$.

\nin(c) (Symmetric Euler element and de Sitter space)
  The symmetry group of de Sitter spacetime is
  $G=\SO_{1,d}(\R)_e$. On its  Lie algebra  
  $\theta(x)=- x^\top$ is a Cartan involution. 
  Let $h=h_{W_1}\in \fg$ be the Euler element
  such that
  \[ e^{t h_{W_1}}p=(\cosh(t)p_0+\sinh(t)p_1,
    \sinh(t)p_0+\cosh(t)p_1,p_2,\ldots,p_d) \]
  and the corresponding wedge reflection is implemented by the linear map 
  \[     \tau_h =\diag(-1,-1,1,1,\ldots,1) \in \SO_{1,d}(\R).\]
    Then
  \[ G^\tau\simeq G_{\be_1}\rtimes \{\1, r_{12}\}
  \simeq\SO_{1,d-1}(\R)_e\rtimes \Z_2, \]
where $r_{12} = \diag(1,-1,-1,1,\ldots, 1)$ is the point
reflection in the $x_1$-$x_2$-plane.
The orbit map of $\be_1 =(0,1,0,\ldots 0)$ induces a diffeomorphism
  $G/G_{\be_1} \to \dS^d$. Note that
  $G_{\be_1}$ is the identity component $G^\tau_e$
  in $G^\tau$. The involution $r_{12}$ is not contained in $G^\tau_e$.
It normalizes $G_{\be_1}$, so that it acts by right multiplication on 
$G/G_{\be_1}$. We thus obtain a $G$-equivariant involution
$\phi(g G_{\be_1}) := g r_{1,2} G_{\be_1}$ on $G/G_{\be_1}$.
The corresponding involution $\psi$ on $\dS^d \cong G/G_{\be_1}$
is $G$-equivariant and maps $\be_1$ to $r_{12}(\be_1) = -\be_1$,
hence satisfies $\psi(x) = -x$ for all $x \in \dS^d$.
As a consequence, the symmetric space $G/G^\tau$ is the 
projective de Sitter space $\dS^{d}/\{\pm\1\}$. 
This is not a causal space
because multiplication by $-1$ reverses the causal structure on $\dS^d$.
\end{Remark}

\subsection{Classifying non-compactly causal structures by Euler elements} 
\label{subsec:4.3}

Let $G$ be a connected Lie group with {\bf simple} Lie algebra $\g$. 
We now describe a bijection between the 
set $\cE(\g)/G$ of $G$-orbits in $\cE(\g)$ and the 
isomorphism classes of non-compactly causal symmetric Lie algebras 
$(\g,\tau,C)$, where $C = C_\fq^{\rm min} \subeq \fq$ is a minimal 
$\Inn_\g(\fh)$-invariant pointed closed convex cone. 
We write 
\[ \cN\cC\cC^{\rm min}(\g) \] 
for the $\Inn(\g)$-orbits in the set of all pairs $(\tau,C)$ for which 
$(\g,\tau,C)$ is an ncc symmetric Lie algebra and $C$ is minimal. 

\begin{Theorem} {\rm(Classification of ncc structures by Euler elements)} 
\label{thm:classif}
Let $\g$ be simple. 
To $h \in \cE(\g)$, we associate 
the ncc symmetric Lie algebra 
$(\g,\tau,C^{\rm min}(h,\theta)) \in \cN\cC\cC^{\rm min}(\g)$, where 
$\tau :=\tau_h \theta$ for a Cartan involution $\theta$ satisfying 
$\theta(h) = -h$ and 
\[ C^{\rm min}(h,\theta) = \oline{\cone}(\Inn_\g(\g^\tau)h).\]
This assignment induces a bijection 
$\Gamma$ from the set $\cE(\g)/\Inn(\g)$ of adjoint orbits of Euler 
elements in $\g$ onto $\cN\cC\cC^{\rm min}(\g)$. 
\end{Theorem}

With the notation from Remark~\ref{rem:2.17}, 
we see in particular that, if $\fa_+ \subeq \fa$ is a closed Weyl chamber
(a fundamental domain for the $\cW$-action), then 
\[ \cE(\g)/\Inn(\g) \cong (\cE(\g) \cap\fa)/\cW \cong \cE(\g) \cap \fa_+ \] 
 is a finite set.  

\begin{Proof} Theorem~\ref{thm:2.3}  implies that 
$(\g,\tau,C^{\rm min}(h,\theta))$ is ncc with $h \in C^{\rm min}(h,\theta)^\circ$. 
If $\theta_1$ is another Cartan involution 
with $\theta_1(h) = -h$, then $\theta_1 \theta(h) = h$. 
Write $\theta_1 = e^{\ad x} \theta e^{-\ad x}$ with $\theta(x) = -x$ 
(cf.\ \cite[Thm.~13.1.7]{HN12}). Then 
\[ h = \theta_1\theta(h) 
= e^{\ad x} \theta e^{-\ad x} \theta(h) 
= e^{2\ad x} \theta \theta(h) 
= e^{2\ad x}h \] 
implies $[x,h] = 0$ 
because $\ad x$ is diagonalizable. Therefore 
\[ \tau_1 := \tau_h \theta_1 = \tau_h e^{\ad x} \theta e^{-\ad x} = 
e^{\ad x} \tau e^{-\ad x}, \] 
and 
\[ e^{\ad x} \: (\g, \tau, C^{\rm min}(h,\theta)) \to (\g, \tau_1, C(h, \theta_1)) \] 
is an isomorphic of ncc symmetric Lie algebras fixing~$h$. 
Therefore the isomorphism class of $(\g,\tau,C^{\rm min}(h,\theta))$ does not depend on the choice of $\theta$. 

If $h_1 = g.h$, then $\theta_1 := g \theta g^{-1}$ 
is a Cartan involution of $\g$ with $\theta_1(h_1) = - h_1$, so that 
\[ \tau_1 := \tau_{h_1} \theta_1 = g \tau g^{-1} \] 
leads to an ncc symmetric Lie algebra $(\g, \tau_1, C(h_1,\theta_1))$ with 
$C(h_1,\theta_1) = g.C^{\rm min}(h,\theta)$. Clearly, 
\[ g \: (\g,\tau,C^{\rm min}(h,\theta)) \to (\g, \tau_1, C(h_1, \theta_1)) \] 
is an isomorphism of causal symmetric Lie algebras. 
This shows that the isomorphism class of $(\g,\tau,C^{\rm min}(h,\theta))$ 
only depends on the orbit $\cO_h \subeq \cE(\g)$. 

If, conversely, $(\g, \tau,C)$ is an ncc symmetric Lie algebra 
with $C$ minimal, and 
$\theta$ a Cartan involution commuting with $\tau$, 
then $C^\circ \cap \fq_\fp$ contains a causal Euler element $h$ 
by Theorem~\ref{thm:uniqueinv}, and minimality implies 
$C = C^{\rm min}(h,\theta)$.
Therefore $\Gamma$ is surjective. To see that it is also injective, 
we recall that any two Euler elements in $C^\circ$ are conjugate under 
$\Inn_\g(\fh)$ (Theorem~\ref{thm:uniqueinv}(b)), 
so that any two Euler elements defining isomorphic 
triples $(\g,\tau,C)$ lie in the same $\Inn(\g)$-orbit. 
\end{Proof}

\begin{Example}\label{ex:3.3} We consider the simple real Lie algebra 
$\g = \fsl_n(\R)$. For\\ $1 \leq p < n$ and $q := n-p$, we obtain an 
Euler element 
\[ h_p := \frac{1}{p+q} \pmat{ q \1_p & 0 \\ 0 & -p \1_q}\] 
(cf.\ Example~\ref{ex:orbitq}). Then 
\[ \tau_{h_p}
\pmat{a & b \\ c & d} =
\pmat{a & -b \\ -c & d} \] 
and $\theta(x) = -x^\top$ is a Cartan involution with $\theta(h_p) = - h_p$. 
For $\tau = \tau_{h_p} \theta$ we then have 
\[ \tau(x) = - \tau_h(x^\top),\qquad 
\tau\pmat{a & b \\ c & d} =
\pmat{-a^\top & c^\top\\ b^\top & -d^\top}. \] 
Therefore $\fh = \so_{p,q}(\R)$ and 
\[ \fq = \Big\{ \pmat{ a & b \\ -b^\top & d} \: 
a^\top =  a, d^\top = d, \tr(a) + \tr(d) = 0\Big\}.\] 
A typical invariant cone in $\fq$ is 
\[ C =\Big\{ \pmat{ a & b \\ -b^\top & d} \in \fq \: 
\pmat{ a & b \\ b^\top & -d} \geq 0 \Big\}.\] 
It contains $h_p$ in its interior. 

Note that the subspace $\fa$ of diagonal matrices in $\fq$ is also maximal 
abelian in $\g$, so that all $G$-orbits in $\cE(\g)$ intersect $\fa$, 
hence also $\fq$ (cf.\ Example~\ref{ex:orbitq}).  
\end{Example}

\begin{Remark}  (a) Let $(\g,\tau)$ be a symmetric Lie algebra of complex type, where 
$\g =\fh_\C$ and $\fh$ is simple hermitian. 
If $\ft \subeq \fh$ is a compactly embedded Cartan 
subalgebra, then $\fa := i \ft \subeq \fq = i \fh$ is maximal 
hyperbolic abelian. As $\Sigma(\g,\fa)$ may have several $3$-gradings, 
there are many Euler elements in $\fa$ which are not contained 
in an $\Inn_\g(\fh)$-invariant cone in $i\fh$. This happens for 
$\fh = \su_{p,q}(\C)$, where we have $p+q-1$ 
$\Inn(\fh)$-orbits of Euler elements in $i \fh$, represented 
by $h_1, \ldots, h_{p+q-1}$ (see Proposition~\ref{prop:orbitq} and 
Theorem~\ref{thm:classif-symeuler}). 
The only Euler elements contained 
in the interior of an invariant cone are those of the form 
$h= i z$, where $z \in \fz(\fh_\fk)$ for a Cartan decomposition 
$\fh = \fh_\fk \oplus \fh_\fp$. This is the case for $h = h_p$. 

\nin (b) The example in (a) shows that, for a simple symmetric Lie algebra 
$(\g,\tau)$, there may be Euler elements in 
$\fq$ which are not contained 
in any hyperbolic $\Inn_\g(\fh)$-invariant cone $C \subeq \fq$, 
i.e., not every Euler element in $\fq$ is causal for $(\g,\tau)$. 
However, the picture changes if we are free to choose~$\tau$. Then Theorem~\ref{thm:classif} implies that  every Euler element $h \in \cE(\g)$ is causal for a suitable 
choice of~$\tau$. 
\end{Remark}

\subsection{The classification of irreducible ncc symmetric Lie algebras}
\label{subsec:4.4} 

We have seen above that irreducible non-compactly causal symmetric
Lie algebras $(\g,\tau)$ are classified by
Euler element $h$ in real simple Lie algebras $\g$ 
and Cartan involutions $\theta$ on $\g$ satisfying
$\theta(h) = -h$ (Section~\ref{subsec:4.3}).
In this case $(\g,\tau)$ with $\tau = \theta \tau_h$ is the corresponding
causal symmetric Lie algebra. Along these lines, one obtains a complete
classification of these structures, which is described in Table~$3$ below. 
It  lists all irreducible non-compactly 
causal symmetric Lie algebras $(\g,\tau)$ according to the subdivision
into the following $4$ types: 
\begin{itemize}
\item Complex type: $\g= \fh_\C$ and $\tau$ is complex conjugation with 
  respect to $\fh$. In this case $\g^c \cong \fh^{\oplus 2}$, so that
  $\rk_\R(\g^c) = 2 \rk_\R(\fh)$. 
\item Cayley type (CT): $\tau = \tau_{h_1}$ for an Euler element $h_1 \in \fh$.
  Then $\rk_\R(\g^c) = \rk_\R(\g) = \rk_\R(\fh)$.
\item Split type (ST):  $\rk_\R(\fh) = \rk_\R(\g^c)$
  and $(\g,\tau)$ is not of  Cayley type.
\item Non-split type (NST):
  $\rk_\R(\g^c) = 2 \rk_\R(\fh)$ and
  $(\g,\tau)$ is not of complex type.
\end{itemize}

\begin{Remark}
(a) If $(\g,\tau)$ is a simple symmetric Lie algebra,
  then either the dual symmetric Lie algebra $\g^c = \fh + i \fq$
  is simple ($(\g,\tau)$ is not of complex type) 
    and if $(\g,\tau)$ is of complex type,
    then $(\g^c,\tau) \cong (\fh \oplus \fh,\tau_{\rm flip})$. 

    \nin (b) For an irreducible ncc symmetric Lie algebra $(\g,\tau)$,
  the Lie algebra $\g$ is simple: Since $\tau = \theta \tau_h$
  for a causal Euler element $h$, this follows from the fact that
  $\theta$ and $\tau_h$, hence also   $\tau$, preserve all simple ideals.

  \nin (c) For an irreducible ncc symmetric Lie algebra $(\g,\tau)$,
  either $\g^c$ is simple hermitian or isomorphic to
  $(\fh \oplus \fh, \tau_{\rm flip})$ (if $\g$ is complex). 
\end{Remark}

In the table below we write $r = \rk_\R(\g^c)$ and $s = \rk_\R(\fh)$.
Further $\fa \subeq \fp$ is maximal abelian of dimension~$r$.
For root systems $\Sigma(\g,\fa)$ of type $A_{n-1}$,
there are $n-1$ Euler elements $h_1, \ldots, h_{n-1}$,
but for the other root systems there are less;
see Theorem~\ref{thm:classif-symeuler} for the concrete list.
For $1 \leq j <n$ we write $j' := \min(j, n-j)$. \\

\hspace{-27mm}
\begin{tabular}{||l|l|l|l|l|l|l|l||}\hline
$\g$ &  $\g^c = \fh + i \fq$ & $r$ & $\fh = \g^{\tau_h\theta}$ 
& $s$ & $\Sigma(\g,\fa)$  & $h$ &  $\g_1(h)$  \\ 
\hline
\hline 
Complex type \phantom{\Big (} &&& &&  &&\\
\hline 
$\fsl_n(\C)$ & $\su_{j,n-j}(\C)^{\oplus 2}$  \phantom{\Big(}& $2j'$
& $\su_{j,n-j}(\C)$ & $j'$ & $A_{n-1}$ & $h_j$ & $M_{j,n-j}(\C)$  \\
 $\sp_{2n}(\C)$  & $\sp_{2n}(\R)^{\oplus 2}$ & $2n$ &  $\sp_{2n}(\R)$ 
& $n$ & $C_n$ & $h_n$ & $\Sym_n(\C)$   \\
$\so_n(\C), n > 4$ &  $\so_{2,n-2}(\R)^{\oplus 2}$ & 4 & $\so_{2,n-2}(\R)$ 
& 2 & $D_{[\frac{n}{2}]}, B_{[\frac{n}{2}]}$  &$h_1$ & $\C^{n-2}$   \\
$\so_{2n}(\C)$ & $\so^*(2n)^{\oplus 2}$ & $2[\frac{n}{2}]$ & $\so^*(2n)$&
$[\frac{n}{2}]$ &$D_{n}$ & $h_{n-1}, h_{n}$ & $\Skew_{n}(\C)$      \\
 $\fe_6(\C)$ &  $(\fe_{6(-14)})^{\oplus 2}$ & $4$ &$\fe_{6(-14)}$ & $2$ &
$E_6$ & $h_1, h_6$ & $M_{1,2}(\bO)_\C$    \\
 $\fe_7(\C)$ &  $(\fe_{7(-25)})^{\oplus 2}$ & $6$ & $\fe_{7(-25)}$ & $3$ & 
$E_7$ & $h_7$ & $\Herm_3(\bO)_\C$     \\
\hline
Cayley type  \phantom{\Big (}&& &&&  &&\\
\hline 
$\su_{r,r}(\C)$ &  $\su_{r,r}(\C)$ & $r$ & $\R \oplus \fsl_r(\C)$& $r$ &
$C_r$ & $h_r$ & $\Herm_r(\C)$ \\
$\sp_{2r}(\R)$  & $\sp_{2r}(\R)$ & $r$ & $\R \oplus \fsl_r(\R)$ & $r$ & 
$C_r$ & $h_r$ & $\Sym_r(\R)$   \\
$\so_{2,d}(\R), d> 2$ &  $\so_{2,d}(\R)$ & $2$ &  
$\R \oplus \so_{1,d-1}(\R)$  & $2$ &   $C_2$ &$ h_2$ & $\R^{1,d-1}$  \\
$\so^*(4r)$ &  $\so^*(4r)$ & $r$ & 
$\R \oplus \fsl_r(\H)$& $r$ & $C_r$ & $h_r$ & $\Herm_r(\H)$ \\
 $\fe_{7(-25)}$ & $\fe_{7(-25)}$ & $3$ & 
$\R \oplus \fe_{6(-26)}$ & $3$  & $C_3$ & $h_3$ & $\Herm_3(\bO)$  \\
\hline
Split type \phantom{\Big (} &&&&&  &&\\
\hline 
$\fsl_{n}(\R)$ &$\su_{j,n-j}(\C)$ & $j'$ &  $\so_{j,n-j}(\R)$ &
$j'$ & 
$A_{n-1}$ & $h_j$ & $M_{j,n-j}(\R)$ \\
$\so_{n,n}(\R)$ & $\so^*(2n)$ & $[\frac{n}{2}]$ & $\so_{n}(\C)$ &
$[\frac{n}{2}]$ &$D_{n}$ & $h_{n-1}, h_{n}$ & $\Skew_n(\R)$  \\
  $\so_{p+1,q+1}(\R)$ &$\so_{2,p+q}(\R)$& $2$ & 
$\so_{1,p}(\R) \oplus \so_{1,q}(\R)$ & $2$
& $ B_{p+1}\, (p<q)$  & $h_1$ &$\R^{p,q}$    \\
$p,q > 1$ && && & $D_{p+1}\, (p = q)$  & &    \\
$\fe_6(\R)$ & $\fe_{6(-14)}$& $2$ & $\fu_{2,2}(\H)$
& $2$ &  $E_6$ & $h_1, h_6$ &
$M_{1,2}(\bO_{\rm split})$   \\
$\fe_7(\R)$ & $\fe_{7(-25)}$ & $3$ &$\fsl_4(\H) = \su^*(8)$ & $3$ & $E_7$ & $h_7$ &
$\Herm_3(\bO_{\rm split})$   \\
\hline 
Non-split type \phantom{\Big (} && &&&  &&\\
\hline 
$\fsl_n(\H)$ &$\su_{2j,2n-2j}(\C)$ & $2 j'$ & $\fu_{j,n-j}(\H)$
& $j'$ & $A_{n-1}$ & $h_j$ & $M_{j,n-j}(\H)$   \\
 $\fu_{n,n}(\H)$ & $\sp_{4n}(\R)$ & $2n$ & $\sp_{2n}(\C)$ & $n$ 
& $C_{n}$ & $h_n$ & $\Aherm_n(\H)$  \\
$\so_{1,d+1}(\R)$  & $\so_{2,d}(\R)$ & $2$ & $\so_{1,d}(\R)$ & $1$ 
& $A_1$ & $h_1$ & $\R^d$ \\
$\fe_{6(-26)}$ & $\fe_{6(-14)}$  &$2$  & $\ff_{4(-20)}$  & $1$ &
$A_2$ & $h_1, h_2$ &
$M_{1,2}(\bO)$  \\
\hline
\end{tabular}\\[2mm] 
{\rm Table 3: Irreducible ncc symmetric Lie algebras 
with corresponding causal\\ Euler elements~$h \in \fa$} 
 
\section{Strongly orthogonal roots} 
\label{sec:5}

In this section we introduce a key technical tool in the structure theory
of causal symmetric spaces: maximal $\tau$-invariant 
sets of long strongly orthogonal roots.
In Section~\ref{subsec:5.1} we recall the construction of such sets
from \cite{Ol91} and discuss its basic properties.
In particular we connect Euler elements with
strongly orthogonal roots and the corresponding $\fsl_2$-subalgebras
(Proposition~\ref{prop:testing}).
An important application of this technique is Theorem~\ref{thm:3.4}
that characterizes irreducible non-compactly causal symmetric Lie algebras
$(\g,\tau,C)$ for which $\fh$ contains Euler elements as those, 
for which the causal Euler element is symmetric.

\subsection{A $\tau$-invariant set of strongly orthogonal roots}
\label{subsec:5.1}

Let $(\g,\tau)$ be an irreducible non-compactly  
causal symmetric Lie algebra and recall that this implies that $\g$ is simple 
(\cite[Prop.~2.13]{NO22b}, \cite[Rem.~3.1.9]{HO97}). 
We fix a causal Euler element $h \in \fq$ and the corresponding 
Cartan involution $\theta = \tau\tau_h$ 
(Theorem~\ref{thm:uniqueinv}); then 
\[ \g^h = \ker(\ad h)= \fh_\fk + \fq_\fp \quad \mbox{ and } \quad 
\fz(\g^h) \cap \fq_\fp = \R h\]
by Lemma~\ref{lem:euler-exist}. 
In particular, any maximal abelian subspace $\fa \subeq \fq_\fp$ contains 
$h$ and is also maximal abelian in $\fq$ and $\fp$, so that 
\begin{equation}
  \label{eq:dima}
 \dim \fa = \rk_\R(\g) = \rk_\R(\g^h).
\end{equation}
Let $\fc \subeq \g$ be a Cartan subalgebra containing $\fa$. Then 
$\fc_\fk := \fc \cap \fk \subeq \fh_\fk$ 
and $\fc$ is invariant under $\tau$ and $\theta$, 
which coincide on $\fc$. 

For the root decomposition of $\g_\C$ with respect to~$\fc_\C$, we then have 
\[ \Sigma =  \Sigma(\g_\C,\fc_\C) \subeq i \fc_\fk^* \oplus \fa^* 
\quad \mbox{ with } \quad 
 -\tau\alpha = \oline \alpha \quad \mbox{ for } \quad 
\alpha \in \Sigma.\] 
Further, $h$ induces a 
$3$-grading of the root system 
\[ \Sigma = \Sigma_{-1} \dot\cup \Sigma_0 \dot\cup \Sigma_1
  \quad \mbox{ with } \quad
  \Sigma_j = \{ \alpha \in \Sigma \: \alpha(h) = j\}.\]
The $\tau$-invariance of $\fc$ implies that $\tau$ acts on 
$\Sigma$, and since $\tau(h) =- h$, we have 
\[ \tau\Sigma_0 = \Sigma_0 \quad \mbox{ and }\quad 
\tau\Sigma_1 = \Sigma_{-1}.\] 
Note that $\Sigma_0 = \Sigma(\fg^h, \fc)$  is the root system of 
the subalgebra~$\g^h$.

According to \cite[Thm.~3.4]{Ol91}, there exists a maximal 
subset $\Gamma = \{ \gamma_1, \ldots, \gamma_r \} 
\subeq \Sigma_1$ of strongly orthogonal long roots, i.e., 
$\gamma_j \pm \gamma_k \not\in\Sigma$ for $j \not=k$, such that 
$-\tau(\Gamma) = \Gamma$
(cf.\ Proposition~\ref{prop:hc}).
Then $\Gamma = \Gamma_0 \dot\cup \Gamma_1$, 
where $\Gamma_0 := \{ \gamma \in \Gamma \: - \tau\gamma = \gamma\}$ 
is the subset of real-valued roots in $\Gamma$, and 
$-\tau$ acts by complex conjugation 
without fixed points on $\Gamma_1$, so that this set has 
an even number of elements. 
For $r_0 := |\Gamma_0|$ and $r_1 := |\Gamma_1|/2$, we obviously 
have 
\[ r = r_0 + 2r_1 \quad \mbox{  and put } \quad 
s := r_0 + r_1.\] 
By \cite[Lemma~4.3]{Ol91}, 
\begin{equation}
  \label{eq:ranks}
\rk_\R(\fh) = s.  
\end{equation}
If $(\g,\tau)$ if of Cayley type 
  then $\fh = \g^{h_1}$ for an Euler element $h_1 \in \h$
(see Section~\ref{subsec:4.4}), hence $\fh$  is of real rank $s$ and 
$\g \cong \g^c$ is of real rank $r$. We therefore have $r = s$, 
which is equivalent to $r_1 = 0$.

The set $\Gamma$ of strongly orthogonal roots specifies 
a subalgebra 
\begin{equation}
  \label{eq:subalgroot}
\fs_\C := \sum_{\gamma \in \Gamma} \g_\C^\gamma + \g_\C^{-\gamma} + \C \gamma^\vee 
\cong \fsl_2(\C)^r 
\end{equation}
of $\g_\C$ invariant under the $\C$-linear extension of $\tau$. 
The involution 
$\tau$ leaves all ideals in $\fs_\C$ corresponding to roots in $\Gamma_0$ 
invariant and induces flip involutions on the ideals corresponding 
to $(-\tau)$-orbits in $\Gamma_1$. 
We thus obtain 
\begin{equation}
  \label{eq:fs-decomp}
\fs  \cong \fsl_2(\R)^{r_0} \oplus \fsl_2(\C)^{r_1}
\quad \mbox{ and } \quad 
 \fs^{\tau} 
\cong \so_{1,1}(\R)^{r_0} \oplus \su_{1,1}(\C)^{r_1}.
\end{equation}
For any Euler element $h_\fs \in \fs \cap \fq_\fp$ 
whose kernel contains no non-zero ideal, the Lie subalgebra 
generated by $h_\fs$ and $\ft_\fq$ is isomorphic to $\fsl_2(\R)^r$. 
This is easily verified by considering the two cases $\fsl_2(\R)$ and 
$\fsl_2(\C)$ separately. 

For the $\fsl_2(\R)$-ideals of $\fs$ we have 
$\fsl_2(\R)^c = \su_{1,1}(\C)$ and for the 
$\fsl_2(\C)$ ideals, we have 
$\fsl_2(\C)^c \cong \su_{1,1}(\C)^{\oplus 2}$, so that 
\[ \fs^c = \fs_\C \cap \g^c \cong \su_{1,1}(\C)^r.\]
In particular $\fs^c \subeq \g^c$ is a subalgebra of full real rank 
$r = \rk_\R(\g^c)$.

The Cayley transform $\kappa_{h} := e^{\frac{\pi i}{2} \ad h}$  
induces a complex structure on $\fh_\fp + i \fq_\fk$ for which $\tau$ 
acts as an antilinear involution. Therefore $\kappa_h$ 
maps $\fh_\fp$ bijectively to $i \fq_\fk$, and thus 
\eqref{eq:ranks} implies that the maximal abelian subspaces 
of $\fq_\fk$ are also of dimension~$s$. Note also that 
$\ad h$ defines a bijection $\fh_\fp \to \fq_\fk$, commuting with $\Inn_\fg(\fh_\fk)$.

In view of \eqref{eq:ranks}, 
$\fs \cap \fq_\fk$ contains a maximal abelian subspace $\ft_\fq$ of 
$\fq_\fk$. With respect to the decomposition of $\fs$ in \eqref{eq:fs-decomp}, 
we may choose 
\[ \ft_\fq = \so_2(\R)^{r_0 + r_1} = \so_2(\R)^s.\] 

\begin{Lemma}\label{lem:equalspec}
 For $x \in \ft_\fq$, we have 
$\rho(\ad x) = \rho(\ad x\res_{\fs}),$ 
where $\rho$ denotes the spectral radius. 
With the basis 
\[ z^j = \Big(0,\ldots, 0, 
\frac{1}{2}\pmat{0 & -1 \\ 1 & 0},0,\cdots, 0\Big), \quad j =1,\ldots, s, \]
in $\so_2(\R)^s$ we have for $x = \sum_{j = 1}^s x_j z^j$ 
\begin{equation}
  \label{eq:rhosl2}
 \rho(\ad x) = \max \{ |x_j| \:  j =  1,\ldots, s \}.
\end{equation}
\end{Lemma}

\begin{Proof}  That $\rho(\ad x\res_{\fs})$ equals 
$\max \{ |x_j| \:  j =  1,\ldots, s \}$ follows by a matrix calculation 
in $\fsl_2(\R)$. As complexification does not change the spectral radius, 
$\rho(\ad x\res_{\fs}) = \rho(\ad x\res_{\fs_\C})$. 
It therefore remains to observe that for semisimple elements 
\[ y = \sum_{j = 1}^r y_j z^j \in \fs_\C \cong \fsl_2(\C)^r,\] we have 
\[ \rho(\ad y) = \rho(\ad y\res_{\fs_\C}).\] 
Clearly, $\rho(\ad y) \geq  \rho(\ad y\res_{\fs_\C}).$ 
From \eqref{eq:hc1} in Proposition~\ref{prop:hc} it follows that, 
for the simple ideals in $\fs_\C$, the Lie algebra 
$\g_\C$ contains only simple modules of dimension~$1$, $2$ and~$3$. 
This implies that $\rho(\ad y) \leq  \rho(\ad y\res_{\fs_\C})$ 
for any semisimple element $y \in \fs_\C$. 
\end{Proof}

\begin{Proposition} \label{prop:testing}
Let $(\g,\tau,C)$ be a simple ncc symmetric Lie algebra. 
Pick a causal Euler element $h \in C^\circ $,
a Cartan involution $\theta$ with $\theta(h) = -h$, so that
  $\tau = \tau_h\theta$, and 
$\ft_\fq \subeq \fq_\fk$ maximal abelian. 
Then $\dim \ft_\fq = s = r_0 + r_1$, 
and the following assertions hold: 
\begin{itemize}
\item[\rm(a)] The Lie algebra $\fl$ generated by $h$ and $\ft_\fq$ 
is reductive. 
\item[\rm(b)] The  commutator algebra $[\fl,\fl]$ is
 isomorphic to $\fsl_2(\R)^s$. 
\item[\rm(c)] $\fz(\fl) = \R h_0$ for some 
hyperbolic element $h_0$ satisfying $\tau(h_0) =- h_0 = \theta(h_0)$ 
which is zero if and only if $\g^c$ is of tube type
or a sum of two ideals of tube type. 
\item[\rm(d)] The Lie algebra $\fl$ is $\tau$-invariant 
  with $\fl^\tau \cong \so_{1,1}(\R)^s$.
 It is also $\theta$-invariant with
    $\fl^\theta  = \ft_\fq = \so_2(\R)^s$.
\end{itemize}
\end{Proposition}

Note that (c) implies that $\fl$ is semisimple, i.e., $h \in [\fl,\fl]$,  
if and only if $\g^c$ is of tube type. 

\begin{Proof} {\bf The complex case:} 
If $\g$ is complex, then $\g \cong \fh_\C$, where 
$\tau$ is complex conjugation, $\fh$ is simple hermitian and $\fq = i \fh$. 
For a Cartan decomposition 
$\fh = \fh_\fk \oplus \fh_\fp$, we then have 
$\fk = \fh_\fk + i \fh_\fp$ and $\fp = \fh_\fp + i \fh_\fk$. 
Now $\ft_\fq = i \fa$, where $\fa \subeq \fh_\fp$ is maximal abelian 
and $s = \rk_\R(\fh)$ and $r = \rk_\R(\g^c) = \rk_\R(\fh \oplus \fh) = 2s$, 
so that $r_0 = 0$. A maximal system 
$\{ \gamma_1, \ldots, \gamma_s\}$ of long strongly orthogonal roots in 
$\Sigma(\fh,\fa)$ now leads to a complex subalgebra 
\[ \fs := \sum_{j = 1}^s \g(\gamma_j) \cong \fsl_2(\C)^s \subeq \g, 
\quad \mbox{ where } \quad 
\g(\alpha) := \g^\alpha + \g^{-\alpha} + [\g^\alpha, \g^{-\alpha}],  \] 
which is invariant under $\ad h$.
Next we recall that $\fh$ is of tube type if  and only if
  the corresponding system of restricted roots is of type
  $C_r$. By Corollary~\ref{cor:hc}   this is equivalent to $h$
  being symmetric and by  Proposition~\ref{prop:3.7}, this is equivalent to
  $h \in \fs$. 
It is easy to see that $h$ generates with $\ft_\fq \cong \so_2(\R)^s$ 
a subalgebra $\fl = \R h + \fl'$ with $\fl' \cong \fsl_2(\R)^s$. 
The assertion thus follows with $r_0 = 0$ and $r_1 = s$. 
We write $h_0$ for the component of $h$ in the center of $\fl$.  
Then  $\tau(h_0) = - h_0$ follows from 
$\tau(h) = - h$. Likewise $\theta(h) = -h$ implies 
$\theta(h_0) = - h_0$, so that $h_0$ is hyperbolic. 

\nin {\bf The real case:} If $\g$ is not complex, then  
the $c$-dual Lie algebra $\g^c := \fh + i \fq$ is 
simple hermitian. 
The subspace $i\ft_\fq \subeq i \fq_\fk = \fq^c_\fp$ is maximal abelian. 
Since all maximal abelian subspaces in $\fq_\fk$ are conjugate under 
the group $\Inn_\g(\fh_\fk)$, we may w.l.o.g.\ assume that 
$i \ft_\fq  \subeq \fs_\C$, where $\fs_\C \cong \fsl_2(\C)^r$ 
is constructed from a $-\tau$-stable maximal set of strongly 
orthogonal roots as in \eqref{eq:subalgroot} above. 
The action of $\tau$ and complex conjugation with respect to $\g$ 
on $\fs_\C$ now show with \eqref{eq:fs-decomp} above that 
\[ \fs := \g \cap \fs_\C  \cong \fsl_2(\R)^{r_0} \oplus \fsl_2(\C)^{r_1}
\quad \mbox{ and } \quad 
 \fs^{\tau} \cong \so_{1,1}(\R)^{r_0} \oplus \su_{1,1}(\C)^{r_1}.\] 
As $\fs^{-\tau} \cap \fk \cong \so_2(\R)^{r_0+r_1}$ 
is contained in a maximal abelian subspace $\ft_\fq \subeq \fq_\fk$, 
it follows that 
\[ \ft_\fq = \fs^{-\tau} \cap\fk = \fq_\fk \cap \fs.\] 

As in the complex case, we see
that $\g^c$ is of tube type if and only if $h \in \fs_\C$
(Corollary~\ref{cor:hc}, Proposition~\ref{prop:3.7}).
Then $h$ is an Euler element in~$\fs$, 
contained in $\fs^{-\tau}$. Inspecting the configurations we obtain 
for the components in the simple ideals of $\fs$, 
it now follows easily that 
the Lie algebra $\fl$ generated by $h$ and $\ft_\fq$ has the 
asserted form. Here the main point is to see that
$\so_2(\R)$ generates with an Euler element in
$i \fsl_2(\R) \subeq \fsl_2(\C)$, a $3$-dimensional Lie subalgebra
isomorphic to $\fsl_2(\R)$. 
\end{Proof}

\subsection{Characterization of modular ncc symmetric Lie algebras}
\label{subsec:5.2}

In this subsection we characterize non-compactly causal
Lie algebras $(\g,\tau, C)$ for which $\fh$ contains an Euler
element by the symmetry of the corresponding causal Euler element.
This condition plays an important role in \cite{NO22b}, where
we study the flows of Euler elements in $\fh$ on the symmetric
space.

\begin{Definition} \label{def:modularss}
 A {\it modular causal symmetric Lie algebra} is a quadruple
  $(\g,\tau, C,h')$, where $(\g,\tau,C)$ is a causal symmetric Lie algebra
  and $h' \in \fh = \g^\tau$  is an  Euler element.
  We somethimes call $(\g,\tau, C)$ {\it modular} if such an Euler
  element $h'$ exists. 
\end{Definition}

\begin{Theorem} \label{thm:3.4} For a simple ncc symmetric Lie algebra $(\g,\tau,C)$ with 
  causal Euler element $h \in \fq_\fp$ satisfying $\tau = \tau_h \theta$,
  the following are equivalent: 
  \begin{itemize}
  \item[\rm(a)] $\fh \cap \cE(\g) \not=\eset$. 
  \item[\rm(b)] $\g^c$ is either simple of tube type or a direct sum of two such ideals. 
  \item[\rm(c)] $h$ is a symmetric Euler element, i.e., 
$-h \in \cO_h = \Inn(\g)h$. 
  \item[\rm(d)] There exists a second Euler element $h_1$ such that 
$\tau_{h_1}(h) = -h$.
\item[\rm(e)] There exists an $\fsl_2(\R)$-subalgebra containing~$h$
  that is invariant under $\tau$ and~$\theta$.
  \item[\rm(f)] $\fh \cap \cO_h \not=\eset$. 
  \end{itemize}
\end{Theorem}

\begin{Proof} (a) $\Rarrow$ (b): Suppose that $h_1 \in \fh$ is an  Euler 
element. Then $h_1$ also is an Euler element in the dual Lie algebra 
$\g^c$. Either $\g^c$ is simple (if $\g$ is not of complex type) 
or $(\g^c, \tau^c) \cong (\fh \oplus \fh, \tau_{\rm flip})$ 
(if $\g$ is of complex type). Now 
\cite[Prop.~3.11(b)]{MN21} implies that, in the first case, $\g^c$ 
is of tube type, and,  in the second case, the same argument 
shows that $\fh$ is of tube type (cf.\ also Corollary~\ref{cor:hc}).

\nin (b) $\Rarrow$ (c): From Proposition~\ref{prop:testing}(c) 
we infer that 
the Lie subalgebra 
$\fl$ generated by $h$ and $\ft_\fq$ is 
isomorphic to $\fsl_2(\R)^s$. 
Therefore (c) follows from the fact that all Euler elements 
in $\fsl_2(\R)$ are symmetric. 

\nin (c) $\Leftrightarrow$ (d): 
The equivalence of (c) and (d) follows  from \cite[Thm.~3.13(b)]{MN21}.

\nin (d) $\Rarrow$ (e):  
\cite[Thm.~3.13]{MN21} implies that we also have 
$\tau_h(h_1) = - h_1$, i.e., $h_1 \in \g^{-\tau_h}$.  
As $h_1$ is hyperbolic, it is conjugate under  
$\Inn_\g(\g^{\tau_h}) = \Inn_\g(\fz_\g(h))$ to an element 
of $\g^{-\tau_h} \cap \fp$ (\cite[Cor.~II.9]{KN96}). 
We may therefore assume that $\theta(h_1) = - h_1$.
Then $\tau = \tau_h \theta$ satisfies $\tau(h_1) = h_1$, i.e., 
$h_1 \in \fh$. \cite[Thm.~3.13]{MN21} further implies that 
$\fb := \Spann \{ h,h_1, [h,h_1]\} \cong \fsl_2(\R)$.
This subalgebra is clearly invariant under $\theta$ and $\tau$.

\nin (e) $\Rarrow$ (f): If $\fb \subeq \g$ is  isomorphic to
$\fsl_2(\R)$, invariant under $\tau$ and $\theta$ and
$h \in \fb$, then $\fb^\tau$ contains an Euler element $h_1$ of
$\fb$, but then $h_1 \in \Inn(\fb)h$ implies that $h_1 \in \cO_h \cap  \fh$.

\nin (f) $\Rarrow$ (a) is trivial. 
\end{Proof}

  \begin{corollary} \label{cor:3.5} 
Let $(\g,\tau) = (\fh_\C,\tau)$ be an irreducible ncc symmetric Lie algebra 
of complex type and $h \in \cE_c(\fq)$ a causal
Euler element. Then the following are equivalent: 
\begin{itemize}
\item[\rm(a)] The real form $\fh$ is of tube type. 
\item[\rm(b)] $h$ is symmetric, i.e., $-h \in \cO_h$. 
\end{itemize}
  \end{corollary}

\begin{Proof} As $\g^c \cong \fh \oplus \fh$, this follows from 
the equivalence of (b) and (c) in Theorem~\ref{thm:3.4}.
\end{Proof}

\begin{Remark} Any pair of Euler elements $(h,h_1)$ which 
are orthogonal in the sense that $\tau_h(h_1) = - h_1$ 
and $\tau_{h_1}(h) = -h$ (cf.\ \cite[Thm.~3.13]{MN21}) 
leads to an embedding of symmetric Lie algebras 
$(\fl,\tau_\fl) \into (\g,\tau)$ 
with $h, h_1 \in \fl\cong \fsl_2(\R)$. 
Here we may w.l.o.g.\ assume that $h_1 \in \fh_\fp$, so that 
$\fl_\fq = [h_1, \fl] = \R h + \R [h_1, h]$. 
The centralizer of $h_1$ in $L := \SL_2(\R)$ is
$L^{h_1} \cong \SO_{1,1}(\R)$. The fact that $\fl$ contains 
an Euler element of $\g$ implies that the adjoint action of 
$L$ on $\g$ descends to a homomorphism 
\[ \Ad(L) \cong \PSL_2(\R) \into \Inn(\g),\] 
which leads to an embedding 
\[ \dS^2 \cong L/L^{h_1} 
\into \Inn(\g)/ \Inn_\g(\fk)^h e^{\ad \fh_\fp} \] 
and also to an embedding of {\it projective $2$-dimensional de Sitter space} 
\[ \PdS^2 := \dS^2/\{\pm \1\} \into \Inn(\g)/ \Inn(\g)^\tau\]
(cf.\ Remark~\ref{rmk:dessym}(c)). 
\end{Remark}

\section{Real crown domains and real tubes} 
\label{sec:6} 

Let $(G, \tau^G)$ be a connected symmetric semisimple Lie group 
with non-compactly causal  symmetric Lie algebra $(\g,\tau,C)$, 
$H \subeq G^\tau$ an open subgroup satisfying $\Ad(H)C = C$
and $M = G/H$ the associated non-compactly causal symmetric  
space. In this section we extend some of the results obtained 
in \cite{NO22b} for the special class of
modular ncc spaces (Definition~\ref{def:modularss}), where $\fh$ contains
an Euler element, to general semisimple non-compactly causal symmetric  Lie algebras. 

Our main result is that, if the cone $C$ is maximal 
 and $(\g,\tau)$ is semisimple without Riemannian 
 ideals, then the connected component
 of $h$ in the intersection of the 
adjoint orbit $\cO_h$ with the 
real tube domain $\cT_{C} = \fh + C^\circ \subeq \g$ 
is the Matsuki crown (see \cite{Ma03}) of the Riemannian symmetric space 
$\cO_h^\fq= \Inn_\g(\fh)h = e^{\ad \fh_\fp}h$, i.e., 
\[ \Inn_\g(\fh) e^{\ad \Omega_{\fq_\fk}}h \quad \mbox{ for } \quad 
\Omega_{\fq_\fk} = \Big\{ x \in \fq_\fk \: \rho(\ad x) < \frac{\pi}{2}\Big\},\] 
where $\rho(\ad x)$ is the spectral radius of $\ad x$ 
(Theorem~\ref{thm:crownchar-gen}) and $\fq_\fk = \fq \cap \fk$ for 
a Cartan decomposition $\g = \fk \oplus \fp$ with~$h \in \fp$.

\begin{Definition} \label{def:3.1}
(a) Let $h \in \fq_\fp \cap C^\circ$ be a causal Euler element, 
Then $\fz_\fh(h) = \fh_\fk$ implies that 
\[\cO_h^\fq := \Inn_\g(\fh)h = e^{\ad \fh_\fp}h\] 
is the non-compact Riemannian symmetric space 
associated to the symmetric Lie algebra~$(\fh,\theta)$. 

\nin (b) For a convex cone $C \subeq \fq$, we write 
\[ \cT_C := \fh + C^\circ\] 
for the corresponding open {\it real tube domain} in $\g$.

\nin (c) We define the {\it Matsuki crown} 
  of the Riemannian symmetric space $\cO_h^\fq$   by 
\[ \cC(\cO_h^\fq) 
:= \Inn_\g(\fh) e^{\ad \Omega_{\fq_\fk}} h
= e^{\ad \fh_\fp}  e^{\ad \Omega_{\fq_\fk}} h, \quad \mbox{ where } \quad 
\Omega_{\fq_\fk} = \Big\{ x \in \fq_\fk \: \rho(\ad x) < \frac{\pi}{2}\Big\}.\] 
Here we have used the polar decomposition 
\[ \Inn_\g(\fh) = \Inn_\g(\fh_\fk) e^{\ad \fh_\fp} \quad \mbox{
and } \quad \Inn_\g(\fh_\fk)\Omega_{\fq_\fk} = \Omega_{\fq_\fk} \] 
so see that 
$\Inn_\g(\fh_\fk) e^{\ad \Omega_{\fq_\fk}} h = e^{\ad \Omega_{\fq_\fk}} h$. 
\end{Definition}

\begin{Remark} \label{rem:matsuki1}
  (The connection with Matsuki's domains) 
To see that the inverse image of our domain $\cC(\cO_h^\fq)$ 
in $G$ is one of the domains considered by Matsuki in \cite[\S 1.2]{Ma03}, 
let us recall his setting: On $G$ we consider 
a Cartan involution~$\theta$ with $\theta(h) = -h$ 
and the involution $\tau = \tau_h\theta$ and consider the the connected 
groups 
\[ H := G^\tau_e \quad \mbox{ and } \quad H' := G^{\tau \theta}_e = G^{\tau_h}_e 
= G^h_e.\] 
Their Lie algebras are 
\[ \fh = \fh_\fk \oplus \fh_\fp \quad \mbox{ and } \quad 
\fh' = \fh_\fk \oplus \fq_\fp = \ker(\ad h).\] 
We choose a maximal abelian subspace $\ft_\fq \subeq \fq_\fk$. 
As $\theta$ fixes $\ft_\fq$, it leaves the corresponding 
weight spaces $\g_\C^\alpha$ invariant, so that 
$\g_\C^\alpha = \fk_\C^\alpha \oplus \fp_\C^\alpha$, and 
Matsuki considers the subset 
\[ \Sigma(\fp_\C,\ft) :=\{ \alpha \in 
\Sigma(\g_\C,\ft) \: \fp_\C^\alpha \not=\{0\}\}\]   
and the domain 
\begin{equation}
  \label{eq:matsdom}
 \ft^+ 
:=\Big \{ y \in \ft \: (\forall \alpha \in \Sigma(\fp_\C,\ft))\ 
|\alpha(y)| < \frac{\pi}{2}\Big\} 
\supeq  \Big\{ y \in \ft \: \rho(\ad y) < \frac{\pi}{2}\Big\}
\end{equation}
which in turn determines the domain 
\[ \cM(H,H') := H \exp(\ft^+) H' \subeq G.\] 
The discussion of the following two examples, combined with 
Lemma~\ref{lem:equalspec} now implies that we have 
equality in \eqref{eq:matsdom}.

\nin (a) For $\g = \fsl_2(\R)$ and $h = \frac{1}{2}\diag(1,-1)$ we have 
$\theta(x) = - x^\top$, 
\[ \tau_h\pmat{a & b \\ c & -a} = \pmat{a &- b \\ -c & -a}  
\quad \mbox{ and } \quad 
 \tau\pmat{a & b \\ c & -a} 
= -\pmat{a &- b \\ -c & -a}^\top
= \pmat{-a &c \\ b & a}.\] 
So $\fk = \so_2(\R) = \fq_\fk \subeq \fq$ and 
$\ft= \fq_\fk = \so_2(\R)$. We conclude that 
$\Sigma(\g_\C,\ft) = \Sigma(\fp_\C,\ft)$. 

\nin (b) For $\g = \fsl_2(\C) \supeq \fh = \su_{1,1}(\C)$, we have 
$\theta(x) = -x^*$, 
\[ \fh = \Big\{ \pmat{ ia & b \\ \oline b & -ia} \: a \in \R, b \in \C\Big\}, 
\quad \fq = i \su_{1,1}(\C) \] 
and 
\[ \fk = \su_2(\C), \quad 
\fq_\fk = \Big\{ \pmat{ 0 & b \\ -\oline b & 0} \:  b \in \C\Big\} 
\supeq \ft = \so_2(\R).\] 
In this case 
$\Sigma(\g_\C,\ft) = \{\pm \alpha\}$ with 
$\alpha\pmat{0 & 1 \\ -1 & 0} = 2i$, 
$\dim_\C (\g_\C^\alpha) = 2$, and both roots occur in $\fp_\C$ for 
$\fp = i \su_2(\C)$. 
\end{Remark}

\begin{Example} \label{ex:supq}
We consider the hermitian simple Lie algebra 
$\g = \su_{p,q}(\C)$. It is of tube 
type if and only if $p = q$ (cf.\ Section~\ref{subsec:3.1}). 
We assume that $p \leq q$, so that 
$r := \rk_\R(\g) = p$. Then 
\[ z := \frac{i}{p+q}\pmat{q \1_p & 0 \\ 0 & -p\1_q} \] 
is an H-element, the subspace of diagonal matrices 
$\ft$ is a compactly embedded Cartan subalgebra with 
\[ \Sigma(\g,\ft) 
= \{ \eps_j - \eps_k \: j \not=k, j,k = 1,\ldots, p+q\} 
\cong A_{p+q-1}.\] 
We consider the adapted positive system 
\[  \Sigma^+ = \{ \eps_j - \eps_k \: 1 \leq j < k \leq p+q\} \] 
with 
\[  \Sigma_1 = \{ \eps_j - \eps_k \: j \leq p <  k\} 
= \{ \alpha \in \Sigma \: -i\alpha(z) > 0\}. \] 
A set of strongly orthogonal roots in $\Sigma_1$ is 
\[ \Gamma = \{ \gamma_j \: j =1,\ldots, p\}, \quad 
\gamma_j := \eps_j - \eps_{p+j}.\] 
In $\g(\gamma_j) \cong \su_{1,1}(\C)$, we have the H-element 
\[ z^j = \frac{i}{2}(E_{jj} - E_{j+p,j+p}).\] 
We then obtain 
\[ z = z^0 + \sum_{j = 1}^p z^j 
\quad \mbox{ with } \quad 
z^0 = \frac{i}{p+q}\pmat{\frac{q-p}{2}\cdot \1_{2p} & 0 \\ 0 & -p\1_{q-p}}. \] 
Note that $z^0 =0$ if and only if $p = q$, i.e., if 
$\g$ is of tube type. 

With the notation $\bx = \diag(x_1, \ldots, x_{p+q})$, we then have 
\begin{align*}
 C_\ft^{\rm max} 
&= \{ -i \bx\in\ft \: (\forall \alpha \in \Sigma_1)\ 
i\alpha(-ix) = \alpha(x) \geq 0\} \\
&= \{ -i \bx \: x_j > x_k\ \mbox{ for } j \leq p, k > p\}
\end{align*}
and 
\[  C_\ft^{\rm min} 
= \cone \{ -i \alpha^\vee \: \alpha \in \Sigma_1\} 
= \cone \{ -i (E_{jj} - E_{kk}) \: j \leq p < k\}.\] 
Considering extreme points of a basis of this cone, we see that 
\[  C_\ft^{\rm min} 
= \cone \Big\{ -i x \: x_j \geq 0, x_k \leq 0, j \leq p < k; 
\sum_j x_j = 0\Big\}.\] 

If $p < q$, then the entry in position $p+1$ of $z^0$ is negative,
so that $z^0 \not \in C_\ft^{\rm min}$. 
\end{Example}

The preceding example shows in particular
  that the following lemma doe not hold for the minimal cone
  $C_\fq^{\rm  min}$; see also Example~\ref{ex:6.8} below.

\begin{Lemma}
  \label{lem:4.7} Let $(\g,\tau)$ be irreducible ncc, 
$h \in C^{\rm max}_\fq$ a causal  Euler element 
and write $h_0$ for its central 
component in the subalgebra $\fl$ generated by 
$h$ and $\ft_\fq$ {\rm(Proposition~\ref{prop:testing})}. 
Then $h_0 \in C^{\rm max}_\fq$. 
\end{Lemma}

\begin{Proof} (a) We start with a discussion concerning 
hermitian Lie algebras, which corresponds to  the ``complex case''. 
Let $\g$ be simple hermitian and 
\[ \fl \cong \R z +  \fsl_2(\R)^r \subeq \g \] 
the reductive subalgebra generated by 
an H-element $z \in \fz(\fk)$ and the subalgebras $\g(\gamma_j)$ 
associated to a  set 
$\{ \gamma_1, \ldots, \gamma_r\}$ of 
strongly orthogonal positive non-compact roots in $\Delta(\g_\C,\ft_\C)$, 
where $\ft \subeq \fk$ is a Cartan subalgebra and 
\[ \Delta_p^+ = \{ \alpha \in \Delta \: i \alpha(z) = 1\}.\] 
Then 
\[ z = z^0 + \sum_{j = 1}^r z^j, \quad 
z^j= \frac{1}{2} \pmat{0 & -1 \\ 1 & 0} \in \fsl_2(\R)\] 
and 
\[ \ft_\fl = \Spann \{ z^j \: j = 0,\ldots, r \} \subeq \ft \cap \fl \] 
is a compactly embedded Cartan subalgebra of~$\fl$. We then have 
\[ z^j \in C_\g^{\rm min} \quad \mbox{ for } \quad j > 0,\] 
(see Section~\ref{subsubsec:2.3.1} for $C_\g^{\rm min}$)
but in general $z^0 \not\in C_\g^{\rm min}$ 
(Example~\ref{ex:supq}). 
We claim that 
\[ z^0 \in C_\ft^{\rm max} := C^{\rm max}_\g \cap \ft 
= \{ x \in \ft \: (\forall \alpha \in \Delta_p^+)\ 
i\alpha(x) \geq 0\}.\] 

From \eqref{eq:hc1} in Proposition~\ref{prop:hc} 
it follows that, for $\alpha \in \Delta_p^+$, we have 
\[ i\alpha(z^j)\in \{\shalf, 1\}
\quad \mbox{ and } \quad 
\sum_{j = 1}^r i\alpha(z^j) \in \{\shalf, 1\}.\] 
From  $i\alpha(z) = 1$ it now follows that 
\[ i \alpha(z^0) \in \Big\{ 0, \frac{1}{2}\Big\}, 
\quad \mbox{ so that }  \quad 
z^0 \in C_\ft^{\rm max}.\] 

\nin (b) (The real situation) 
Now we prove the lemma. Let $(\g,\tau)$ be simple ncc not of complex type; 
the complex type is covered by (a). 
Then $\g^c$ is simple hermitian and 
\[  \g = \fh \oplus  \fq \subeq \g_\C = \g^c \oplus i \g^c.\] 
We then have maximal invariant cones 
\[  C^{\rm  max}_\fq \subeq \fq \quad \mbox{ and } \quad 
  C^{\rm  max}_{i\g^c} \subeq i\g^c,\] 
and \cite[Lemma~7.10]{Ol91} (see also \cite[Thm.~VIII.1]{KN96}) implies that 
\[  C^{\rm  max}_\fq  = \fq \cap  C^{\rm  max}_{i\g^c}.\] 

As the Euler element $h \in C^{\rm  max,\circ}_\fq \cap \fq_\fp$ also is an Euler 
element in $i\fk^c \subeq i \g^c$. The central component 
$h_0 \in \fz(\fl)$ coincides with the central component 
of $h$, considered as an Euler element in $i\g^c$ (as in (a)). 
Therefore (a) implies that 
\[ h_0 \in \fq \cap  C^{\rm  max}_{i\g^c} = C^{\rm  max}_\fq.\qedhere\] 
\end{Proof}

We shall need the following lemma 
(\cite[Lemma~4.8]{NO22b}). 

\begin{Lemma} \label{lem:closedorbit-gen} 
Let $(G,\tau^G)$ be a connected reductive Lie group. 
If $D \subeq \cT_{C_\fq} = \fh + C_\fq^\circ \subeq \fq$ is compact 
and $H \subeq G^\tau$ an open subgroup with $\Ad(H)C_\fq = C_\fq$, then 
$\Ad(H)D$ is closed in $\g$.   
Moreover, there exists a smooth $\Ad(H)$-invariant function $\psi \: \cT_{C_\fq}
\to (0,\infty)$ such that $z_n \to z_0 \in \partial \cT_{C_\fq}$ 
for $z_n \in \cT_{C_\fq}$ implies $\psi(z_n) \to \infty$.
\end{Lemma}

The following theorem has a twin in 
\cite[Thm.~4.10]{NO22b}, where we consider only the 
modular case, where $\fh$ contains an Euler element, 
and this stronger assumption permits us to work with a general 
cone~$C_\fq$. Here we work with general ncc spaces, where $\fh$ may not 
contain an Euler element, but in this case we have to assume that 
$C_\fq$ is maximal.

\begin{Theorem} \label{thm:crownchar-gen}  
{\rm(Crown Theorem for causal Euler elements)} 
Let $(\g,\tau)$ be a semisimple ncc symmetric Lie algebra 
without $\tau$-invariant Riemannian ideals 
and $h \in \fq$ a causal Euler element. 
Let
\[ H = \Inn_\g(\fh) \subeq G = \Inn(\g) \quad \mbox{ and } \quad
  C := C^{\rm max}_\fq \subeq \fq \]
be the maximal pointed generating 
$\Ad(H)$-invariant cone containing~$h$.
Then the connected component of $h$ in the open subset 
$\cO_{h} \cap \cT_{C}$ of $\cO_{h}$ 
is the Matsuki crown $\cC(\cO_h^\fq) = \Ad(H) e^{\ad \Omega_{\fq_\fk}} h$. 
\end{Theorem}

In \cite[\S 7]{MNO22a} we show that $\cO_h \cap \cT_{C_q^{\rm max}}$
is actually connected. 

\begin{Proof} Our assumptions on $(\g,\tau)$ imply 
that all simple $\tau$-invariant ideals are irreducible ncc 
and that $C$ is the product of the maximal $\Ad(H)$-invariant 
cones in the irreducible factors. Therefore both sets whose equality 
is to be shown are adapted to the decomposition into irreducible factors 
and we may therefore assume that $(\g,\tau)$ is irreducible 
and that $G = \Inn(\g)$ has trivial center
(cf.~\cite[Prop.~2.14]{NO22b}). 
Then $K = G^\theta$ is compact and the polar decomposition 
of $H$ implies that $H = H_K e^{\ad \fh_\fp}$ with $H_K = H^h$ 
(Corollary~\ref{cor:hhcompact}).

We will derive the theorem from the following three claims: 
\begin{itemize}
\item[\rm(a)] $\cC(\cO_h^\fq)$ is connected and open in $\cO_h$. 
\item[\rm(b)] $\cC(\cO_h^\fq) \subeq \cT_{C}$. 
\item[\rm(c)] $\cC(\cO_h^\fq)$ is relatively closed in $\cO_h \cap 
\cT_{C}$.
\end{itemize}
As $\cC(\cO_h^\fq)$ is connected by definition 
(a), (b) and (c) imply that it is the connected component of $h$ in 
the open subset $\cO_{h} \cap \cT_{C}$. 

\nin (a) To see that $\cC(\cO_h^\fq)$ is connected, we use the polar decomposition 
$H = H_K \exp(\fh_\fp)$. Then $\Ad(H_K)\Omega_{\fq_\fk} = \Omega_{\fq_\fk}$ 
and $\Ad(H_K)h = \{h\}$ (cf.\ Lemma~\ref{lem:hpm1}). 
This implies that 
\[ \cC(\cO_h^\fq) 
= \Ad(H) e^{\ad \Omega_{\fq_\fk}} h 
= e^{\ad \fh_\fk} e^{\ad \Omega_{\fq_\fk}} \Ad(H_K) h 
= e^{\ad \fh_\fk} e^{\ad \Omega_{\fq_\fk}}h,\] 
which is obviously connected. 

Next we use \cite[Lemma~C.3(a)]{NO22b} to 
see that the exponential map 
\[ \Exp \: \fq_\fk \to \cO_h, \quad x \mapsto e^{\ad x} h \] 
is regular in $x\in \Omega_{\fq_\fk}$ because $\rho(\ad x) < \pi/2 < \pi$. 
Now \cite[Lemma~C.3(b)]{NO22b} implies that the map 
\[ \Phi \: H \times \Omega_{\fq_\fk} \to \cO_h, \quad 
(g,x) \mapsto \Ad(g) e^{\ad x} h \] 
is regular in $(g,x)$ because $\Spec(\ad x) \subeq (-\pi/2,\pi/2)i$ 
does not intersect $\big(\frac{\pi}{2} + \Z \pi\big)i$. 
This implies that the differential of $\Phi$ is surjective in each point 
of $H \times \Omega_{\fq_\fk}$; hence its image is open.

\nin (b) We  observe that 
both sides of (b) are $\Ad(H)$-invariant, and 
\[ \cC(\cO_h^\fq)
= \Ad(H) e^{\ad \Omega_{\fq_\fk}} h
= \Ad(H) e^{\ad \Omega_{\ft_\fq}} h\]  
for a maximal abelian subspace $\ft_\fq \subeq \fq_\fk$ and 
$\Omega_{\ft_\fq} := \Omega_{\fq_\fk} \cap \ft_\fq$. 
Here we use that $H_K$ fixes $h$ and $\Ad(H_K)\ft_\fq = \fq_\fk$. 
Therefore it suffices to show 
$e^{\ad x} h \in \cT_{C}$ for 
\[ x \in \Omega_{\ft_\fq} = \Big\{ y \in \ft_\fq \: 
\rho(\ad y)  < \frac{\pi}{2}\Big\}.\] 

From Proposition~\ref{prop:testing} we infer that 
the causal Euler element $h$ and $\ft_\fq$ generate a 
$\tau$-invariant reductive subalgebra $\fl$ 
with 
\[ [\fl,\fl] \cong 
\fsl_2(\R)^s  \quad \mbox{ and } \quad 
\fl^\tau \cong \so_{1,1}(\R)^s\]
in which $h$ is a causal Euler element, 
contained in the interior of the 
pointed generating cone 
$C_\fl := C \cap \fl\subeq \fl^{-\tau}$ 
which is invariant under $\Inn(\fl^\tau)$. 
Let $s := r_0 + r_1$. 

For elements of $\fsl_2(\R)$, we recall the notation from 
\eqref{eq:sl2ra} and \eqref{eq:sl2rb} above:
\begin{equation}
  \label{eq:sl2a-sep}
h^0 = \frac{1}{2} \pmat{1 & 0 \\ 0 & -1},\quad 
e =  \pmat{0 & 1 \\ 0 & 0}, \quad 
f =  \pmat{0 & 0 \\ 1 & 0}.
\end{equation}
For a suitable isomorphism 
$[\fl,\fl] \to  \fsl_2(\R)^{r_0} \oplus \fsl_2(\C)^{r_1}$, 
we obtain
\[ h = h_0 + \sum_{j = 1}^s h_j 
\quad \mbox{ and } \quad 
\ft_\fq = \so_2(\R)^s,\]
where we write $h_j$ for the Euler elements $h^0$ in the $j$th summand, 
For 
\[ x 
= \sum_{j = 1}^s x_j \frac{e_j -f_j}{2} \in \ft_\fq,\] 
we then obtain with 
\eqref{eq:turn} 
\begin{equation}
  \label{eq:40b}
 p_\fq(e^{\ad x} h) 
= \cosh(\ad x)(h) 
= h_0 + \sum_{j=1}^s \cos(x_j) h_j.
\end{equation}
Since $\rho(\ad x) < \frac{\pi}{2}$ for $x \in \Omega_{\ft_\fq}$ 
and $i x_j \in \Spec(\ad x)$, 
we have $|x_j| < \frac{\pi}{2}$ for $j =1,\ldots, s$, 
hence $\cos(x_j) \in (0,1]$. 
As $h_0 \in C_\fl$ (Lemma~\ref{lem:4.7}) 
and $h_j \in C_\fl$ for $j = 1,\ldots, s$, we have 
\[ p_\fq(e^{\ad x}h) \in C_{\fa_\fs}^\circ 
\quad \mbox{ for } \quad 
C_{\fa_\fs} := C \cap \fa_\fs, \quad 
\fa_\fs := \Spann \{ h_j \: j =0,\ldots, s\}.\] 
Now $h \in C^\circ \cap \fa_\fs$ implies 
$C_{\fa_\fs}^\circ =
C^\circ\cap \fa_s \subeq C_\fl$ (Lemma~\ref{lem:coneint}), 
so that 
\[ p_\fq(e^{\ad x}h) \in C^\circ.\]  
This proves~(b).

\nin (c) We have to show that 
$\cC(\cO_h^\fq)$ is relatively closed in $\cO_{h}\cap \cT_{C}$. 
If $D \subeq \Omega_{\ft_\fq}$ is compact, 
then $\Ad(H) e^{\ad D} h$ is closed in $\g$ 
by Lemma~\ref{lem:closedorbit-gen}, and by (a) it is contained in~$\cT_{C}$. 

Now suppose that the sequence 
$\Ad(h_n) e^{\ad x_n} h$, 
$h_n \in H$, $x_n \in \Omega_{\ft_\fq}$, converges to some 
element in $\cT_{C}$ which is not contained in $\cC(\cO_h^\fq)$. 
As we may assume that the bounded sequence $x_n \in \Omega_{\ft_\fq}$ converges 
in $\ft_\fq$, it converges by the preceding paragraph to a boundary point 
$y \in \partial \Omega_{\ft_\fq}$. Writing 
\[ y = \sum_{j = 1}^s y_j  \frac{e_j - f_j}{2},\] 
we claim that there exists a $j$ with $|y_j| = \frac{\pi}{2}$. 
As $\rho(\ad y\res_{\fl}) = \max \{ |y_j| \: j =1,\ldots,s \},$ 
this follows from $\rho(\ad y\res_{\fl}) = \rho(\ad y)$ 
(Lemma~\ref{lem:equalspec}). 
Now \eqref{eq:40b} and 
\begin{equation}
  \label{eq:maxinter}
 C \cap \Spann \{ h_0, h_1, \ldots, h_s\} 
\subeq \R h_0 + \sum_{j = 1}^s [0,\infty) h_j.
\end{equation}
imply that 
\[ e^{\ad y} h \in  \partial \cT_{C^{\rm max}_\fq}. \]
For the $H$-invariant function 
$\psi$ from Lemma~\ref{lem:closedorbit-gen}, 
this leads to 
\[ \psi(\Ad(h_n) e^{\ad x_n} h) =\psi(e^{\ad x_n} h) \to \infty, \]
contradicting the convergences of the sequence 
$\Ad(h_n) e^{\ad x_n} h$ in $\cT_{C^{\rm max}_\fq}$. 
\end{Proof}

\begin{Remark} In the special case where 
$(\g,\tau)$ is Riemannian,  $\fq = \fq_\fp$ and 
$\fh = \fh_\fk$, the real crown domain $\cC(\cO_h^\fq)$ reduces to a point. 
Hence there is no interesting analog of the preceding theorem 
in the Riemannian case, and we therefore assume that $(\g,\tau)$ 
contains no Riemannian summands. 
\end{Remark}

\begin{Example} \label{ex:6.8}
  We consider the Lie algebra $\g = \su_{2,1}(\C)$ 
which is hermitian, but not of tube type. In particular 
$(\g_\C, \tau)$, $\tau(z) = \oline z$, is ncc of complex type, 
but $\g$ contains no Euler element. 
Concretely, we have 
\[ \su_{2,1}(\C) 
= \Big\{ \pmat{ a & b \\ b^* & - \tr a} \: a \in \fu_2(\C), b \in \C^2\Big\}.\] 
For the Cartan involution $\theta(x) = - x^*$, we obtain 
\[ \fk =\Big\{ \pmat{ a & 0 \\ 0 & - \tr a} \: a \in \fu_2(\C)\Big\} 
\cong \fu_2(\C)\] 
with 
\[ \fz(\fk) = \R i h_c, \quad h_c = \frac{1}{3}\diag(1,1,-2).\] 
The subspace $\ft \subeq \fk$ of diagonal matrices is a compactly embedded 
Cartan subalgebra of $\g$. 
In $\Delta = \Delta(\g_\C,\ft_\C)$ we have 
\[ \Delta_k = \{ \pm (\eps_1 - \eps_2) \}, \quad 
\Delta_p^+ = \{ \alpha \in \Delta_p \: \alpha(h_c) = 1 \} 
= \{ \eps_1 - \eps_3,\eps_2 - \eps_3 \} \] 
and $\Gamma := \{ \eps_1 - \eps_3 \}$ is a maximal system of strongly 
orthogonal roots. 
In $\fp$ the set $\Gamma$ leads to the maximal abelian subspace 
\[ \fa = \R h \quad \mbox{ with }\quad h := \frac{1}{2}\pmat{
0 & 0 & 1 \\ 
0 & 0 & 0 \\ 
1 & 0 & 0} \] 
which generates with $\fa$ the Lie algebra 
\[ \fs = \ft + [\fs,\fs] \cong \gl_2(\R) 
\quad \mbox{ with } \quad 
[\fs,\fs] 
= \Bigg\{ \pmat{ a & 0 & b  \\ 0 & 0 & 0 \\ \oline b & 0 & -a} 
\: a \in i \R, b \in \C\Bigg\} \cong \su_{1,1}(\C).\] 
In $\fs$ the element $ih_c$ decomposes into 
$ih_c^0 + i h_c^1$, where $ih_c^0$ is central and $ih_c^1 \in [\fs,\fs]$. 
Concretely, we have 
\[ h_c^0 = \frac{1}{6}\diag(-1,2,-1), \quad 
h_c^1 = \frac{1}{3}\diag(1,0,-1).\] 
In $i \ft$ we have the cone 
\begin{align*}
 C^{\rm max} &:= \{ \diag(x_1, x_2, -x_1 - x_2) \: 
2x_1 +  x_2 \geq 0, 2 x_2 + x_1 \geq 0\} \\
&\supeq 
C^{\rm min} = \cone(\diag(1,0,-1), \diag(0,1,-1)).
\end{align*}
We have 
\[ p_{i\fg}(e^{\frac{\pi i}{2} \ad h}h_c) 
= \cos\big(\frac{\pi}{2}\ad h\big) h_c = h_c^0 
\in \partial C^{\rm max} \setminus C^{\rm min}.\]
Therefore the real crown domain in 
$\Ad(G_\C)h_c$ is {\bf not} contained 
in the tube domain $\g + i C_{\rm min,\g} \subeq \g_\C$. 
\end{Example}

\section{Components of the stabilizer group $G^h$} 
\label{sec:7}

This section is devoted to an analysis of the group
$\pi_0(G^h)$ of connected components of the centralizer $G^h$
of an Euler element~$h$ in a simple real Lie algebra.
With a polar decomposition
$G^h = K^h \exp(\fh_\fp)$, this group equals $\pi_0(K^h)$.
As $K/K^h \cong \cO_h^K := \Ad(K)h$ is a compact symmetric space, we discuss
this problem in Section~\ref{subsec:7.1} in the context of
compact symmetric spaces, where
$\pi_0(K^h)$ appears as a quotient of $\pi_1(K/K^h)$
in the long exact homotopy sequence
\[ \pi_1(K) \to \pi_1(K/K^h) \onto \pi_0(K^h) \to \pi_0(K) = \1. \] 

In Section~\ref{subsec:7.2} we explore this situation further,
using that $\cO_h^K$ actually is a symmetric R-space. Here
the strongly orthogonal roots come in handy and permit us
in Theorem~\ref{thm:7.8}  to show that
$G^h$ is connected if $(\g,\tau)$ is either of complex type or
non-split type (cf.\ Section~\ref{subsec:4.4}),
and if it is of split type or Cayley type, then
it either is trivial or $\Z_2$.
In Section~\ref{subsec:7.3} we finally collect some consequences
of this result such as the identity 
\[ \Inn_{\g_\C}(\g^c) \cap \Inn_{\g_\C}(\g)  = \Inn_{\g_\C}(\fk)^h e^{\ad \fh_\fp}.\]

\subsection{The fundamental group of a compact 
symmetric space} 
\label{subsec:7.1}

Let $G$ be a connected symmetric Lie group with {\bf compact} 
Lie algebra~$\g$, $\tau$ an involutive automorphism of $G$,
$\g = \fh \oplus \fq$ the corresponding eigenspace decomposition,
$H \subeq G^\tau$ an open subgroup, $X = G/H$ the corresponding 
compact symmetric space and $q_X \: G \to X$ the quotient map 
with the base point $e_X := q(e)$. 
We pick a maximal abelian subspace 
$\ft_\fq \subeq \fq$ and enlarge it to a Cartan subalgebra 
$\ft \subeq \g$, so that $\ft$ is $\tau$-invariant and 
$\ft = \ft_\fh \oplus \ft_\fq$. 

We write $q_G \: \tilde G \to G$ for the simply connected covering. 
For a subgroup $B \subeq G$, we write $B^* := q_G^{-1}(B)$ for its inverse 
image in $\tilde G$. 
Then $X = G/H \cong \tilde G/H^*$, which leads to an isomorphism 
$\tilde X \cong \tilde G/H^*_e$ and thus to 
$\pi_1(X) \cong \pi_0(H^*)$ 
(\cite[Cor.~11.1.14]{HN12}). 

In $G$ we have the tori 
\[ T_Q := \exp(\ft_\fq) \subeq T := \exp(\ft).\] 

Now $T_X := \Exp(\ft_\fq) = q_X(T_Q) \subeq X$ is a maximal flat torus 
whose fundamental group is the lattice 
\[ \Gamma_X := \{ x \in \ft_\fq \: \Exp(x) = e_X\}.\] 
We likewise have $T_{\tilde X} \subeq \tilde X$ and a discrete 
subgroup $\Gamma_{\tilde X} \subeq \Gamma_X$. 
On the group level we likewise define 
\[ \Gamma_G := \{ x \in \ft \: \exp_G x = e\} 
\cong \pi_1(T) \quad \mbox{ and } \quad 
\Gamma_{\tilde G} := \{ x \in \ft \: \exp_{\tilde G} x = e\} 
\cong \pi_1(T^*).\] 

In the following lemma note that $Z(G) =\{e\}$ if and only if $G = \Inn(\fg)$.

\begin{Lemma} \label{lem:9.1} The following assertions hold: 
  \begin{itemize}
  \item[\rm(a)] The inclusion $T \into G$ induces an isomorphism 
$\Gamma_G/\Gamma_{\tilde G} \to \pi_1(G)$. 
  \item[\rm(b)] If $Z(G) = \{e\}$, then 
$\pi_1(G) \cong Z(\tilde G)$ and 
$\Gamma_G = \{ x \in \ft \: e^{\ad x} = \1\}$. 
  \item[\rm(c)] $G = \exp_G \fh \exp_G \fq$. 
  \item[\rm(d)] $H^* = H^*_e \exp(\{ x \in \ft_\fq \: e^{2 \ad x} = \1\}) 
\subeq H^*_e (H^* \cap T_Q)$. 
  \item[\rm(e)] $H = H_e (H \cap T_Q)$. 
  \end{itemize}
\end{Lemma}

\begin{Proof} (a) follows from \cite[Cor.~14.2.10]{HN12} 
(see also \cite[Thm.~VII.6.7]{He78}).

\nin (b) If $Z(G) = \{e\}$, then $G$ is semisimple and 
$Z(\tilde G) = \ker q_G \cong \pi_1(G)$. 
   
\nin (c) First we assume that $G$ is semisimple, hence compact. 
As a closed subgroup of~$G$, the group $H$ has only 
finitely many connected components and 
$\hat X := G/H_e$ is still compact. 
The surjectivity of the Riemannian exponential 
function of $\hat X$ (\cite[Thm.~I.10.3/4]{He78}) 
implies $\hat X = \Exp(\fq)$, and therefore 
$G = \exp(\fq) H_e = \exp \fq \exp \fh$. 
Here we use that the exponential function of the compact group $H_e$ is surjective.

Now we turn to the general case. 
As the commutator group $G'$ with the semisimple Lie algebra $\g' = [\g,\g]$ 
is compact, the preceding argument yields 
\[ G' = \exp(\fq \cap \g') \exp (\fh \cap \g').\] 
For the, possibly non-compact,  group $G$, we now obtain 
\[ G = Z(G)_e G' 
= \exp(\fh \cap \fz(\g)) \exp(\fq \cap \fz(\g)) 
\exp(\fq \cap \g') \exp(\fh \cap \g') 
= \exp(\fq) \exp(\fh).\] 

\nin (d)  (cf.~\cite[Thm.~VII.9.1]{He78}) If $h \in H^*$, then 
$q_G(h) \in G^\theta$ implies that 
$z := h\theta(h)^{-1} \in \ker(q_G) \subeq Z(\tilde G)$. 
In view of (c), there exists an $x \in\fq$ and 
$g \in H^*_e = \tilde G^\tau$ with 
$h = \exp(x)g$, so that 
\[ z = h\theta(h)^{-1} = \exp(2x) \in Z(\tilde G).\] 
Now we pick $h_1 \in \tilde G^\tau$ 
with $x_1 := \Ad(h_1)x \in \ft_\fq$ 
(\cite[Thm.~II.8]{KN96})  
and find 
\[ z = h_1 z h_1^{-1} 
= \exp(2 \Ad(h_1)x) = \exp(2x_1). \]
Then $ \1 = \Ad(z) =e^{2 \ad x_1}$, and (d) follows. 

\nin (e) follows immediately from (d)  by applying $q_G$. 
\end{Proof}

\begin{Proposition}
  \label{prop:x1} The homomorphism 
$\iota \: \pi_1(T_X) \cong \Gamma_X \to \pi_1(X)$ induced by the inclusion 
$T_X \into X$ is surjective 
and induces an isomorphism 
\[ \Gamma_X/\Gamma_{\tilde X} \to \pi_1(X).\]
\end{Proposition}

\begin{Proof} First we show that $\iota$ is surjective. 
We shall use the exact sequence 
\begin{equation}
  \label{eq:ex1}
 \pi_1(G) \to \pi_1(X) \ssmapright{\delta} \pi_0(H),
\end{equation}
which is part of the long exact homotopy sequence of the 
principal $H$-bundle $G \to X$. 
Every element in $\pi_0(H)$ can be represented 
by an element $g = \exp x$ in $T_Q \cap H$ (Lemma~\ref{lem:9.1}(e)), 
and then $x \in \Gamma_X$. This shows that 
$\delta \circ \iota$ is surjective. 

It remains to see that $\ker\delta \subeq \im\iota$. 
So let $[\gamma] \in \ker\delta$. 
By \eqref{eq:ex1},  $[\gamma]$ comes from a loop in $G$, 
hence from a loop in $T$ 
(Lemma~\ref{lem:9.1}(a)). We may therefore assume that 
$\gamma(t) = \exp(tx)H$ with $x \in \ft$ satisfying 
$\exp(x) = e$.  
Writing $x = x_\fh + x_\fq$ with $x_\fh \in \ft_\fh$ and 
$x_\fq \in \ft_\fq$, we then have 
\[ \gamma(t) = \exp(tx_\fq) H = \Exp(t x_\fq) \quad \mbox{ with } \quad 
\Exp(x_\fq) = \exp(x_\fq)H = e_X.\] 
This implies $x_\fq \in \Gamma_X$ 
and shows that $\iota$ is surjective. 

That $\ker\iota$ is $\Gamma_{\tilde X}$ follows 
immediately from the fact that, for 
$x \in \Gamma_X$, the curve 
$\gamma_x(t) := \Exp_X(tx)$ lifts to a loop in 
$\tilde X$ if and only  if $\Exp_{\tilde X}(x) = e_{\tilde X}$, 
which means that $x \in \Gamma_{\tilde X}$.    
\end{Proof}

\begin{Remark} \cite[Thm.~VII.8.5]{He78} also provides a description 
of the discrete subgroup $\Gamma_{\tilde X} \subeq \ft_\fq$ in terms of 
root data. We have 
\begin{equation}
  \label{eq:gammarel1}
\Gamma_{\tilde X} 
= \frac{1}{2} \Gamma_{T_Q^*} 
\quad \mbox{ for } \quad 
\Gamma_{T_Q^*} := \{ x \in \ft_\fq \: \exp_{\tilde G} x = e\}.
\end{equation}
In fact, the simply connected covering space 
$\tilde X \cong \tilde G/H^*_e$ can be identified with 
the identity component $\tilde G^{-\theta}_e$ of 
the set $\tilde G^{-\theta} = \{ g \in \tilde  G \: \theta(g) = g^{-1}\}$ 
on which $\tilde G$ acts transitively by $g.x := gx\theta(g)^{-1}$ 
and the stabilizer of $e$ is the connected subgroup~$\tilde G^{\theta}$. 
This leads to the realization $T_{\tilde X} = \Exp_{\tilde X}(\ft_\fq) \subeq \tilde G$ 
with $\Exp_{\tilde X}(x) = \exp(2x)$, and this proves \eqref{eq:gammarel1}.
Moreover, by  \cite[Thm.~VII.8.5]{He78}, the subgroup $\Gamma_{\tilde X} \subeq \ft_\fq$ 
is generated by 
the elements of the form 
\[ \frac{2\pi i}{\la \alpha, \alpha \ra} A_\alpha, \quad 
\alpha \in \Sigma, \] 
where $A_\alpha \in i\ft_\fq$ is the unique element satisfying 
$\kappa(A_\alpha, x) = \alpha(x)$ for $x \in \ft_\fq$ 
and the Cartan--Killing form~$\kappa$.
\end{Remark}

\begin{Remark} The recent preprint 
\cite{BoGa22} deals with a closely related problem: 
  the determination of $\pi_0(G(\R))$ for a connected complex
  reductive group $G(\C)$. Let $G := G(\C)$ and write $\sigma$ for the
  antiholomorphic involution with $G(\R) = G^\sigma$.
  We also write $\theta$ for the antiholomorphic involution on $G$ for which
  $U := G^\theta$ is compact and assume that $\sigma$ commutes with $\theta$, so that
  $\theta$ induces a Cartan involution on $G(\R)$.
  Then polar decomposition implies that
  \[ \pi_0(G(\R)) = \pi_0(U^\sigma),\]
  so that it is about the component group of $U^\sigma$, where $U$ is a connected
  compact Lie group.   
\end{Remark}
  
\subsection{Application to symmetric R-spaces} 
\label{subsec:7.2}

\begin{Remark} 
(a) If $h \in \fp$ is an Euler element and
$\tau_h$ is the corresponding involution, then
$\fp$ decomposes into eigenspaces
$\fp^{\tau_h}$ and $\fp^{-\tau_h}$ and
the $K$-orbit $\cO^K_h =\Ad(K)h \subeq \fp$
is $\tau_h$-invariant. Moreover, $\tau_h$ induces on this
orbit an involution turning it into a so-called
{\it extrinsic symmetric space}, for the definition see \cite{Fe80}.
In \cite{Fe80} Ferus shows that all
extrinsic symmetric submanifolds embed into
such spaces.

\nin (b) The compact symmetric spaces of the form $\cO^K_h$ 
are precisely the symmetric R-spaces, when considered as 
homogeneous spaces of $K$ (\cite{Lo85}). 
We refer to \cite{MNO22a} 
for their interpretation as real flag manifolds 
$G/P^-$ on which $K$ acts transitively with point stabilizer 
$K \cap P^- = K^h$. 
\end{Remark}

Let $G = \Inn(\g)$, $K = \Inn_\g(\fk)$, 
and consider the symmetric R-space 
\[ X := \Ad(K)h \subeq \fp \] 
defined by an Euler element in $h \in \fp$. We want to determine 
the group $\pi_0(G^h) \cong \pi_0(K^h)$ of connected components of $G^h$, resp., 
$K^h$. 

We consider the torus $T_X := e^{\ad \ft_\fq} h \subeq X$ 
with the fundamental group 
\[ \Gamma_X = \{ x \in \ft_\fq \: e^{\ad x}h= h \}.\] 
We shall need the following piece of the long exact homotopy 
sequence of the \break $K^h$-principal bundle 
$K \to X$: 
\[ \pi_2(X) \into \pi_1(K^h) \to \pi_1(K) 
\smapright{\pi_1(q)}  \pi_1(X) \to \pi_0(K^h) \to \1,\] 
where $q \: K \to X\cong K/K^h$ is the orbit map. 
Then the surjectivity of the homomorphism $\pi_1(X) \to \pi_0(K^h)$ implies that 
\[ \pi_0(K^h) \cong \coker(\pi_1(q)).\]
With Proposition~\ref{prop:x1} we see that 
$\pi_1(X) \cong \Gamma_X/\Gamma_{\tilde X}$, so that we obtain a surjective 
map 
\begin{equation}
  \label{eq:delta1}
\delta \: \Gamma_X \to \pi_0(K^h), \quad x \mapsto [e^{\ad x}].
\end{equation}
We shall now determine the range of this map 
with the aid of Proposition~\ref{prop:testing}, 
where we have seen that 
$h$ and $\ft_\fq$ generate a reductive Lie algebra 
\[ \fl \cong \R h_0 \oplus \fsl_2(\R)^s
\supeq \ft_\fq = \so_2(\R)^s, \qquad s = r_0 + r_1.\]

\begin{Lemma} \label{lem:9.4} The following assertions hold: 
  \begin{itemize}
  \item[\rm(a)] The torus $T_Q := \exp(\ft_\fq)$ satisfies 
$T_Q^h = \exp(\Gamma_X) = Z(L')$, where $L' \subeq L$ denotes the 
commutator subgroup. 
  \item[\rm(b)] Write $L = Z(L)_e L_1 \cdots L_s$, where 
$L_j$ is the integral subgroup corresponding to the $j$-th 
simple ideal $\fl_j$ in $\fl$. 
Then $Z(L) = Z(L)_e Z(L_1) \cdots Z(L_s)$, 
where the group $Z(L_j)$ is either trivial or 
contains two elements. In particular we have 
\[ T_Q^h= Z(L_1) \cdots Z(L_s).\] 
  \item[\rm(c)]  The subgroup $Z(L_1) \cdots Z(L_{r_0})\subeq T_Q^h$ 
maps surjectively onto $\pi_0(K^h)$. 
  \end{itemize}
\end{Lemma}

\begin{Proof} (a) In $\fsl_2(\R)$ and $\fsl_2(\C)$ any element 
$x \in \so_2(\R)$ satisfying $e^{\ad x} h = h$ for an Euler element $h$ 
also satisfies $e^{\ad x} = \1$. Therefore 
$T_Q^h \subeq Z(L')$. The converse inclusion follows from 
$Z(L') \subeq \exp(\ft_\fq) = T_Q$, which implies that $Z(L') \subeq T_Q^h$. 

\nin (b) As the center $Z(L_j)$ commutes for each $j$ with all of $L$, we have 
$Z(L) = Z(L)_e Z(L_1) \cdots Z(L_s)$. The subgroup $Z(L_j)$ 
is the image of the center of a group of the type 
$\SL_2(\R)$, 
hence contains at most two elements. 
Now $Z(L_1)\cdots Z(L_s) = Z(L')$, combined with (a), implies~(b).

\nin (c) For $j > r_0$, the ideal $\fl_j$ is contained 
as $\fsl_2(\R)$ in a complex $\tau$-invariant Lie algebra 
$\fs_j \cong \fsl_2(\C)$. If $S_j := \la \exp \fs_j \ra$, 
then 
$K_j := K \cap S_j \cong \SU_2(\C)$ or $\PSU_2(\C)$ and 
$\Ad(K_j)h$ is a $2$-sphere, hence simply connected. Therefore 
$K_j^h$ is connected and therefore $Z(L_j)$ does not contribute to $\pi_0(K^h)$. 
This implies (c). 
\end{Proof}

\begin{Remark}
  \label{rem:9.5} (a) The fact that $h \in \fl$ is an Euler element of $\g$ 
restricts the possibilities for irreducible $\fl$-submodules of $\g$ 
significantly. 
Writing 
$V_k^j$ for the $k$-dimensional simple complex module of the $j$th  ideal $\fl_j$, 
we see that the only non-trivial complex simple $\fl$-submodules of $\g_\C$ 
are the adjoint modules $V_3^j$, tensor products 
$V_2^{j_1} \otimes V_2^{j_2}$ for $j_1 \not=j_2$, 
and, if $h_0 \not=0$, also $2$-dimensional modules 
$V_2^j$ on which $h_0$ acts by $\pm \frac{1}{2}\1$. 
On the other two types the central element $h_0$ acts trivially.

Let $z_j \in Z(L_j)$ be generators. Then 
$z_{j_1} z_{j_2}$ acts trivially on $V_3^{j_1}$ and $V_2^{j_1} \otimes V_2^{j_2}$, 
but non-trivially on $V_2^{j_1} \otimes V_2^{j_3}$ for $j_3 \not= j_1, j_2$. 

\nin (b) 
Write $h \in \fl$ as $h = h_0 + h_1$ with $0 \not=h_0 \in \fz(\fl)$ and 
$h_1 \in [\fl,\fl]$. Then the structure of the restricted root system implies 
that $h_1$ defines a $5$-grading of $\g$: 
\[ \g = \g_{-1}(h_1) \oplus \g_{-1/2}(h_1) \oplus \g_0(h_1) 
\oplus \g_{1/2}(h_1) \oplus \g_1(h_1),\] 
where $h_0$ commutes with $\g_{\pm 1}(h_1)$. Therefore 
\[ \g_{\rm tt} := \fz_\g(h_0) 
= \g_{-1}(h_1) \oplus (\g_0(h_0) \cap \g_0(h_1))\oplus \g_1(h_1) \subeq \g \] 
is a $3$-graded $\tau$-invariant subalgebra in which $h_1$ is an Euler element. 
Let $r_0', r_1'$ and $s' := r_0' + r_1'$ be the corresponding numbers. 
We claim that 
\begin{equation}
  \label{eq:rsremain}
 r_0 = r_0', \quad r_1 = r_1' \quad \mbox{ and } \quad s' = s.
\end{equation}
In fact, as $h_0$ is hyperbolic 
$r' = \rk_\R(\g_{\rm tt}^c) = \rk_\R(\g^c) = r$. 
Further $\ft_\fq \subeq \fl \subeq \g_{\rm tt}$ shows that 
$s' = \rk_\R(\fh_{\rm tt}) =s$. This implies that 
$r_1' = r'- s' = r-s = r_1$ and $r_0' = s' - r_1' = s-r_1 = r_0$. 
\end{Remark}

\begin{Theorem} {\rm(The group $\pi_0(G^h)$)}  \label{thm:7.8}
Let $G = \Inn(\g)$.   For the group $\pi_0(K^h) \cong \pi_0(G^h)$, the following 
assertions hold: 
\begin{itemize}
\item[\rm(a)] If $(\g,\tau)$ is of {\bf complex type} or {\bf non-split type}, 
then $G^h$ is connected. 
\item[\rm(b)] If $(\g,\tau)$ is of {\bf Cayley type} or  {\bf split type}, 
then $G^h$ is connected if $r$ is odd. If $r$ is even, then it is not connected 
and $\pi_0(G^h) \cong \Z_2$ in the following cases of Cayley type: 
  \begin{itemize}
  \item[$\bullet$] $r = 2$ and $\g = \so_{2,n}(\R)$ with $n \geq 3$ odd.
  \item[$\bullet$] $r \geq 4$ even and $\g = \sp_{2r}(\R)$, 
    \end{itemize}
and the following cases of split type:  
    \begin{itemize}
  \item[$\bullet$] $\g = \fsl_{2n}(\R)$ with $h = \frac{1}{2}(\1_n,-\1_n)$ 
and $n$ even. 
  \item[$\bullet$] $\g = \so_{p,q}(\R)$ with $p,q > 2$ and $p+q$ odd. 
  \item[$\bullet$] $\g = \so_{2n,2n}(\R)$. 
  \end{itemize}
\end{itemize}
\end{Theorem}

\begin{Proof}  (a) If $(\g,\tau)$ is of {\bf complex type}, then 
 $\g= \fh_\C$ and $\tau$ is antilinear. 
Here $\fc = \fa_\C$, where $i\fa \subeq \fh_\fk$ is a compactly embedded Cartan subalgebra of $\fh$. For every root $\alpha \in \Sigma(\g,\fc)$ we have 
$(-\tau)\alpha = \oline \alpha$. As the roots are complex linear on 
$\fa_\C$, no root is fixed by $-\alpha$, and thus $r_0 = 0$ 
and $s = r_1$. Hence $K^h$ is connected by Lemma~\ref{lem:9.4}(c). 

If $(\g,\tau)$ is of {\bf non-split type}, then 
$s = \rk_\R(\fh)$ and $r = \rk_\R(\g^c) = 2s$, so that 
$r_0 = 0$ and $r_1 = s$.  
Therefore Lemma~\ref{lem:9.4}(c) implies that 
$K^h$ is connected. 
  
\nin (b) We first discuss {\bf Cayley type} Lie algebras. 
Let 
\[ r := \rk_\R(\g) \quad \mbox{ and } \quad s = \rk_\R(\fh) = \rk_\R(\g^h).\] 
Then $s \leq r$ and there exists a subalgebra 
$\fs \cong \fsl_2(\R)^r$ containing $h$, so that we must have $s = r$, 
and therefore $r_1 = 0$. 

If $r = 1$, then $\g = \fsl_2(\R)$ and $K^h = Z(G)$ is trivial. 

If $r > 1$, then the restricted root system $\Sigma(\g,\fa)$ 
is of type $C_r$ ($\g$ is hermitian of tube type), 
which implies that the simple $L$-submodules of $\g$ 
are the adjoint modules $\fl_j$ and the tensor products 
$V_{j_1,j_2} = V_2^{j_1} \otimes V_2^{j_2}$. 
Note that Cayley type algebras are hermitian of tube type, 
so that $h_0 = 0$ follows from Proposition~\ref{prop:testing}(c). 
For any product $z := z_{j_1} \cdots z_{j_k} \in Z(L)$ with 
$j_1 < \ldots < j_k$ and $k < r$, we then find a tensor product 
$V_{j_1,j_2}$ on which $z$ acts non-trivially, and 
$z_1 \cdots z_r = \1$. This shows that $Z(L) \cong \Z_2^{r-1}$. 

The restricted root system $\Sigma(\g,\fa)$ is of type $BC_r$ 
with the subsystem $\Sigma(\g^h,\fa)$ of type $A_{r-1}$. 
The corresponding Weyl group is the symmetric group $S_r$ 
which acts by permutations of the central involutions 
$z_1, \ldots, z_r$ of~$L$. 
Since the Weyl group elements are induced by the normalizer of $\fa$ 
in the connected group $\Inn_\g(\fh_\fk)$, it follows that 
$z_1, \ldots, z_r$ all lie in the same connected component of~$K^h$. 
As these elements are involutions, 
all even products $z_{j_1} \cdots z_{j_k}$ are contained in $K^h_e$. 
It follows that $|\pi_0(K^h)| \leq 2$ and, 
$z_1 \cdots z_r = \1$ further implies that 
$\pi_0(K^h)$ is trivial if $r$ is odd. 
So the only cases that have to be inspected in detail arise 
for $r$ even. 

\nin {\bf Case 1: $r = 2$.} Then $\g = \so_{2,n}(\R)$ for some $n \geq 3$. 
In this case 
$G \cong \PSO_{2,n}(\R)_e.$ 
Let $G^* := \SO_{2,n}(\R)_e$ with maximal compact subgroup 
$K^* = \SO_2(\R) \times \SO_n(\R)$. Here $h \in \so_{2,n}(\R)$ 
is a diagonalizable rank-$2$ element whose 
$\pm 1$-eigenspaces are isotropic, f.i., 
$\be_2 \pm \be_3$ (up to conjugacy).  Hence its centralizer 
leaves the plane $F := \R \be_2 + \R \be_3$ invariant and we thus obtain a 
homomorphism $(K^*)^h \to \OO_{1,1}(\R)$. 
The centralizer of the Lorentz boost in $\so_{1,1}(\R)$ 
is $\SO_{1,1}(\R)$ which is not connected and contains $-\id_{\R^2}$. 
As $\diag(-1,-1,-1,-1,1,\ldots) \in (K^*)^h$ maps to $-\id_{\R^2}$, 
it follows that $(K^*)^h$ has at least $2$-connected components. 

The invariance of $F$ under $(K^*)^h$ also shows that 
it leaves $\R \be_2$ and $\R \be_3$ invariant, so that 
\[ (K^*)^h 
\subeq {\rm S}(\OO_1(\R) \times \OO_1(\R)) \times 
{\rm S}(\OO_1(\R) \times \OO_{n-1}(\R))
\cong \{ \pm\1_{\R^2}\}  \times 
{\rm S}(\OO_1(\R) \times \OO_{n-1}(\R)), \] 
where the group on the right has $4$ connected components. 
As the group on the right maps surjectively onto $\OO_{1,1}(\R)$ 
and $(K^*)^h$ onto $\SO_{1,1}(\R)$,  
we see that $(K^*)^h$ has exactly $2$ components. 

If $n$ is odd, then $-\1 \not\in G^*$, so that $G \cong G^*$ 
and $K \cong K^*$. It follows that $G^h$ has two connected components. 

If $n$ is even, then $-\1 \in \SO_{2,n}(\R)_e$ and 
$G \cong \SO_{2,n}(\R)_e/\{\pm \1\}$. Then 
$-\1 \not \in (K^*)^h_e$  (consider the restriction to $F$) 
and the fact that $(K^*)^h$ has $2$ connected components 
imply that $K^h$ is connected.\begin{footnote}{The group $G = \PSO_{2,n}(\R)_e$ 
acts by causal automorphisms on the causal compactification 
$\hat M \cong (\bS^1 \times \bS^{n-1})/\{\pm \1\}$ 
of $n$-dimensional Minkowski space $M = \R^{1,n-1}$. 
If $\hat M$ is orientable, then the connected group preserves the 
orientation. As the subgroup $G^h \subeq \R^\times_+ 
\OO_{1,n-1}(\R)$ fixes $0$ and 
acts by linear maps, this implies that 
$G^h \subeq \R^\times_+ \SO_{1,n-1}(\R)_e$, and since the 
latter group is connected, it follows that 
$G^h \cong \SO_{1,n-1}(\R)_e$ is connected. The manifold $\hat M$ is 
orientable if and only if the antipodal map is orientation 
preserving on $\bS^{n-1}$, i.e., if $n$ is even. 
If this is not the case, then $\hat M$ is not orientable 
and $G^h \cong \R^\times_+\OO_{1,n-1}(\R)^\up$ is not connected. 
}\end{footnote}

\nin {\bf Case 2: $r \geq 3$.} Let $E = \g_1(h)$ be the euclidean Jordan algebra 
for which $\g$ is the conformal Lie algebra. 

\begin{itemize}
\item For $E = \Sym_r(\R)$, we consider 
$G^* := \Sp_{2r}(\R)$ with $(G^*)^h \cong \GL_r(\R)$, 
which has $2$ connected components and contains 
$Z(G^*) = \{ \pm \1\}$. Therefore 
$G^h \cong (G^*)^h/\{\pm \1\}$ is connected if and only if 
$-\1 \not\in \GL_r(\R)_e$, which is equivalent to 
$\det(-\1) = (-1)^r = -1$. This corresponds to $r$ odd. 
If $r$ is even, then $G^h$ is not connected. 
\item For $E = \Herm_r(\C)$ we consider 
$G^* := \SU_{r,r}(\C)$ with 
\[ (G^*)^h \cong \{ g \in \GL_r(\C) \: \det(g) \in \R\}.\] 
This group has two connected components, corresponding to the 
sign of the real-valued determinant.  It contains 
\[ Z(G^*) 
= \{\zeta \1_{2r} \: \zeta \in \C^\times, \zeta^{2r} = 1 \} 
\cong  \{\zeta \1_r \: \zeta \in \C^\times, 
\det(\zeta\1_r) = \zeta^r \in \{\pm 1\} \}
\cong C_{2r}.\] 
If $\zeta \in \C^\times$ is an $r$th root of $-1$, then 
$\zeta \1_r \in Z(G^*) \cap (G^*)^h$ with $\det(\zeta\1_r) < 0$ 
shows that both connected components of $(G^*)^h$ intersects 
$Z(G^*)$. Therefore 
$G^h = \Ad((G^*)^h) \cong (G^*)^h/Z(G^*)$ is connected. 
\item For $E = \Herm_r(\H)$ we consider 
\[ G^* := \SO^*(4r) 
:= \{ g \in \SU_{2r,2r}(\C) \: g^\top A g = A\} 
\quad \mbox{ for } \quad A = \pmat{ 0 & \1 \\ \1 & 0}.\] 
Then 
\[ K^* := G^* \cap \U_{4r}(\C) 
=\Big\{ \pmat{ a & 0 \\ 0 & a^{-\top}} \: a \in \U_{2r}(\C) \Big\} 
\cong \U_{2r}(\C).\] 
In the Lie algebra 
\[ \so^*(4r) = 
\Big\{ \pmat{a & b \\ b^* & -a^\top} \: a\in \fu_{2r}(\C), b \in \Skew_{2r}(\C)\Big\} 
\] 
an Euler element is given by 
\[ h = \frac{1}{2} \pmat{ 0 & J \\ -J & 0} 
\quad \mbox{ for } \quad 
J := \pmat{ 0 & -\1 \\ \1 & 0} 
\in \Skew_{2r}(\C).\] 
An element $k = \diag(a, a^{-\top})$, 
$a \in K^*$, commutes with $h$ if and only if 
$J = a J a^\top = a J \oline a^{-1}$, i.e., 
$a J = J \oline a$. One readily checks that this is equivalent to 
$a$ being of the form $\pmat{\alpha & \beta \\ -\oline\beta & \oline\alpha}$, 
i.e., to $a \in \U_r(\bH) \subeq \U_{2r}(\C)$. 
We conclude that $(K^*)^h \cong \U_r(\H)$ is connected. 
This implies that $G^h = \Ad((G^*)^h)$ is connected. 
\item For $E = \Herm_3(\bO)$ and $\g = \fe_{7(-25)}$, the rank $r = 3$ is odd, so that 
$G^h$ is connected by the discussion above. 
\end{itemize}

Now we turn to {\bf split type} Lie algebras. Then 
\[ s = \rk_\R(\fh) = \rk_\R(\g^c) = r, 
\quad \mbox{  so that } \quad r_0 = r = s \quad \mbox{ and } \quad 
r_1 = 0 \] 
(see Table 3).
As in (c), we see that $G^h$ is connected if $r$ is odd and 
that it has at most $2$ connected components. 

\begin{itemize}
\item For $\g = \fsl_{2n}(\R)$ and $G^* = \SL_{2n}(\R)$, we have 
\[ (G^*)^h = {\rm S}(\GL_n(\R) \times \GL_n(\R)),\] 
which has two connected components. 
As $-\1 \in (G^*)^h$ is contained in the identity component 
if and only if $n$ is even, it follows that 
$G^h$ is connected if and only if $n$ is odd. 
\item For $\g = \so_{p,q}(\R)$, we consider $G^* = \SO_{p,q}(\R)_e$, $p,q > 2$,   
with maximal compact subgroup 
$K^* = \SO_p(\R) \times \SO_q(\R)$. Here $h \in \so_{p,q}(\R)$ 
is a diagonalizable rank-$2$ element whose 
$\pm 1$-eigenspaces are one-dimensional isotropic, f.i., 
$\be_p \pm \be_{p+1}$.  Hence its centralizer $(G^*)^h$  
leaves the plane $F := \R \be_p + \R \be_{p+1}$ invariant. 
This shows that 
\[ (K^*)^h 
\subeq {\rm S}(\OO_1(\R) \times \OO_{p-1}(\R)) \times 
{\rm S}(\OO_1(\R) \times \OO_{q-1}(\R))
\cong \OO_{p-1}(\R) \times \OO_{q-1}(\R). \] 
The fact that $(K^*)^h$ preserves the $1$-dimensional subspaces 
generated by \break $\be_p \pm \be_{p+1}$ shows that 
$(K^*)^h$ has $2$-connected components and that 
its restriction to the Minkowski plane 
$F$ contains~$-\1$. 

If $p+q$ is odd, then $-\1 \not\in G^*$, so that $G \cong G^*$ 
and $G^h$ has two connected components. 

If $p+q$ is even, then $-\1 \in G^*$ and 
$G \cong G^*/\{\pm \1\}$. Then 
$-\1 \not \in (K^*)^h_e$ (consider the restriction to $F$) 
implies that $K^h$ is connected. 
\item We realize the split real form 
$\so_{2n,2n}(\R) \subeq \so_{4n}(\C)$ as 
\begin{align*}
 \so_{2n,2n}(\R)
 &= \Big\{ X \in \gl_{4n}(\R) \: 
X I + I X^\top = 0\Big\} \\
&= \Big\{ \pmat{a & b \\ c & - a^\top} \: b,c \in \Skew_{2n}(\R), 
a \in \gl_{2n}(\R)\Big\}
\end{align*}
for 
\[ I = \pmat{0 & \1_{2n} \\ \1_{2n} & 0}.\]
This exhibits the Euler element 
\[ h := \frac{1}{2} \pmat{ \1 & 0 \\ 0 & -\1} 
\quad \mbox{ with } \quad 
\g_1(h) \cong \Skew_{2n}(\R)\] 
and we obtain for $G^* := \SO_{2n,2n}(\R)_e$ the subgroup 
\[ (G^*)^h = \Big\{ \pmat{ a & 0 \\ 0 & a^{-\top}} \: a \in \GL_{2n}(\R)\Big\} 
\cong \GL_{2n}(\R) \] 
with $2$ connected components. 
Further, $Z(G^*) = \{ \pm \1\} \subeq (G^*)^h_e$, 
so that $G^h = \Ad((G^*)^h)$ also has $2$ connected components. 
\item The split real form $\g = \fe_7(\R)$ of type $E_7$: 
Then $\g^c$ is the hermitian real form of $\fe_7(\C)$, 
hence of real rank $r = 3$. Therefore $r_0 = r = 3$ 
is odd, so that $G^h$ is connected by the discussion above. 
\qedhere\end{itemize}
\end{Proof}


\subsection{The maximal compact subgroup of  $H  = G \cap G^c$}
\label{subsec:7.3}

In this subsection we collect some consequences 
of our discussion of $\pi_0(G^h)$ in the preceding subsection.
In particular, we show that, for the c-dual
Lie algebra $\g^c = \fh + i \fq$, we have 
\[ \Inn_{\g_\C}(\g^c) \cap \Inn_{\g_\C}(\g)  = \Inn_{\g_\C}(\fk)^h e^{\ad \fh_\fp}.\]

\begin{Proposition} \label{prop:hk=kh}
Let $(\g,\tau,C)$ 
be a semisimple non-compactly causal 
symmetric Lie algebra, 
where $\tau = \tau_h \theta$, $h \in \fq_s \cap C^\circ$
is a causal Euler element,  
\[ G := \Inn_{\g_\C}(\g)\cong \Inn(\g), \quad K := G^\theta = \Inn_\g(\fk),\quad \mbox{ 
and }  \quad G^c := \Inn_{\g_\C}(\g^c).\]
 Then $H := G \cap G^c$ is $\tau$-invariant and  satisfies 
\[ H = K^h \exp(\fh_\fp) \quad \mbox{ and } \quad 
H \cap K = K^h.\] 
\end{Proposition}

Note that $\tau = \tau_h \theta$ and $G = \Inn(\g)$ imply 
$G^h \subeq G^{\tau}$ because $K^h \subeq K^{\tau_h} = K^\tau \subeq G^\tau$
and $G^h = K^h \exp(\fq_\fp)$. 
Therefore the  preceding proposition  shows in particular that
\begin{equation}
  \label{eq:H-g-tau}
 G \cap G^c \subeq G^\tau.
\end{equation}

\begin{Proof} First we write $\g$ as a direct sum
  $\g_k \oplus \g_r \oplus \g_s$ as in \eqref{eq:gdeco},
  where $\g_s$ is the sum of all simple ideals not commuting with $h$
(the strictly ncc part), $\g_r$ is the sum of all non-compact simple
ideals commuting with $h$ on which $\tau = \theta$
(the non-compact Riemannian part),
and $\g_k$ is the sum of all compact ideals (they commute with $h$).
As $h \in \fq_s$, the corresponding subgroups
$G_k$ and $G_r$ are contained in $G^h$ and  
$\fh_p \cap \g_k = \{0\} = \fh_p \cap \g_r$.
Therefore we may assume that $\g = \g_s$ and, by decomposition
into simple ideals, even that
$(\g,\tau)$ is irreducible non-compactly causal.

We start with the polar decomposition 
$G^h = K^h \exp(\fp^{-\tau_h})$ which implies that 
\[ H_K := H \cap K = G^c \cap K = G \cap K^c = K \cap K^c.\] 
Using the strongly orthogonal roots and the subgroup 
$L \subeq G$ (cf.\ Lemma~\ref{lem:9.4}), 
we find with Lemma~\ref{lem:9.1}(d), applied
with the compact group $K$ ($G$ in the lemma) and the subgroup $K^h$
($H$ in the lemma), that 
\begin{equation}
  \label{eq:khpol}
 K^h = K^h_e (K^h \cap Z(L))
\quad \mbox{ with } \quad 
K^h \cap Z(L) \subeq T_Q = \exp(\ft_\fq).
\end{equation}

We claim that all generators $z_j \in Z(L_j)$ are contained in $G\cap G^c$. 
In fact, we have 
$\fl_j \cong \fsl_2(\R)$, 
$\fl_j^c \cong \su_{1,1}(\C)$ and $\fl_{j,\C} = \fsl_2(\C)$. 
Clearly $-\1 \in \SL_2(\R) \cap \SU_{1,1}(\C)$ holds in $\SL_2(\C)$, so that
 $z_j \in L_j \cap L_j^c$. 
This implies that $T_Q^h \subeq Z(L) \subeq L \cap L^c$ is contained 
in $H = G \cap G^c$, and hence by \eqref{eq:khpol} that 
$K^h \subeq H.$ 
Since we also have $H_K \subeq K^h$ (a consequence of $\Ad(H_K)$ fixing 
$h \in C$) (Lemma~\ref{lem:hpm1}), 
we obtain $H_K = K^h$. 
\end{Proof}

\begin{Lemma}\label{lem:cover-order} 
If $H' \subeq G^\tau$ is another open subgroup satisfying 
\begin{equation}
  \label{eq:adhc}
\Ad(H') C_\fq^{\rm max} = C_\fq^{\rm max},  
\end{equation}
then $H' \subeq H = K^h\exp(\fh_\fp)$ and we obtain an 
equivariant covering of ordered symmetric spaces 
\begin{equation}
  \label{eq:covncc}
 M' := G/H' \to M = G/H, \quad gH' \mapsto gH.
\end{equation}
\end{Lemma}

\begin{Proof} In the polar decomposition 
$H' = H'_K \exp(\fh_\fp)$ we have $H'_K \subeq K^\tau = K^{\tau_h}$. 
Since every element $g \in G^{\tau_h}$ normalizes 
$\g_0(h)$ and preserves $\g_1(h) + \g_{-1}(h)$, 
which are inequivalent representations of the Lie algebra $\g_0(h)$, 
we either have $\Ad(g)\g_{\pm 1}(h) = \g_{\pm 1}(h)$ or 
$\g_{\mp 1}(h)$. In the first case $\Ad(g)h = h$, and in the second 
$\Ad(g)h=-h \in -C_\fq^{\rm max}$. Therefore \eqref{eq:adhc} implies 
$H'_K \subeq K^h$, so that $H' \subeq H = G_\cD$. We therefore 
have an equivariant covering of homogeneous spaces as in~\eqref{eq:covncc}.
\end{Proof}

As an important application of the preceding discussion, we obtain
the following result on the fundamental group of $M$, connecting
it to connected components of the stabilizer of $h$, and {\bf not}
to connected components of $H$, as usual.

\begin{Proposition} \label{prop:7.12}
  Let $q_G \: \tilde G \to G$ denote the simply connected covering group. 
Then $q_G$ induces the universal covering 
\[ q_M \: \tilde M := \tilde G/\tilde G^\tau  \to M = G/H
  \quad \mbox{ for } \quad H = K^h \exp(\fh_\fp) \]
and 
\[ \pi_1(M) \cong \pi_0(\tilde G^h).\] 
\end{Proposition} 

\begin{Proof} The subgroup $\tilde G^\tau$ is connected because 
$\tilde G$ is simply connected 
(\cite[Thm.~IV.3.4]{Lo69}). 
Thus $\tilde M$ is the simply connected covering of $M = G/H \cong 
\tilde G/q_G^{-1}(H)$. 
This further implies that 
\[ \pi_1(M) \cong  \pi_0(q_G^{-1}(H))\] 
(\cite[Cor.~11.1.14]{HN12}). 
As $H = K^h \exp(\fh_\fp)$,  the inclusion 
$\ker(q_G) \subeq \tilde K^h$ leads to the polar 
decompositions 
\[ q_G^{-1}(H) = q_G^{-1}(H_K) \exp(\fh_\fp)= q_G^{-1}(K^h) \exp(\fh_\fp)
= \tilde K^h \exp(\fh_\fp) \] 
and $\tilde G^h = \tilde K^h \exp (\fq_\fp)$, 
which imply that $\pi_0(\tilde G^h) \cong \pi_0(\tilde K^h) \cong 
\pi_0(q_G^{-1}(H)) \cong \pi_1(M)$.
\end{Proof}

\appendix

\section{Some calculations in $\fsl_2(\R)$} 
\label{subsec:sl2b}

In this subsection we collect some formulas concerning the 
$3$-dimensional Lie algebra $\fsl_2(\R)$ that we shall use below.  
For $\g  = \fsl_2(\R)$, we fix the Cartan involution 
$\theta(x) = - x^\top$, so that 
\[ \fk = \so_2(\R) \quad \mbox{ and }\quad 
\fp = \{ x \in \fsl_2(\R) \: x^\top = x \}.\]
The basis elements 
\begin{equation}
  \label{eq:sl2ra}
h^0 := \frac{1}{2} \pmat{1 & 0 \\ 0 & -1},\quad 
e^0 =  \pmat{0 & 1 \\ 0 & 0}, \quad
f^0 =  \pmat{0 & 0 \\ 1 & 0}
\end{equation}
and 
\begin{equation}
  \label{eq:sl2rb}
h^1 = \frac{1}{2} \pmat{0 & 1 \\ 1 & 0} = \frac{1}{2}(e^0 + f^0),\quad 
e^1 =  \frac{1}{2}\pmat{-1 & 1 \\ -1 & 1}, \quad 
f^1= \frac{1}{2} \pmat{-1 & -1 \\ 1 & 1}
\end{equation}
satisfy
\[  [h^j, e^j] = e^j, \quad [h^j, f^j] = -f^j, \quad 
[e^j, f^j] = 2 h^j \quad \mbox{ and }\quad \theta(e^j) = -f^j
\quad \mbox{ for } \quad j=1,2.\] 
For the involution 
\[ \tau\pmat{a & b \\ c & d} = \pmat{a & -b \\ -c & d} 
\ \ \mbox{ we have } \ \  \fh = \g^\tau = \R h^0 \ \  \mbox{ and } \ \ 
\fq = \g^{-\tau} = \R h^1 + \R (e^0 - f^0)\]
and 
\[ C = [0,\infty) e^0 + [0,\infty) f^0  \] 
is a hyperbolic $\Inn(\fh)$-invariant cone in $\fq$, containing 
$h^1$ as a causal Euler element. 

The subspace $\ft_\fq := \R (e^0 - f^0) = \so_2(\R)$ of $\fq$ 
is maximal elliptic. For 
\[ x_0 := \frac{\pi}{4} (e^0 - f^0) = \frac{\pi}{4} \pmat{0 & 1 \\ -1 & 0} \in \ft_\fq \]
and 
\[ g_0 := \exp(x_0) = \pmat{\cos\big(\frac{\pi}{4}\big) & \sin\big(\frac{\pi}{4}\big)\\
-\sin\big(\frac{\pi}{4}\big) & \cos\big(\frac{\pi}{4}\big)} 
= \frac{1}{\sqrt 2} \pmat{1 & 1 \\ -1 & 1},\]
we then have 
\begin{equation}
  \label{eq:firstconj2}
 \Ad(g_0) h^1 = h^0 \quad \mbox{ and } \quad \Ad(g_0) h^0 = - h^1,
\end{equation}
More generally, we have for $t \in \R$ 
\begin{equation}
  \label{eq:turn}
 e^{\frac{t}{2}\ad(e^0 - f^0)} h^0 = \cos(t) h^0 - \sin(t) h^1  
\end{equation}
because
\[\Big[\frac{e^0-f^0}{2}, h^0\Big] =  -h^1,  \quad 
\Big[\frac{e^0-f^0}{2}, h^1\Big] =  h^0.\]

\section{Some general facts on invariant cones} 
\label{app:a}

\begin{Lemma} \label{lem:coneint} 
Let $E$ be a finite dimensional real vector space, 
$C \subeq E$ a closed convex cone and $E_1 \subeq E$ a linear subspace. 
If the interior $C^\circ$ of $C$ intersects $E_1$, then 
$C^\circ \cap E_1$ coincides with the relative interior $C_1^\circ$ 
of the cone $C_1 := C \cap E_1$ in~$E_1$.
\end{Lemma}

\begin{Proof} Clearly, $C^\circ \cap E_1 \subeq C_1^\circ$. 
Pick $x_0 \in C^\circ \cap E_1$. 
If $x \in C_1 = C \cap E_1$ and $t \in [0,1)$, then 
$x_t := x_0 + t(x-x_0) = (1-t) x_0 + t x \in C^\circ$ 
shows that $C^\circ \cap E_1$ is dense in $C_1$. 
If, in addition, $x \in C_1^\circ$,
then there exists an $s > 1$ with $x_s \in C_1$. 
Then the argument from above shows that 
\[ x = x_1 = x_0 + \frac{1}{s} (x_s - x_0) \in C^\circ.
\qedhere\]
\end{Proof}

\begin{Lemma} {\rm(Cone Extension Lemma)} \label{lem:coneext} 
Let $E= E_1 \oplus E_2$ 
be a finite dimensional real vector space and 
$p_1 \: E \to E_1$ the projection along~$E_2$. 
Further, let $K \subeq \GL(E_2)\subeq \GL(E)$ be a compact subgroup
and $H \subeq \SL(E_1)\subeq \SL(E)$ a subgroup. 
If $C_1 \subeq E_1$ is  pointed generating $H$-invariant closed convex cone 
and $h \in E$ with $p_1(h) \in C_1^\circ$, 
then there exists a $H \times K$-invariant pointed generating closed convex cone 
$C \subeq E$ such that 
\begin{itemize}
\item[\rm(a)] $p_1(C) = C \cap E_1 = C_1$. 
\item[\rm(b)] $C^\circ \cap E_1 = C_1^\circ$. 
\item[\rm(c)] $h \in C^\circ$. 
\end{itemize}
\end{Lemma}

\begin{Proof} As $H \subeq \SL(E_1)$,  the characteristic function 
\[ \phi \: C_1^\circ  \to (0,\infty),\quad 
\phi(x) = \int_{C_1^\star} e^{-\alpha(x)}\, d\alpha \] 
is smooth and $H$-invariant with the property that 
$x_n \to x_0$ with $x_n \in C_1^\circ$ and $x_0 \in \partial C_1$ 
implies $\phi(x_n) \to \infty$ (\cite[Thm.~V.5.4]{Ne00}). 
Let 
\[ B := \{ x \in C_1 \: \phi(x) \leq 2\phi(p_1(h)) \}.\] 
This is a closed convex $H$-invariant subset with $C_1
= \cone(B) = \oline{\R_+ B}$. Further, let $D \subeq E_2$
be a $K$-invariant compact $0$-neighborhood 
such that $h - p_1(h) \in  \Int_{E_2}(D)$.  
Then $B + D$ is a  {$H \times K$-invariant}
closed convex subset not containing $0$, 
and thus $C := \cone(B + D)$ is a \break $H\times K$-invariant pointed generating 
invariant cone which satisfies 
\[ C \cap E_1 \subeq p_1(C) \subeq C_1 \subeq C \cap E_1.\] 
Moreover, $h \in \Int_{E_1}(B) + \Int_{E_2}(D) \subeq C^\circ$ and 
\[ C^\circ  \cap E_1 \subeq C_1^\circ 
= (0,\infty) B \subeq C^\circ.\qedhere\]
\end{Proof}

\begin{Example} The condition $H \subeq \SL(E_1)$ is crucial, as the following 
example shows. We consider 
\[ E = \R^2, \quad K = \{\1\}\quad \mbox{ and }  \quad 
H = \{ \diag(t,1) \: t > 0\}.\] 
Then every open neighborhood $U\subeq E$ of a point $(t,0) \in E_1$ 
has the property that $\{0\} \times \R$ is contained in the closed 
convex cone generated by $H.U$. Therefore 
the cone $C = [0,\infty) \times \{0\}$ does not extend to a pointed 
generating $H$-invariant cone $\hat C$ containing 
$(0,\infty) \times \{0\}$ in its interior. 
However, there are $H$-invariant pointed generating invariant cones 
$\hat C$ with $\hat C \cap E_1 = C$, but they contain $C$ in their boundary. 
\end{Example}

\begin{Proposition}\label{prop:project}
{\rm (\cite[Prop.~2.11]{Ne10})}
Let $K$ be a compact group acting continuously on the
finite-dimensional real vector space $E$ by the representation 
\break $\pi \: K \to \GL(E)$ and $p(v) := \int_K \pi(k)v\, d\mu_K(k)$
the corresponding fixed point projection,
where $\mu_K$ is a normalized Haar measure on~$K$. 
If $\Omega \subeq E$ is an open or closed  
  $K$-invariant convex subset, then
\[ p(\Omega) = \Omega \cap E^K.\] 
\end{Proposition}

The preceding proposition implies in particular for any convex invariant
subset with non-empty interior the relation 
\[ p(\Omega^\circ)  = \Omega^\circ \cap E^K \subeq (\Omega \cap E^K)^\circ.\]
That we actually have equality follows from
Lemma~\ref{lem:coneint}. 

\section{Lorentzian symmetric spaces} 
\label{subsec:lorentz}

Time-oriented Lorentzian symmetric spaces are in particular
causal symmetric spaces. Not all such spaces are reductive,
as the biinvariant Lorentzian structures on the $4$-dimensional
(solvable) oscillator group shows (\cite{HN93}). If, however,
$(\g,\tau, \beta)$ is reductive and Lorentzian
($\beta$ denoting the Lorentzian form on $\fq$), then we may assume that
$\fz(\g) \subeq \g^{-\tau}$. Accordingly
\[ \g = \fz(\g) \oplus \bigoplus_{j = 1}^n (\g_j, \tau_j), \]
where each $(\g_j, \tau_j)$ is irreducible and the corresponding
direct sum decomposition
\[ (\fq,\beta)= (\fz(\g),\beta_\fz) \oplus \bigoplus_{j = 1}^n (\fq_j, \beta_j) \]
is orthogonal. Therefore at most one summand contains timelike vectors
and the other summands are space-like, hence correspond to Riemannian
symmetric spaces. So two types occur:
\begin{itemize}
\item[(CL)] The central type, where $\beta_\fz$ is Lorentzian and all forms
  $\beta_j$ are negative definite. 
\item[(SL)] The simple type, where some $\beta_{j_0}$ is Lorentzian
  and all other summands are negative definite.  
\end{itemize}

For the classification of Riemannian symmetric spaces,
we refer to Helgason's monograph \cite{He78}. These determine
the spaces of central type and to understand the other type,
one needs to know the irreducible Lorentzian spaces,
whose classification we recall below.

\begin{Theorem} If $(\g,\tau)$ is  an irreducible semisimple symmetric Lie algebra 
and the corresponding $d$-dimensional  symmetric space $M$ is Lorentzian, then it is 
locally isomorphic to de Sitter space $\dS^d$ or anti-de Sitter space~$\AdS^d$. 
Accordingly, $(\g,\fh)$ is isomorphic to 
\[ (\so_{1,d}(\R), \so_{1,d-1}(\R)) \quad \mbox{ or } \quad 
(\so_{2,d-1}(\R), \so_{1,d-1}(\R)).\] 
\end{Theorem}

\begin{Proof} 
Let $(\g,\tau)$ be an irreducible semisimple symmetric Lie algebra. 
We are interested in a description of all Lorentzian symmetric spaces 
$G/H$. To see if $(\g,\tau)$ is Lorentzian, we choose a Cartan involution 
$\theta$ of $\g$ commuting with $\tau$. We then have 
\[ \fq = \fq_\fk \oplus \fq_\fp,\] 
where the Cartan--Killing form of $\g$ is negative definite on $\fq_\fk$ and 
positive definite on $\fq_\fp$. That a corresponding symmetric space 
$G/H$ is Lorentzian is equivalent to one of the two subspaces 
$\fq_\fk$ or $\fq_\fp$ to be one-dimensional, because this implies that 
a suitable multiple of the Cartan--Killing form on $\fq$ is a Lorentzian 
form. As $(\g,\tau)$ is Lorentzian if and only if the dual pair 
$(\g^c = \fh \oplus i\fq, \tau^c)$ is Lorentzian, and 
$(i\fq)_\fk = i \fq_\fp$ and 
$(i\fq)_\fp = i \fq_\fk$, we may assume that $\fq_\fp$ is one-dimensional. 
Then $\theta$ defines a dissecting involution on the associated 
$1$-connected symmetric space $G/H$, where 
$G$ is $1$-connected and $H = G^\tau$. Then the classification of 
irreducible symmetric spaces with dissecting involutions 
in \cite{NO20} implies that 
$G/H$ is locally isomorphic to a quadric. In particular 
$\g \cong \so_{p,q}(\R)$ and $\fh \cong \so_{p-1,q}(\R)$. 
The Lorentzian property now implies that 
$q = 1$ or $p = 2$, which corresponds to 
de Sitter space and Anti-de Sitter space.   
\end{Proof}

\begin{Remark} In addition to irreducible Lorentzian symmetric spaces,
  there is also the class of indecomposable solvable symmetric spaces;
  classified by Cahen and Wallach in \cite{CW70}.
  According to \cite[Thm.~3]{CW70}, all indecomposable Lorentzian symmetric
  Lie algebras are either semisimple or solvable. 
  A detailed exposition of the classification can be found
  in \cite[\S 3.3]{KO08}. The corresponding symmetric Lorentzian
  Lie algebras $(\g,\tau,\beta)$, where $\beta \: \g \times \g \to \R$
  is an invariant non-degenerate symmetric  bilinear form, have
  the following structure.
  We start with a quadratic space
  $(\fa,\kappa_\fa)$ of signature $(p+2q,p)$ and orthogonal basis
  \[ \be_1, \ldots, \be_{2q}, \quad \be_1',\ldots, \be_{2p}' \]
  with
  \[ \kappa_\fa(\be_j, \be_j) = 1, \quad \kappa_\fa(\be'_j, \be'_j) =
  \begin{cases}
    1 & \text{ for } j \leq p \\
    -1 & \text{ for } j > p.
  \end{cases}\]
  We also have a skew-symmetric endomorphism
  $D \in\so(\fa,\kappa_\fa)$ acting by
  \[
  D \be_j = \lambda_j \be_{j + q}, \quad 
  D \be_{j+q} = -\lambda_j \be_{j}, \quad
  D \be'_j = \mu_j \be'_{j + p}, \quad 
  D \be'_{j+p} = \mu_j \be'_{j} \]
  with
  \[    0 <  \lambda_1 \leq \cdots \leq \lambda_q,\quad 
   0 < \mu_1 \leq  \cdots \leq \mu_p .\] 
   Now
   \[ \g = \R \oplus \fa \oplus \R \]
   with the Lie bracket
   \[ [(z,a,t),(z',a',t')] := (\kappa_\fa (Da,a'), t Da' - t' Da, 0) \]
   and the invariant symmetric bilinear form
   \[ \beta((z,a,t), (z',a',t')) = zt' + z' t + \kappa_\fa (a,a').\]
   The involution $\tau$ is given by
   \[ \tau(z,a,t) = (-z,\tau_\fa(a),-t),\]
   and 
   \[    \fa^{-\tau} = \Spann \{ \be_1, \ldots, \be_q,
   \be_1', \ldots, \be_p' \}, \quad
   \fa^{\tau} = \Spann \{ \be_{q+1}, \ldots, \be_{2q},
   \be_{p+1}', \ldots, \be_{2p}' \}. \]
   For $p = 0$ these Lie algebras are called
   {\it oscillator algebras} and the form $\beta$ on $\g$
   is Lorentzian because $\kappa_\fa$ is positive definite.
   In general the form $\beta$ is positive definite on
   $\fa^{-\tau} = \fa \cap \fq$, so that the restriction of
   $\beta$ to $\fq = \g^{-\tau}$ is Lorentzian, but the form
   $\beta$ on $\g$ has signature $(p+1 + 2q, p+1)$. 
   \end{Remark}

In addition to Lorentzian symmetric spaces, 
also natural generalizations of conformal spacetime-geometries 
have been discussed from the physics perspective 
in \cite{MdR07}, where the conformal compactifications of simple 
euclidean Jordan algebras are studied. 
\medskip

\noindent 
 {\bf Acknowledgment:} The research of V. Morinelli VM was 
partially supported by a  Humboldt Research Fellowship for Experienced Researchers; the University of Rome through the
MIUR Excellence Department Project, the  ``Tor Vergata'' CUP E83C18000100006 and ``Tor Vergata''   ``Beyond Borders'' CUP E84I19002200005,  Fondi di Ricerca Scientifica d'Ateneo 2021, OAQM, CUP E83C22001800005, and the European Research Council Advanced Grant 669240 QUEST. The research of K.-H. Neeb was partially supported by DFG-grant NE 413/10-1. The research of G.~\'Olafsson was partially
supported by Simons grant 586106.

\end{document}